\documentstyle[12pt]{article}
\begin{document}
\baselineskip=12pt
\pagestyle{plain}
\noindent \today \\
To appear in Advances in Physics \\
\bigskip
\begin{center}
\begin{large}
{\bf THEORY OF PHASE ORDERING KINETICS}
\end{large}
\bigskip
\medskip

A. J. BRAY \\
Department of Theoretical Physics \\
The University, Manchester M13 9PL

\bigskip
\medskip

{\bf ABSTRACT}
\end{center}
\medskip
\noindent The theory of phase ordering dynamics -- the growth of
order through domain coarsening when a system is quenched from the
homogeneous phase into a broken-symmetry phase -- is reviewed,
with the emphasis on recent developments. Interest will focus on
the scaling regime that develops at long times after the quench.
How can one determine the growth laws that describe the time-dependence
of characteristic length scales, and what can be said
about the form of the associated scaling functions? Particular
attention will be paid to systems described by more complicated order
parameters than the simple scalars usually considered, e.g.\ vector
and tensor fields. The latter are needed, for example, to describe
phase ordering in nematic liquid crystals, on which there have been a
number of recent experiments. The study of topological defects
(domain walls, vortices, strings, monopoles) provides a unifying
framework for discussing coarsening in these different systems.

\tableofcontents
\section{INTRODUCTION}
\label{SEC:INTRO}
The theory of phase ordering kinetics, or `domain coarsening' following
a temperature quench from a homogeneous phase into a two-phase region,
has a history going back more than three decades to the pioneering work
of Lifshitz and coworkers \cite{Lifshitz,LS,W}. Since that time, many
excellent reviews have appeared, including those by Gunton et al.\
\cite{Gunton}, Binder \cite{Binder}, Furukawa \cite{Furukawa}, and
Langer \cite{Langer}. In the present article I will not, therefore,
attempt to cover the complete history of the field. Rather, I will
concentrate on some of the recent developments, over the last five years
or so, which in my view are interesting and represent significant advances
or new directions of research, e.g.\ the recent interest in systems with
non-scalar order parameters.
The fundamental concepts and background material necessary for the
understanding and appreciation of these new developments will,
nevertheless, be discussed in some detail. It follows that, while this
article does not aim to be a complete or comprehensive account, it does
aspire to be self-contained and comprehensible to non-experts. By adopting
a fairly pedagogical approach, I hope that the article may also serve as a
useful introduction to the field.

In order to keep the length of the article within reasonable bounds, I
will concentrate primarily on theoretical developments, although important
results from experiment and simulations will be cited as appropriate.
For the same reason, I apologise in advance to all those authors whose work
has not been explicitly discussed.

Systems quenched from a disordered phase into an ordered phase
do not order instantaneously. Instead, the length scale of ordered
regions grows with time as the different broken symmetry phases
compete to select the equilibrium state. To fix our ideas, it is helpful
to consider the simplest, and most familiar, system: the ferromagnetic
Ising model. Figure 1 shows the spontaneous magnetization as a function
of temperature. The arrow indicates a temperature quench, at
time $t=0$, from an initial temperature $T_I$ above the critical point
$T_C$ to a final temperature $T_F$ below $T_C$. At $T_F$ there are two
equilibrium phases, with magnetization $\pm M_0$. Immediately after the
quench, however, the system is in an unstable disordered state corresponding
to equilibrium at temperature $T_I$. The theory of phase ordering kinetics
is concerned with the dynamical evolution of the system from the initial
disordered state to the final equilibrium state.

Part of the fascination of the field, and the reason why it remains a
challenge more than three decades after the first theoretical papers appeared,
is that, in the thermodynamic limit, final equilibrium is never achieved!
This is because the longest relaxation time diverges with the system size
in the ordered phase, reflecting the broken ergodicity. Instead, a network
of domains of the equilibrium phases develops, and the typical length
scale associated with these domains increases with time $t$.
This situation is illustrated in Figure 2, which shows a Monte Carlo
simulation of a two-dimensional Ising model, quenched from $T_I=\infty$
to $T_F=0$. Inspection of the time sequence may persuade the reader
that domain growth is a {\em scaling} phenomenon -- the domain patterns
at later times look statistically similar to those at earlier times, apart
from a global change of scale. This `dynamic scaling hypothesis'
will be formalized below.

For pedagogical reasons, we have introduced domain-growth
in the context of the Ising model, and will continue to use magnetic
language for simplicity. A related phenomenon that has been studied for many
decades, however, by metallurgists, is the spinodal decomposition of binary
alloys, where the late stages of growth are known as `Ostwald ripening'.
Similar phenomena occur in the phase separation of fluids or binary liquids,
although in these cases the phase separation is accelerated by the earth's
gravitational field, which severely limits the temporal duration of the
scaling regime. The gravitational effect can be moderated by using
density-matched binary liquids and/or performing the experiments under
microgravity. All of the above systems, however,
contain an extra complication not present in the Ising ferromagnet.
This is most simply seen by mapping an AB alloy onto an Ising model.
If we represent an A atom by an up spin, and a B atom by a down spin,
then the {\em equilibrium} properties of the alloy can be modelled very
nicely by the Ising model. There is one important feature of the alloy,
however, that is not captured by the Ising model with conventional
Monte-Carlo dynamics. Flipping a single spin in the Ising model
corresponds to converting an A atom to a B atom (or vice versa), which
is inadmissible. The dynamics must conserve the number of A and B atoms
separately, i.e. the magnetization (or `order parameter') of the Ising
model should be {\em conserved}. This will influence the form of the
coarse-grained equation of motion, as discussed in section \ref{SEC:MODELS}
and lead to slower growth than for a non-conserved order parameter.

In all the systems mentioned so far, the order parameter (e.g.\ the
magnetization of the Ising model) is a scalar. In the last few years,
however, there has been increasing interest in systems with more
complex order parameters. Consider, for conceptual simplicity, a
planar ferromagnet, in which the order parameter is a vector confined
to a plane. After a quench into the ordered phase, the magnetization
will point in different directions in different regions of space, and
singular lines (vortex lines) will form at which the direction is not
well defined. These `topological defects' are the analog of domain walls
for the scalar systems. We shall find that, quite generally, an
understanding of the relevant topological defects in the system,
combined with the scaling hypothesis, will take us a long way towards
understanding the forms of the growth laws and scaling functions for
phase ordering in a wide variety of systems.

The article is organised as follows. The following section introduces
most of the important concepts, presents dynamical models appropriate
to the various phase-ordering systems, and analyses these models using
simple physical arguments. Section \ref{SEC:DEFECTS} broadens the
discussion to more general phase-ordering systems, with non-scalar
order parameters, and introduces the key concept of
`topological defects' which, in later sections, will provide
a unifying framework for analytic treatments. Section \ref{SEC:SOLUBLE}
involves a temporary excursion to the realm of exactly soluble models.
These models, although of interest in their own right, unfortunately lack
many of the physical features of more realistic models. In section
\ref{SEC:APPROXSF}, approximate analytical treatments are presented for
the more physical models introduced in sections \ref{SEC:MODELS} and
\ref{SEC:DEFECTS}. Finally, sections \ref{SEC:SHORT}, \ref{SEC:GROWTH} and
\ref{SEC:RG} present some exact results for the short-distance behaviour
and the growth laws for these systems, and make some observations
concerning universality classes for the dynamics of phase ordering.

\bigskip

\section{DYNAMICAL MODELS}
\label{SEC:MODELS}
It is convenient to set up a continuum description in terms
of a coarse-grained order-parameter field (e.g.\ the `magnetization density')
$\phi({\bf x},t)$, which we will initially take to be a scalar field.
A suitable Landau free-energy functional to describe the ordered phase is
\begin{equation}
\label{EQN:HAMILTONIAN}
F[\phi] = \int d^dx\,\left(\frac{1}{2}\,(\nabla \phi)^2 + V(\phi)\right)\ ,
\end{equation}
where the `potential' $V(\phi)$ has a double-well structure, e.g.\
$V(\phi) = (1-\phi^2)^2$. We will take the minima of $V(\phi)$ to
occur at $\phi=\pm 1$, and adopt the convention that $V(\pm 1)=0$.
The potential $V(\phi)$ is sketched in Figure 3. The two minima of
$V$ correspond to the two equilibrium states, while the
gradient-squared term in (\ref{EQN:HAMILTONIAN}) associates an
energy cost to an interface between the phases.

In the case where the order parameter is not conserved, an appropriate
equation for the time evolution of the field $\phi$ is
\begin{eqnarray}
\partial \phi/\partial t & = & -\delta F/\delta \phi \nonumber \\
                         & = & \nabla^2\,\phi - V'(\phi)\ ,
\label{EQN:MODELA}
\end{eqnarray}
where $V'(\phi) \equiv dV/d\phi$. A kinetic coefficient $\Gamma$, which
conventionally multiplies the right-hand side of (\ref{EQN:MODELA}), has
been absorbed ibto the timescale. Eq.\ (\ref{EQN:MODELA}), a simple
`reaction-diffusion' equation, corresponds to simple gradient descent,
i.e.\ the rate of change of $\phi$ is proportional to the gradient
of the free-energy functional in function space.
This equation provides a suitable coarse-grained
description of the Ising model, as well as alloys that undergo an
order-disorder transition on cooling through $T_C$, rather than phase
separating. Such alloys form a two-sublattice structure, with each
sublattice occupied predominantly by atoms of one type. In Ising model
language, this corresponds to antiferromagnetic ordering. The
magnetization is no longer the order parameter, but a `fast mode',
whose conservation does not significantly impede the dynamics of the
important `slow modes'.

When the order parameter is conserved, as in phase separation, a
different dynamics is required. In the alloy system, for example, it is
clear physically that A and B atoms can only exchange locally (not over
large distances), leading to diffusive transport of the order parameter,
and an equation of motion of the form
\begin{eqnarray}
\partial \phi/\partial t & = & \nabla^2\,\delta F/\delta \phi \nonumber \\
                         & = & -\nabla^2\,[\nabla^2\,\phi - V'(\phi)]\ ,
\label{EQN:MODELB}
\end{eqnarray}
which can be written in the form of a continuity equation, $\partial_t \phi
=-\nabla \cdot j$, with current $j = -\lambda \nabla (\delta F/\delta \phi)$.
In (\ref{EQN:MODELB}), we have absorbed the transport coefficient $\lambda$
into the timescale.

Eqs.\ (\ref{EQN:MODELA}) and (\ref{EQN:MODELB}) are sometimes called the
Time-Dependent-Ginzburg-Landau (TDGL) equation and the
Cahn-Hilliard equation respectively. A more detailed
discussion of them in the present context can be found in the lectures
by Langer \cite{Langer}. The same equations with additional
Langevin noise terms on the
right-hand sides are familiar from the theory of critical dynamics, where
they are `model A' and `model B' respectively in the classification of
Hohenberg and Halperin \cite{HH}.

The absence of thermal noise terms in (\ref{EQN:MODELA}) and
(\ref{EQN:MODELB}) indicates that we are effectively working at $T=0$.
A schematic Renormalization Group (RG) flow diagram for
$T$ is given in Figure 4, showing the three RG fixed points at $0$, $T_C$
and $\infty$, and the RG flows. Under coarse-graining, temperatures above
$T_C$ flow to infinity, while those below $T_C$ flow to zero. We therefore
expect the final temperature $T_F$ to be an irrelevant variable (in the
scaling regime) for quenches into the ordered phase. This can be shown
explicitly for systems with a conserved order parameter
\cite{Bray89,Bray90}. For this case the thermal fluctuations at $T_F$
simply renormalize the bulk order parameter and the surface tension
of the domain walls: when the characteristic scale of the domain pattern is
large compared to the domain wall thickness (i.e.\ the bulk correlation length
in equilibrium), the system behaves {\em as if} it were $T=0$, with the
temperature dependence entering through $T$-dependent model parameters.

In a similar way, any short-range correlations present at $T_I$ should be
irrelevant in the scaling regime, i.e.\ all initial temperatures are
equivalent to $T_I=\infty$. Therefore we will take the {\em initial
conditions} to represent a completely disordered state. For example,
one could choose the `white noise' form
\begin{equation}
\langle \phi({\bf x},0)\,\phi({\bf x'},0) \rangle = \Delta\,
\delta({\bf x}-{\bf x'})\ , \nonumber
\label{EQN:SRIC}
\end{equation}
where $\langle \cdots \rangle$ represents an average over an ensemble of
initial conditions, and $\Delta$ controls the size of the initial fluctuations
in $\phi$. The above discussion, however, indicates that the precise form of
the initial conditions should not be important, as long as only short-range
spatial correlations are present.

The challenge of understanding phase ordering dynamics, therefore,
can be posed as finding the nature of the late-time solutions of
deterministic differential equations like (\ref{EQN:MODELA}) and
(\ref{EQN:MODELB}), subject to
random initial conditions. A physical approach to this formal
mathematical problem is based on studying the structure and
dynamics of the topological defects in the field $\phi$. This is
approach that we will adopt. For scalar fields, the topological
defects are just domain walls.


\subsection{THE SCALING HYPOTHESIS}
\label{SEC:SCALINGHYP}
Although originally motivated by experimental and simulation
results for the structure factor and pair correlation function
\cite{BS,Marro,Furu78,Furu79}
for ease of presentation it is convenient to introduce the scaling
hypothesis first, and then discuss its implications for growth laws and
scaling functions. Briefly, the scaling hypothesis states that there exists,
at late times, a single characteristic length scale $L(t)$ such that
the domain structure is (in a statistical sense) independent of time
when lengths are scaled by $L(t)$. It should be stressed that scaling
has not been proved, except in some simple models such as the one-dimensional
Glauber model\cite{Glauber} and the $n$-vector model with $n = \infty$
\cite{CZ}. However, the evidence in its favour is compelling
(see, e.g., Figure 5).

We shall find, in section \ref{SEC:GROWTH},
that the scaling hypothesis, together with
a result derived in section \ref{SEC:SHORT} for the tail of the structure
factor, is sufficient to determine the form of $L(t)$ for most cases of
interest.

Two commonly used probes of the domain structure are the equal-time
pair correlation function
\begin{equation}
C({\bf r},t) = \langle \phi({\bf x}+{\bf r},t)\,\phi({\bf x},t)\rangle\ ,
\end{equation}
and its Fourier transform, the equal-time
structure factor,
\begin{equation}
S({\bf k},t) = \langle \phi_{\bf k}(t)\,\phi_{\bf -k}(t) \rangle\ .
\end{equation}
Here angle brackets indicate an average over initial conditions.
The structure factor can, of course, be measured in scattering experiments.
The existence of a single characteristic length scale, according to the
scaling hypothesis, implies that the pair correlation function and the
structure factor have the scaling forms
\begin{eqnarray}
C({\bf r},t) & = & f(r/L)\ ,\nonumber \\
S({\bf k},t) & = & L^d\,g(kL)\ ,
\label{EQN:STRUCT}
\end{eqnarray}
where d is the spatial dimensionality, and $g(y)$ is the Fourier transform
of $f(x)$. Note that $f(0)=1$, since (at $T=0$) there is perfect order
within a domain.

At general temperatures $T<T_c$, $f(0) = M^2$, where $M$
is the equilibrium value of the order parameter. (Note that the {\em scaling
limit} is defined by $r \gg \xi$, $L \gg \xi$, with $r/L$ arbitrary, where
$\xi$ is the equilibrium correlation length). Alternatively, we can
extract the factor $M^2$ explicitly by writing  $C({\bf r},t)=M^2\,f(r/L)$.
The statement that $T$ is irrelevant then amounts to asserting that any
remaining temperature dependence can be absorbed into the domain scale $L$,
such that the function $f(x)$ is independent of $T$.

The scaling forms (\ref{EQN:STRUCT}) are well supported by simulation data
and experiment. As an example, Figure 5 shows the  scaling plot for $f(x)$
for the 2-D Ising model, with $x=r/t^{1/2}$.

For future reference, we note that the different-time correlation function,
defined by $C({\bf r},t,t') =
\langle \phi({\bf x}+{\bf r},t)\,\phi({\bf x},t') \rangle$,
can also be written in scaling form. A simple generalization of
(\ref{EQN:STRUCT}) gives \cite{Furukawa89,Furukawa89a}
\begin{equation}
C({\bf r},t,t') = f(r/L,r/L')\ ,
\label{EQN:TWOTIME}
\end{equation}
where $L$, $L'$ stand for $L(t)$ and $L(t')$. Especially interesting is the
limit $L \gg L'$, when (\ref{EQN:TWOTIME}) takes the form
\begin{equation}
C({\bf r},t,t') \to (L'/L)^{\bar{\lambda}}\,h(r/L)\ ,\ \ \ \ \ L \gg L'\ ,
\label{EQN:lambdabar}
\end{equation}
where the exponent $\bar{\lambda}$,
first introduced by Fisher and Huse in the
context of non-equilibrium relaxation in spin glasses \cite{FH88},
is a non-trivial
exponent associated with phase ordering kinetics \cite{NB90}.
It has recently been measured in an experiment on twisted nematic
liquid crystal films \cite{Mason93}. The {\em autocorrelation}
function, $A(t) = C({\bf 0},t,t')$ is therefore a function only of the
ratio $L'/L$, with $A(t) \sim (L'/L)^{\bar{\lambda}}$ for $L \gg L'$.

In the following sections, we explore the forms of the scaling functions
in more detail. For example, the linear behaviour of $f(x)$, for small
scaling variable $x$ in Figure 5, is a generic feature for scalar fields,
both conserved and non-conserved. We shall see that it is a simple
consequence of the existence of `sharp' (in a sense to be clarified),
well-defined domain walls in the system. A corollary that we shall
demonstrate  is that the structure factor scaling function $g(y)$
exhibits a power-law tail, $g(y) \sim y^{-(d+1)}$ for $y \gg 1$, a
result known as `Porod's law' \cite{Porod51,Debye}.
In section \ref{SEC:GROWTH}  we shall show that
this result, and its generalization to more complex fields, together with
the scaling hypothesis, are sufficient to determine the growth law
for $L(t)$.

\subsection{DOMAIN WALLS}
\label{SEC:WALLS}
It is instructive to first look at the properties of a flat
equilibrium domain wall. From (\ref{EQN:MODELA}) the wall profile
is the solution of the equation
\begin{equation}
d^2 \phi/dg^2 = V'(\phi)\ , \label{EQN:PROFILE}
\end{equation}
with boundary conditions $\phi(\pm \infty) = \pm 1$, where $g$ is a
coordinate normal to the wall. We can fix the `centre' of the wall
(defined by $\phi=0$) to be at $g=0$ by the extra condition $\phi(0)=0$.
Integrating (\ref{EQN:PROFILE}) once,
and imposing the boundary conditions, gives
$(d\phi/dg)^2 = 2V(\phi)$. This result can be used in
(\ref{EQN:HAMILTONIAN}) to give the
energy per unit area of wall, i.e.\ the surface tension, as
\begin{equation}
\sigma = \int_{-\infty}^\infty dg\,(d\phi/dg)^2
= \int_{-1}^{1}d\phi\,\sqrt{2V(\phi)}\ .
\label{EQN:SIGMA}
\end{equation}
Note that, for scalar fields, the two terms in (\ref{EQN:HAMILTONIAN})
contribute equally to the wall energy.

The profile function $\phi(g)$ is sketched in Figure 6.
For $g \to \pm\infty$, linearizing
(\ref{EQN:PROFILE}) around $\phi=\pm 1$ gives
\begin{equation}
1 \mp \phi \sim \exp(-[V''(\pm 1)]^{1/2}|g|)\ ,\ \ \ \ g \to \pm\infty\ ,
\label{EQN:PROFILETAIL}
\end{equation}
i.e.\ the order parameter saturates exponentially fast away from the
walls. It follows that the excess energy is localized in the domain walls,
and that the driving force for the domain growth is the wall curvature,
since the system energy can only decrease through a reduction in the
total wall area. The growth mechanism is rather different, however, for
conserved and nonconserved fields.

\medskip

\subsection{NONCONSERVED FIELDS: THE ALLEN-CAHN EQUATION}
\label{SEC:AC}
The existence of a surface tension implies a force per unit
area, proportional to the mean curvature, acting at each point on the
wall. The calculation is similar to that of the excess pressure inside
a bubble. Consider, for example, a spherical domain of radius $R$, in
three dimensions. If the force per unit area is $F$, the work done by
the force in decreasing the radius by $dR$ is $4\pi F R^2 dR$. Equating
this to the decrease in surface energy, $8\pi\sigma RdR$, gives
$F=2\sigma/R$. For model A dynamics, this force will cause the walls to
move, with a velocity proportional to the local curvature. If the
friction constant for domain-wall motion is $\eta$, then this
argument gives $\eta\,dR/dt = -2\sigma/R$. For general dimension $d$,
the factor `2' on the right is replaced by $(d-1)$.

It is interesting to see how this result arises directly from the equation
of motion (\ref{EQN:MODELA}). We consider a single spherical domain of
(say) $\phi=-1$ immersed in a sea of $\phi=+1$.
Exploiting the spherical symmetry, (\ref{EQN:MODELA}) reads
\begin{equation}
\frac{\partial \phi}{\partial t} = \frac{\partial^2 \phi}{\partial r^2} +
\frac{d-1}{r}\frac{\partial \phi}{\partial r} - V'(\phi)\ .
\label{EQN:SPHERE}
\end{equation}
Provided the droplet radius $R$ is much larger than the interface width
$\xi$, we expect a solution of the form
\begin{equation}
\phi(r,t) = f\left(r - R(t)\right)\ .
\end{equation}
Inserting this in (\ref{EQN:SPHERE}) gives
\begin{equation}
0 = f'' + [(d-1)/r + dR/dt]f' - V'(f)\ .
\label{EQN:SPHERE2}
\end{equation}
The function $f(x)$ changes from -1 to 1 in a small region of width $\xi$
near $x=0$. It's derivative is, therefore, sharply peaked near $x=0$ (i.e.\
near $r=R(t)$). Multiplying (\ref{EQN:SPHERE2}) by $f'$ and integrating
through the interface gives
\begin{equation}
0 = (d-1)/R + dR/dt\ ,
\label{EQN:RADIUS}
\end{equation}
where we have used $f'=0$ far from the interface, and $V(f)$ has the same
value on both sides of the interface (in the absence of a bulk driving
force, i.e.\ a magnetic field). Integrating (\ref{EQN:RADIUS}) gives
$R^2(t) = R^2(0) - 2(d-1)t$, i.e.\ the collapse time scales with the
initial radius as $t \sim R^2(0)$. Equation (\ref{EQN:RADIUS}) is identical
to our previous result obtained by considering the surface tension as
the driving force, provided the surface tension $\sigma$
and friction constant $\eta$ are equal. This we show explicitly below.

The result for general curved surfaces was derived by Allen and
Cahn \cite{AC}, who noted that, close to a domain wall, one can write
$\nabla \phi = (\partial\phi/\partial g)_t\,\hat{g}$,
where $\hat{g}$ is a unit vector normal to the wall
(in the direction of increasing $\phi$), and so
$\nabla^2 \phi = (\partial^2\phi/\partial g^2)_t
+ (\partial\phi/\partial g)_t\,\nabla \cdot \hat{g}$. Noting also the
relation $(\partial \phi/\partial t)_g
= - (\partial \phi/\partial g)_t\,(\partial g/\partial t)_\phi$,
(\ref{EQN:MODELA}) can
be recast as
\begin{equation}
- (\partial \phi/\partial g)_t\,(\partial g/\partial t)_\phi
= (\partial\phi/\partial g)_t\,\nabla \cdot \hat{g}
+ (\partial^2\phi/\partial g^2)_t - V'(\phi)\ .
\label{EQN:AC1}
\end{equation}
Assuming that, for gently curving walls, the wall profile is given by the
equilibrium condition (\ref{EQN:PROFILE}), the final two terms in
(\ref{EQN:AC1}) cancel. Noting also
that $(\partial g/\partial t)_\phi$ is just the wall velocity $v$ (in the
direction of increasing $\phi$), (\ref{EQN:AC1}) simplifies to
\begin{equation}
v = -\nabla \cdot \hat{g} = -K\ ,
\label{EQN:AC}
\end{equation}
the `Allen-Cahn equation', where $K \equiv \nabla \cdot \hat{g}$ is
$(d-1)$ times the mean curvature. For brevity, we will call $K$ simply
the `curvature'. An alternative derivation of (\ref{EQN:AC}) follows the
approach used for the spherical domain, i.e.\ we multiply
Eq.\ (\ref{EQN:AC1}) by $(\partial \phi/\partial g)_t$ and integrate
(with respect to $g$) through the interface. This gives the same result.

Equation (\ref{EQN:AC}) is an important result,
because it establishes that the motion
of the domain walls is determined (for non-conserved fields) purely by the
local curvature. In particular, the detailed shape of the potential is not
important: the main role of the double-well potential $V(\phi)$ is to
establish (and maintain) well-defined domain walls. (Of course, the
well depths must be equal, or there would be a volume driving force).
We shall exploit this insensitivity to the potential, by choosing a
particularly convenient form for $V(\phi)$, in section \ref{SEC:APPROXSF}.

For a spherical domain, the curvature $K$ is $(d-1)/R$, and
(\ref{EQN:AC}) reduces to (\ref{EQN:RADIUS}).
Our explicit treatment of the spherical domain verifies the
Allen-Cahn result, and, in particular, the independence from the
potential of the interface dynamics.

A second feature of (\ref{EQN:AC}) is that the surface tension $\sigma$
(which {\em does} depend on the potential) does not explicitly appear. How
can this be, if the driving force on the walls contains a factor $\sigma$?
The reason, as we have already noted, is that one also needs to consider
the {\em friction constant} per unit area of wall, $\eta$.
The equation of motion for the walls in this dissipative system is
$\eta v = -\sigma K$. Consistency with (\ref{EQN:AC})
requires $\eta = \sigma$. In fact, $\eta$ can be calculated independently,
as follows. Consider a plane wall moving uniformly (under the influence
of some external driving force) at speed $v$. The rate of energy
dissipation per unit area is
\begin{eqnarray}
dE/dt & = & \int_{-\infty}^\infty dg\,\frac{\delta F}{\delta \phi}\,
\frac{\partial \phi}{\partial t} \nonumber \\
      & = & -\int_{-\infty}^\infty dg\,
\left(\frac{\partial \phi}{\partial t}\right)^2\ ,
\label{EQN:DISS1}
\end{eqnarray}
using (\ref{EQN:MODELA}).
The wall profile has the form $\phi(g,t) = f(g-vt)$, where
the profile function $f$ will, in general, depend on $v$. Putting this
form into (\ref{EQN:DISS1}) gives
\begin{equation}
dE/dt  =  -v^2 \int dg\,(\partial \phi/\partial g)^2 = -\sigma v^2\ ,
\label{EQN:DISS2}
\end{equation}
where the definition (\ref{EQN:SIGMA}) of the surface tension $\sigma$
was used in the final step, and the profile function $f(x)$
replaced by its $v=0$ form to lowest order in $v$. By definition,
however, the rate of energy dissipation is the product of the frictional
force $\eta v$ and the velocity, $dE/dt = -\eta v^2$. Comparison with
(\ref{EQN:DISS2}) gives $\eta = \sigma$. We conclude that,
notwithstanding some contrary suggestions in the literature,
the Allen-Cahn equation is completely consistent with the idea that
domain growth is driven by the surface tension of the walls.

\subsection{CONSERVED FIELDS}
For conserved fields the interfaces cannot move independently. At late
times the dominant growth mechanism is the transport of the order
parameter from interfaces of high curvature to regions of low curvature
by diffusion through the intervening bulk phases. To see how this
works, we first linearize (\ref{EQN:MODELB}) in one of the bulk phases,
with say $\phi \simeq 1$. Putting $\phi = 1 + \tilde{\phi}$ in
(\ref{EQN:MODELB}), and linearizing in $\tilde{\phi}$, gives
\begin{equation}
\partial \tilde{\phi}/\partial t = -\nabla^4 \tilde{\phi}
+ V''(1) \nabla^2 \tilde{\phi}\ .
\label{EQN:BULKLINEAR}
\end{equation}
Since the characteristic length scales are large at late times, the
$\nabla^4$ term is negligible, and (\ref{EQN:BULKLINEAR}) reduces to
the diffusion equation, with diffusion constant $D=V''(1)$. The
interfaces provide the boundary conditions, as we shall see. However,
we can first make a further simplification. Due to the conservation
law, the interfaces move little during the time it takes the diffusion
field $\tilde{\phi}$ to relax. If the characteristic domain size is $L$,
the diffusion field relaxes on a time scale $t_D \sim L^2$. We shall see
below, however, that a typical interface velocity is of order $1/L^2$,
so the interfaces only move a distance of order unity (i.e. much less
than $L$) in the time $t_D$. This means that the diffusion field relaxes
quickly compared to the rate at which the interfaces move, and is essentially
always in equilibrium with the interfaces. The upshot is that the diffusion
equation can be replaced by Laplace's equation, $\nabla^2 \tilde{\phi}=0$,
in the bulk.

To derive the boundary conditions at the interfaces, it is convenient to
work, not with $\tilde{\phi}$ directly, but with the chemical potential
$\mu \equiv \delta F/\delta\phi$. In terms of $\mu$, (\ref{EQN:MODELB})
can be written as a continuity equation,
\begin{eqnarray}
\partial \phi/\partial t  & = & - \nabla \cdot j \label{EQN:CONTINUITY}\\
                       j  & = & - \nabla \mu  \label{EQN:CURRENT}\\
                     \mu  & = & V'(\phi) - \nabla^2 \phi\ .
\label{EQN:mu}
\end{eqnarray}
In the bulk, $\mu$ and $\tilde{\phi}$ are proportional to each other,
because (\ref{EQN:mu}) can be linearized to give
$\mu = V''(1)\tilde{\phi} - \nabla^2 \tilde{\phi}$, and the $\nabla^2$
term is again negligible. Therefore $\mu$ also obeys Laplace's equation,
\begin{equation}
\nabla^2 \mu = 0\ ,
\label{EQN:LAPLACE}
\end{equation}
in the bulk.

The boundary conditions are derived by analysing (\ref{EQN:mu}) near an
interface. As in the derivation of the Allen-Cahn equation, we
consider surfaces of constant $\phi$ near the interface and introduce
a Cartesian coordinate system at each point, with a coordinate $g$ normal
to the surface (and increasing with increasing $\phi$). Then (\ref{EQN:mu})
becomes (compare Eq.\ (\ref{EQN:AC1}),
\begin{equation}
\mu = V'(\phi) - (\partial \phi/\partial g)_t\,K
               - (\partial^2\phi/\partial g^2)_t\
\end{equation}
near the interface, where $K=\nabla \cdot \hat{g}$ is the curvature.
The value of $\mu$ at the interface can be obtained (just as in our
treatment of the spherical domain in section \ref{SEC:AC}), by
multiplying through by $(\partial \phi/\partial g)_t$, which is sharply
peaked at the interface, and integrating over $g$ through the interface.
Noting that $\mu$ and $K$ vary smoothly through the interface, this gives
the completely general relation
\begin{equation}
\mu\Delta\phi = \Delta V-\sigma K\
\label{EQN:GT1}
\end{equation}
at the interface, where $\Delta \phi$ is the change in $\phi$ across
the interface, and $\Delta V$ is the difference in the
minima of the potential for the two bulk phases.
In deriving (\ref{EQN:GT1}), we have used
$(\partial \phi/\partial g)_t \to 0$ far from the interface, and made the
identification $\int dg (\partial \phi/\partial g)_t^2 = \sigma$, as in
(\ref{EQN:SIGMA}), with $\sigma$ the surface tension. Simplifying to the
case where the minima have equal depth (we shall see that the general case
introduces no new physics), and taking the minima to be at $\phi=\pm 1$ as
usual, gives $\Delta V=0$ and $\Delta \phi=2$. Then (\ref{EQN:GT1}) becomes
\begin{equation}
\mu = -\sigma K/2\ .
\label{EQN:GT2}
\end{equation}
This (or, more generally, Eq.\ (\ref{EQN:GT1})) is usually known as the
Gibbs-Thomson boundary condition. Note that we have assumed that the order
parameter takes its equilibrium value ($\pm 1$) in both bulk phases.
This is appropriate to the late stages of growth in which we are primarily
interested.

To summarize, (\ref{EQN:GT2}) determines $\mu$ on the interfaces in terms
of the curvature. Between the interfaces, $\mu$ satisfies the Laplace
equation (\ref{EQN:LAPLACE}). The final step is to use (\ref{EQN:CURRENT})
to determine the motion of the interfaces. An interface moves with a
velocity given by the imbalance between the current flowing into and
out of it:
\begin{equation}
v\,\Delta\phi = j_{\rm out} - j_{\rm in} = -[\partial \mu/\partial g]
  = -[\hat{g}\cdot\nabla \mu]\ ,
\label{EQN:VELOCITY}
\end{equation}
where $v$ is the speed of the interface in the direction of increasing
$\phi$, $g$ is the usual coordinate normal to interface, $[\cdots]$
indicates the discontinuity across the interface,  and we have
assumed as usual that $\phi \simeq \pm 1$ in the bulk phases.

To illustrate how (\ref{EQN:LAPLACE}), (\ref{EQN:GT2}) and
(\ref{EQN:VELOCITY}) are used, we consider again the case of a single
spherical domain of negative phase ($\phi=-1$) in an infinite sea of positive
phase ($\phi=+1$). We restrict ourselves to $d=3$ for simplicity.
The definition of $\mu$, Eq.\ (\ref{EQN:mu}), gives $\mu=0$
at infinity. Let the domain have radius $R(t)$. The solution of
(\ref{EQN:LAPLACE}) that obeys the boundary conditions $\mu=0$
at infinity and (\ref{EQN:GT2}) at $r=R$, and respects the spherical
symmetry is (using $K=2/R$ for $d=3$)
$\mu=-\sigma/r$ for $r \ge R$.
Inside the domain, the $1/r$ term must be absent to avoid an unphysical
singularity at $r=0$. The solution of (\ref{EQN:LAPLACE}) in this region
is therefore $\mu={\rm const}$. The boundary condition (\ref{EQN:GT2})
gives $\mu=-\sigma/R$.

To summarize,
\begin{eqnarray}
\mu & = &  - \sigma/R\ ,\ \ \ \ \ \ \ r \le R \nonumber \\
    & = &  - \sigma/r\ ,\ \ \ \ \ \ \ r \ge R\ .
\end{eqnarray}
Using (\ref{EQN:VELOCITY}), with $\Delta\phi=2$,  then gives
\begin{equation}
dR/dt = v = - \frac{1}{2}[\partial \mu/\partial r]^{R+\epsilon}_{R-\epsilon}
          = -\sigma/2R^2\ ,
\label{EQN:COPRADIUS}
\end{equation}
and hence $R^3(t) = R^3(0) - 3\sigma t/2$. We conclude that a domain of
initial radius $R(0)$ evaporates in a time proportional to $R^3(0)$.
This contrasts with the $R^2(0)$ result obtained for a non-conserved
order parameter. In the non-conserved case, of course, the domain
simply shrinks under the curvature forces, whereas for the conserved
case it evaporates by the diffusion of material to infinity.

We now briefly discuss the case where the potential minima have unequal
depths, as sketched in Figure 7. Consider first a planar interface
separating the two equilibrium phases, with order parameter values
$\phi_1$ and $\phi_2$. Since no current flows, ${\bf j} = -\nabla\mu=0$
gives $\mu=constant$. From the definition (\ref{EQN:mu}) of $\mu$, and
the fact that $\nabla^2\phi$ vanishes far from the interface, it follows that
$\mu = V'(\phi_1) = V'(\phi_2)$. On the other hand, the Gibbs-Thomson
boundary condition (\ref{EQN:GT1}) for a flat interface ($K=0$) gives
$\mu = \Delta V/\Delta\phi$. Combining these two results gives
\begin{equation}
V'(\phi_1) = V'(\phi_2) = \Delta V/\Delta\phi\ ,
\end{equation}
leading to the common tangent construction, shown in Figure 7, that
determines $\phi_1$ and $\phi_2$ as the points where the common tangent
touches the potential. If one now repeats the calculation for a spherical
drop, with a domain with $\phi=\phi_1$ immersed in a sea with $\phi=\phi_2$,
one obtains the equation of motion for the radius,
$dR/dt = -2\sigma/(\Delta\phi)^2 R^2$, a simple generalisation of
(\ref{EQN:COPRADIUS}). Henceforth, we will consider only the case of
degenerate minima.

\subsection{GROWTH LAWS}
\label{SEC:L(t)}
The scaling hypothesis suggests a simple intuitive derivation of
the `growth laws' for $L(t)$, which are really just generalizations
of the calculations for isolated spherical domains.
For model A, we can estimate both sides of
the Allen-Cahn equation (\ref{EQN:AC}) as follows.
If there is a single characteristic scale
$L$, then the wall velocity $v \sim dL/dt$, and the curvature $K \sim 1/L$.
Equating and integrating gives $L(t) \sim t^{1/2}$ for non-conserved
scalar fields.

For conserved fields (model B), the argument is slightly more subtle.
We shall follow the approach of Huse \cite{Huse86}.
{}From (\ref{EQN:GT2}), the chemical potential
has a typical value $\mu \sim \sigma/L$ on
interfaces, and varies over a length scale of order $L$. The current,
and therefore the interface velocity $v$, scale as
$\nabla \mu \sim \sigma/L^2$, giving $dL/dt \sim \sigma/L^2$ and
$L(t) \sim (\sigma t)^{1/3}$. A  more compelling argument for this
result will be given in section \ref{SEC:GROWTH}. We note, however, that
the result is supported by evidence from computer simulations
\cite{Huse86,Sims} (which usually require , however, some extrapolation
into the asymptotic scaling regime) as well as a Renormalization
Group (RG) treatment \cite{Bray89,Bray90}. In the limit that one phase occupies
an infinitesimal volume fraction, the original Lifshitz-Slyozov-Wagner
theory convincingly demonstrates a $t^{1/3}$ growth. This calculation,
whose physical mechanism is the evaporation of material (or magnetization)
from small droplets and condensation onto larger droplets, will be discussed
briefly in the following subsection.

It is interesting that these growth laws can also be obtained using naive
arguments based on the results for single spherical domains \cite{Langer}.
For nonconserved dynamics, we know that a domain of radius $R$ collapses
in a time of order $R^2$. Therefore, crudely speaking, after time $t$ there
will be no domains smaller than $t^{1/2}$, so the characteristic domain
size is $L(t) \sim t^{1/2}$. Of course, this is an oversimplification, but
it captures the essential physics. For conserved dynamics, the same line
of argument leads to $t^{1/3}$ growth. In fact, this approach can be
used rather generally, for a variety of systems \cite{BRunpub}, and gives
results which agree, in nearly all cases, with the exact growth laws that
will be derived in section \ref{SEC:GROWTH}.

\subsection{THE LIFSHITZ-SLYOZOV-WAGNER THEORY}
\label{SEC:LSW}
In their seminal work, Lifshitz and Slyozov, and independently Wagner,
derived some exact results in the limit that the minority phase occupies
a negligible volume fraction \cite{LS,W}. in particular, they showed that
the characteristic size of the minority phase droplets increases like
$t^{1/3}$.

We begin by considering again a single spherical droplet of minority phase
($\phi=-1$), of radius $R$, immersed in a sea of majority phase, but now we
let the majority phase have order parameter $\phi=\phi_0 < 1$ at
infinity, i.e.\ the majority phase is `supersaturated'
with the dissolved minority species. If the minority droplet is large enough
it will grow by absorbing material from the majority phase. Otherwise it will
shrink by evaporation as before. A `critical radius' $R_c$ separates these two
regimes.

With the convention that $V(\pm 1)=0$, the boundary condition (\ref{EQN:GT1})
at $r=R$ becomes $(1+\phi_0)\mu = V(\phi_0) -2\sigma/R$, while the boundary
condition at infinity is $\mu = V'(\phi_0)$. Solving the Laplace equation
for $\mu$ with these boundary conditions gives
\begin{eqnarray}
\mu & = & V'(\phi_0) + \left(\frac{V(\phi_0)}{1+\phi_0} - V'(\phi_0)\right)\,
\frac{R}{r} - \frac{2\sigma}{1+\phi_0}\,\frac{1}{r}\ ,\ \ \ \ \ r \ge R\ , \\
& = & \frac{V(\phi_0)}{1+\phi_0} - \frac{2\sigma}{1+\phi_0}\,\frac{1}{R}\ ,
\ \ \ \ \ r \le R\ .
\end{eqnarray}
Eq.\ (\ref{EQN:VELOCITY}) gives the interface velocity,
\begin{equation}
\frac{dR}{dt} = \left(\frac{V(\phi_0)}{(1+\phi_0)^2}
- \frac{V'(\phi_0)}{1+\phi_0}\right)\,\frac{1}{R}
- \frac{2\sigma}{(1+\phi_0)^2}\,\frac{1}{R^2}\ .
\end{equation}

Now consider the limit of small supersaturation, $\phi_0 = 1-\epsilon$
with $\epsilon \ll 1$. To leading non-trivial order in $\epsilon$, the
velocity is
\begin{equation}
\frac{dR}{dt} = \frac{\sigma}{2R}\,\left(\frac{1}{R_c} - \frac{1}{R}\right)\ ,
\label{EQN:LSW1}
\end{equation}
where $R_c = \sigma/V''(1)\epsilon$ is the critical radius.

In the Lifshitz-Slyozov-Wagner (LSW) theory, an assembly of drops is
considered. Growth proceeds by evaporation of drops with $R<R_c$ and
condensation on to drops with $R>R_c$.
The key idea is to use the time-dependent
supersaturation $\epsilon(t)$ as a kind of mean-field, related to the
time-dependent critical radius via $R_c(t) = \sigma/V''(1)\epsilon(t)$, and
to use (\ref{EQN:LSW1}) with the time-dependent $R_c$ for the dynamics
of a given drop.

So far the discussion has been restricted to spatial dimension $d=3$.
However, a result of the form (\ref{EQN:LSW1}) can be derived (with a
$d$-dependent numerical constant multipying the right-hand side) for general
$d>2$. The next step is to write down a scaling distribution of droplet radii,
\begin{equation}
n(R,t) = \frac{1}{R_c^{d+1}}\,f\left(\frac{R}{R_c}\right)\ ,
\label{EQN:LSW2}
\end{equation}
obeying the continuity equation
\begin{equation}
\frac{\partial n}{\partial t} + \frac{\partial}{\partial R}
\left( v(R) n(R) \right) = 0\ ,
\label{EQN:LSW3}
\end{equation}
where $v(R)$ is just the velocity $dR/dt$.

Suppose the spatial average of the order parameter is $1-\epsilon_0$. At late
times the supersaturation $\epsilon(t)$ tends to zero, giving the constraint
\begin{equation}
\epsilon_0 = 2V_d \int_0^\infty dR\,R^d n(R,t)
  = 2V_d \int_0^\infty dx\,x^d f(x)\ ,
\label{EQN:LSW4}
\end{equation}
where $V_d$ is the volume of the $d$-dimensional unit sphere.
Eq.\ (\ref{EQN:LSW4}) fixes the normalisation of $f(x)$. A linear equation for
$f(x)$ can be derived by inserting the scaling form (\ref{EQN:LSW2}) into
the continuity equation (\ref{EQN:LSW3}) and using (\ref{EQN:LSW1}) for
$v(R)$, which we write in the form (valid for general $d>2$)
\begin{equation}
v(R) = \frac{\alpha_d}{R}\left(\frac{1}{R_c}-\frac{1}{R}\right)\ ,
\label{EQN:LSWv}
\end{equation}
where $\alpha_3 = \sigma/2$. This procedure gives
\begin{equation}
\frac{\dot{R_c}}{R_c^{d+2}}\left\{(d+1)f + xf'\right\}
 = \frac{\alpha_d}{R_c^{d+4}}\left\{\left(\frac{2}{x^3} - \frac{1}{x^2}\right)f
   + \left(\frac{1}{x}-\frac{1}{x^2}\right)f'\right\}\ ,
\label{EQN:LSW5}
\end{equation}
where $\dot{R_c} \equiv dR_c/dt$ and $f' \equiv df/dx$. A consistent solution
requires that the $R_c$-dependence drop out from this equation. This means
that $R_c^2\dot{R_c} = \alpha_d \gamma$, a constant, giving
\begin{equation}
R_c(t) = (3\alpha_d\gamma t)^{1/3}\ ,
\label{EQN:LSWR_c}
\end{equation}
Integrating (\ref{EQN:LSW5}) then gives
\begin{equation}
\ln f(x) = \int^x \frac{dy}{y}\,\frac{2-y-\gamma(d+1)y^3}{\gamma y^3 -y +1}\ .
\label{EQN:LSWint}
\end{equation}
It is clear that $f(x)$ cannot be non-zero for arbitrarily large $x$, or
one would have the asymptotic behaviour $f(x) \sim x^{-(d+1)}$, and the
normalisation integral (\ref{EQN:LSW4}) would not exist. Therefore $f(x)$
must vanish for $x$ greater than some value $x_{max}$, which must be the
first pole of the integrand on the positive real axis. The existence of
such a pole requires $\gamma \le 4/27$. Lifshitz and Slyozov argue that
the only physically acceptable solution is $\gamma = \gamma_0 = 4/27$,
corresponding to a double pole at $x_{max}=3/2$. The argument is as
follows. In terms of the scaled radius $x=R/R_c$, Eqs.\ (\ref{EQN:LSWv})
and (\ref{EQN:LSWR_c}) imply
\begin{eqnarray}
\frac{dx}{dt} & = & \frac{1}{3\gamma t}\,\left(\frac{1}{x}
- \frac{1}{x^2} - \gamma x \right) \nonumber \\
 & = & \frac{1}{3\gamma t}\,g(x)\ ,
\label{EQN:xflow}
\end{eqnarray}
the last equality defining the function $g(x)$. The form of $g(x)$ is
sketched in Figure 8, where the arrows indicate the flow of $x$ under
the dynamics (\ref{EQN:xflow}).  From Figure 8(a) it is clear that
for $\gamma < \gamma_0$ all drops with $x>x_1$ will asymptotically
approach the size $x_2 R_c(t)$, which tends to $\infty$ with $t$.
Therefore it will not be possible to satisfy the condition
(\ref{EQN:LSW4}) which imposes the conservation of the order parameter.
Similarly, Figure 8(c) shows that for $\gamma > \gamma_0$, all points
move to the origin and the conservation condition again cannot be
satisfied. The only possibility is that $\gamma$ tends to $\gamma_0$
asymptotically from above (it cannot reach $\gamma_0$ in finite time,
otherwise all drops with $x>3/2$ would eventually arrive at $x=3/2$ and
become stuck, much like the case $\gamma < \gamma_0$).

Evaluating the integral (\ref{EQN:LSWint}) with $\gamma=\gamma_0=4/27$
gives the scaling function for the droplet size distribution:
\begin{equation}
f(x) = const.\, x^2 \left(3+x\right)^{-1-4d/9}
\left(\frac{3}{2} - x\right)^{-2-5d/9}\exp\left(-\frac{d}{3-2x}\right)
\label{EQN:LSW_f}
\end{equation}
for $x<3/2$, and $f(x) =0$ for $x \ge 3/2$. The constant prefactor is fixed
by the normalisation integral (\ref{EQN:LSW4}).

Lifshitz and Slyozov have shown that the above scaling solution is obtained
for generic initial conditions in the limit of small volume fraction $v$
of the minority phase. The general-$d$ form (\ref{EQN:LSW_f}) for $f(x)$
has been derived by Yao et al.\ \cite{Yao}. Note that the method described
above only works for $d>2$: it is easy to show that the constant $\alpha_d$
in (\ref{EQN:LSWv}), which sets the time scale for the growth via
(\ref{EQN:LSWR_c}), vanishes linearly with $(d-2)$ for $d \to 2$,
reflecting the singular nature of the Green's function for the Laplacian
in $d=2$. The general expression is $\alpha_d = (d-1)(d-2)\sigma/4$.
Working in the limit of strictly vanishing $v$, Rogers and Desai
\cite{RogersDesai} found the same scaling form (\ref{EQN:LSW_f})
for $d=2$, but with $R_c \sim (t/\ln t)^{1/3}$. For small, but non-zero
$v$, Yao et al.\ \cite{Yao} find $R_c \sim (t/|\ln v|)^{1/3}$.
The two results correspond to taking the limit $v \to 0$ before
\cite{RogersDesai} or after \cite{Yao} the limit $t \to \infty$.

Many groups have attempted, with varying degrees of success, to extend
the LSW treatment to non-zero $v$, either by expanding in $v$ (actually,
$\sqrt{v}$)
\cite{Marqusee,Tokuyama}, or by the use of physically motivated
approximation schemes and/or numerical methods
\cite{Yao,Voorhees,Beenakker,Ardell,Tsumuraya,Brailsford,Marder}.
When $v$ is of order unity, however, such that both phases are continuous,
different techniques are required, discussion of which will be postponed to
section \ref{SEC:APPROXSF}.

\subsection{BINARY LIQUIDS}
\label{SEC:BINLIQS}
The phase separation of binary liquids is a phenomenon of considerable
experimental interest. Model B is inappropriate for this system, since
it takes no account of the transport of the order parameter by
hydrodynamic flow. Here we briefly review the modifications to model B
needed to describe binary liquids.

The principal new ingredient is `advection' of the order parameter by
the fluid. The appropriate modification of (\ref{EQN:MODELB}) is
\begin{equation}
\partial \phi/\partial t + {\bf v}\cdot\nabla\phi = \lambda \nabla^2 \mu\ ,
\end{equation}
where ${\bf v}$ is the (local) fluid velocity, and we have reinstated
the transport coefficient $\lambda$. The velocity obeys the Navier-Stokes
equation which, with the simplification that the
fluid is incompressible, reads
\begin{equation}
\rho\,\left(\frac{\partial{\bf v}}{\partial t}
+ ({\bf v}\cdot\nabla){\bf v}\right) = \eta \nabla^2 {\bf v} - \nabla p
-\phi \nabla \mu\ ,
\label{EQN:NS}
\end{equation}
where $p$ is the pressure, $\eta$ the viscosity, and the density $\rho$ is
constant. The final term in (\ref{EQN:NS}) arises from the free energy
change per unit volume $\phi\,\delta\mu$ that accompanies the transport
of a fluid region with order parameter $\phi$ over a distance for
which the change in the chemical potential is $\delta \mu$ :
chemical potential gradients act as a driving force on the fluid.

In the overdamped limit appropriate to most experimental systems,
the left side of (\ref{EQN:NS}) can be set to zero.
The velocity is then `slaved to the order parameter' \cite{KO,Ohta}.
The resulting linear equation for ${\bf v}$ can be
solved in Fourier space:
\begin{equation}
v_{\alpha}({\bf k}) = \frac{1}{\eta k^2}\left(-ik_{\alpha}p({\bf k})
                       +F_{\alpha}({\bf k})\right)\ ,
\label{EQN:v}
\end{equation}
where ${\bf F} = -\phi{\nabla}\mu$. The incompressibility condition
${\bf k}\cdot{\bf v}({\bf k})=0$ determines the pressure.
Putting the result into (\ref{EQN:v}) gives, with the summation
convention for repeated indices,
\begin{eqnarray}
v_{\alpha}({\bf k}) & = & T_{\alpha\beta}({\bf k})\,F_{\beta}({\bf k})
\nonumber \\
T_{\alpha\beta}({\bf k}) & = & \frac{1}{\eta k^2}\left(\delta_{\alpha\beta}
                                -\frac{k_\alpha k_\beta}{k^2}\right)\ ,
\end{eqnarray}
where $T$ is the `Oseen' tensor. In real space (for $d=3$)
\begin{equation}
T_{\alpha\beta}({\bf r}) = \frac{1}{8\pi\eta r}\left(\delta_{\alpha\beta}
                                +\frac{r_\alpha r_\beta}{r^2}\right)\ .
\label{EQN:OSEEN}
\end{equation}

Putting everything together gives the equation of motion in real space,
\begin{equation}
\partial{\phi}/\partial t = \lambda \nabla^2 \mu - \int d{\bf r}'\,
[\nabla \phi(r)\cdot T({\bf r}-{\bf r}')\cdot\nabla' \phi({\bf r}')]\,
\mu({\bf r'})\ ,
\label{EQN:BINLIQS}
\end{equation}
where we recall that $\mu({\bf r}) = \delta F/\delta\phi({\bf r})$.
In deriving the final form (\ref{EQN:BINLIQS}), integration by parts
(exploiting the transverse property,
$T_{\alpha\beta}({\bf k})k_\beta = 0$, of the Oseen tensor)
was used to convert $\phi({\bf r}')\nabla'\mu({\bf r}')$ to
$-\mu({\bf r}')\nabla'\phi({\bf r}')$ inside the integral.

It should be emphasized that, as usual, thermal fluctuations have been
neglected in (\ref{EQN:BINLIQS}).  We have previously argued, on rather
general grounds, that these are negligible at late times, because the
coarsening is controlled by a strong coupling renormalisation group fixed
point (see, in particular, the discussion in section \ref{SEC:RG}).
For binary liquids, however, a rather subtle situation can arise when
the nominally dominant coarsening mechanism does not operate. Then thermal
fluctuations do contribute. We will enlarge on this below.

We can use dimensional arguments to estimate the sizes of the two terms
on the right-hand side of (\ref{EQN:BINLIQS}). Using $\mu \sim \sigma/L$,
$T \sim 1/\eta L$, and $\nabla\phi \sim 1/L$, gives $\lambda\sigma/L^3$
for the first (`diffusive') term and $\sigma/\eta L$ for the second
(`advective') term. Advective transport of the order parameter therefore
dominates over diffusion for $L \gg (\lambda\eta)^{1/2}$. To determine
$L(t)$ in this regime, we use the expression for the fluid velocity,
\begin{equation}
{\bf v}({\bf r}) = \int d{\bf r}'\,[T({\bf r}-{\bf r}')\cdot
\nabla\phi({\bf r}')]\,\mu({\bf r}')\ .
\label{EQN:vREAL}
\end{equation}
Using the same dimensional arguments as before, and also $v \sim L/t$,
gives $L(t) \sim \sigma t/\eta$, a result first derived by Siggia
\cite{Siggia}. This result has been confirmed by experiments
\cite{BINLIQEXPTS} and by numerical simulations
\cite{BINLIQSIMS,ShinOono,Alexander}.
Because the inertial terms are negligible compared to the viscous force
here, we will call this the `viscous hydrodynamic' (or just `viscous')
regime.

Under what conditions is it correct to ignore the `inertial' terms on
the left-hand side of (\ref{EQN:NS})?  Using dimensional arguments again,
we see that these terms are of order $\rho L/t^2$. Comparing this to the
driving term $\phi\nabla \mu \sim \sigma/L^2$ on the right, (the viscous
term $\eta \nabla^2 v$ is of the same order in the viscous regime), and
using the result derived above, $t \sim \eta L/\sigma$, for this regime,
shows that the inertial terms are negligible when $L \ll \eta^2/\sigma\rho$.
At sufficiently late times, when this inequality is violated, the inertial
terms will therefore be important. In this `inertial' regime, $L(t)$ is
determined by equating the inertial terms, which scale as $\rho L/t^2$,
to the driving term $\phi\nabla\mu$, which scales as $\sigma/L^2$ (and
the viscous term is negligible) to give $L \sim (\sigma t^2/\rho)^{1/3}$.
The $t^{2/3}$ growth in the inertial regime was first predicted by
Furukawa \cite{FuruInertial}.

To summarise, there are in principle three growth regimes for phase separation
in binary liquids, after a deep quench, with the growth laws
\begin{eqnarray}
L(t) & \sim & (\lambda\sigma t)^{1/3}\ ,\ \ \ \ \ L \ll (\lambda\eta)^{1/2}\ ,
\ \ \ ({\rm `diffusive'})\ , \\
     & \sim & \sigma t/\eta\ ,\ \ \ \ \
(\lambda\eta)^{1/2} \ll L \ll \eta^2/\rho\sigma\ ,\ \ ({\rm `viscous\
hydrodynamic'}) \\
     & \sim & (\sigma t^2/\rho)^{1/3}\ ,\ \ \ \ \ L \gg \eta^2/\rho\sigma\ ,
\ \ \ \ \ ({\rm `inertial\ hydrodynamic'})\ .
\end{eqnarray}
These results basically follow from dimensional analysis.
The `inertial hydrodynamic' regime has not, to my knowledge, been
observed experimentally and we will not consider it further. However, a
$t^{2/3}$ regime has been observed at late times in simulations of
two-dimensional binary liquids \cite{Alexander,2DBINLIQS}.

Siggia \cite{Siggia} has discussed the physical origin of the linear growth
in the `viscous hydrodynamic' regime. He argues that the essential mechanism
is the hydrodynamic transport of fluid along the interface driven by the
surface tension. This mechanism, however, can only operate if both phases
are continuous. If, by contrast, the minority phase consists of independent
droplets (which occurs for volume fractions less than about 15\%), this
mechanism tends to make the droplets spherical but does not lead to any
coarsening (it is easy to show, for example, that the hydrodynamic term in
(\ref{EQN:BINLIQS}) vanishes for a single spherical droplet).
In the absence of thermal fluctuations, therefore, the
Lifshitz-Slyozov evaporation-condensation mechanism determines the growth
even beyond the nominal crossover length given above. Thermal fluctuations,
however, facilitate a second coarsening mechanism, namely droplet
coalescence driven by Brownian motion of the droplets.  Again, Siggia has
given the essential argument. The mobility $\mu$ of a droplet of size $L$
is of order $1/\eta L$, so the diffusion constant is given by the Einstein
relation as $D=\mu k_B T \sim k_BT/\eta L$, where $k_B$ is Boltzmann's
constant. The time for the droplet to diffuse a distance of order $L$ (and
coalesce with another droplet) is $t \sim L^2/D \sim \eta L^3/k_BT$, which
gives $L \sim (k_BTt/\eta)^{1/3}$.

The presence of two different mechanisms leading to the same growth
exponent suggests a `marginal operator' in the theory, and the
Renormalisation Group treatment of section \ref{SEC:RG} lends support
to this view. The RG approach also shows how, in this case, a nominally
irrelevant variable (temperature) can affect the late-stage growth in a
non-trivial way. From a physical point of view, it seems plausible that
the presence of competing mechanisms will lead to a late-stage morphology
that depends on the ratio of the amplitudes derived for the two mechanisms,
i.e.\ that scaling functions will depend continuously on this ratio.
This could tested by numerical simulations, where the transport coefficient
$\lambda$ and viscosity $\eta$ can be independently varied. In real binary
liquids, however, these coefficients are related, and the ratio of amplitudes
for the two mechanisms depends only on the volume fraction $v$
(see \cite{Siggia} and the discussion in section \ref{SEC:RGBINLIQS}).
Note that, even without hydrodynamics, the scaling functions will depend
on $v$, since the morphology does. The role of the Lifshitz-Slyozov
mechanism can be enhanced by going to small $v$, since the growth
rate due to evaporation-condensation is independent of $v$ for small $v$.
By contrast, the coalescence rate increases with $v$. To see this, we refine
the argument given in the previous paragraph \cite{AndrewNote}.

Let $R$ be a typical droplet radius. Then the droplet number density is
$n \sim v/R^3$. The time for a droplet to diffuse a distance of order its
radius is $t_R \sim R^2/D$. The volume swept out by the drop in time $t$
(for $t>t_R$) is of order $R^3\,t/t_R \sim RDt$. In a `coalescence time'
$t_c$ the expected number of drops in this  volume is of order unity,
i.e.\ $nRDt_c \sim 1$, giving $t_c \sim R^2/vD \sim \eta R^3/vk_BT$,
where we used $D \sim k_BT/\eta R$ in the last step. This implies that
$R$ grows with time as $R \sim (vk_BTt/\eta)^{1/3}$, a result first given
by Siggia \cite{Siggia}.

\section{TOPOLOGICAL DEFECTS}
\label{SEC:DEFECTS}
The domain walls discussed in the previous section are the simplest
form of `topological defect', and occur in systems described by scalar
fields \cite{Kleman}. They are surfaces, on which the order parameter
vanishes, separating domains of the two equilibrium phases. A domain
wall is topologically stable: local changes in the order parameter
can move the wall, but cannot destroy it. For an isolated flat wall,
the wall profile function is given by the solution of
(\ref{EQN:PROFILE}), with the appropriate boundary conditions,
as discussed in section \ref{SEC:WALLS} (and sketched in Figure 6).
For the curved walls present in the phase ordering process,
this will still be an approximate solution locally, provided the typical
radius of curvature $L$ is large compared to the intrinsic width
(or `core size'), $\xi$, of the walls. (This could be defined from
(\ref{EQN:PROFILETAIL}) as
$\xi = [V''(1)]^{-1/2}$, say). The same condition, $L \gg \xi$, ensures
that typical wall separations are large compared to their width.

Let us now generalize the discussion to vector fields. The `$O(n)$ model'
is described by an $n$-component vector field $\vec{\phi}({\bf x},t)$,
with a free energy
functional $F[\vec{\phi}]$ that is invariant under global rotations of
$\vec{\phi}$. A suitable generalization of (\ref{EQN:HAMILTONIAN}) is
\begin{equation}
F[\vec{\phi}] = \int d^dx\,\left(\frac{1}{2}\,(\nabla\vec{\phi})^2
                  + V(\vec{\phi})\right)\ ,
\label{EQN:VECTORF}
\end{equation}
where $(\nabla\vec{\phi})^2$ means
$\sum_{i=1}^d\sum_{a=1}^n(\partial_i\phi^a)^2$ (i.e.\ a scalar product over
both spatial and `internal' coordinates), and $V(\vec{\phi})$ is `mexican
hat' (or `wine bottle') potential, such as $(1-\vec{\phi}^2)^2$, whose
general form is sketched in Figure 9. It is clear that $F[\vec{\phi}]$
is invariant under global rotations of $\vec{\phi}$ (a continuous symmetry),
rather than just the inversion symmetry ($\phi \to -\phi$, a discrete
symmetry) of the scalar theory. We will adopt the convention that $V$ has
its minimum for $\vec{\phi}^2=1$.

For non-conserved fields, the simplest dynamics (model A) is a
straightforward generalization of (\ref{EQN:MODELA}), namely
\begin{equation}
\partial\vec{\phi}/\partial t = \nabla^2 \vec{\phi} - dV/d\vec{\phi}\ .
\label{EQN:VECTORA}
\end{equation}
For conserved fields (model B), we simply add another $(-\nabla^2)$ in front
of the right-hand side.

Stable topological defects for vector fields can be generated, in analogy
to the scalar case, by seeking stationary solutions of (\ref{EQN:VECTORA})
with appropriate boundary conditions. For the $O(n)$ theory in
$d$-dimensional space, the requirement that all $n$ components of
$\vec{\phi}$ vanish at the defect core defines
a surface of dimension $d-n$ (e.g.\ a domain wall is a
surface of dimension $d-1$: the scalar theory corresponds to $n=1$).
The existence of such defects therefore requires $n \le d$. For $n=2$
these defects are points (`vortices') for $d=2$ or lines (`strings', or
`vortex lines') for $d=3$. For $n=3$, $d=3$ they are points (`hedgehogs',
or `monopoles'). The field configurations for these defects are
sketched in Figures 10(a)-(d). Note that the forms shown are radially
symmetric with respect to the defect core: any configuration obtained
by a global rotation is also acceptable. For $n<d$, the field
$\vec{\phi}$ only varies in the $n$ dimensions `orthogonal' to the
defect core, and is uniform in the remaining $d-n$ dimensions
`parallel' to the core.

For $n<d$, the defects are spatially extended. Coarsening occurs by
a `straightening out' (or reduction in typical radius of curvature)
as sharp features are removed, and by the shrinking and disappearance
of small domain bubbles or vortex loops. These processes reduce the
total area of domain walls, or length of vortex line, in the system.
For point defects ($n=d$), coarsening occurs by the mutual annihilation
of defect-antidefect pairs. The antidefect for a vortex (`antivortex')
is sketched in Figure 10(e). Note that the antivortex in {\em not}
obtained by simply reversing the directions of the arrows in 10(b):
this would correspond to a global rotation through $\pi$. Rather,
the vortex and antivortex have different `topological charges': the
fields rotates by $2\pi$ or $-2\pi$ respectively on encircling the
defect. By contrast, an antimonopole {\em is} generated by reversing
the arrows in 10(d): the reversed configuration cannot be generated
by a simple rotation in this case.

For the radially symmetric defects illustrated in 10(b)--(d), the field
$\vec{\phi}$ has the form $\vec{\phi}({\bf r}) = \hat{r}\,f(r)$, where
$\hat{r}$ is a unit vector in the radial direction, and $f(r)$ is the
profile function. Inserting this form into (\ref{EQN:VECTORA}),
with the time derivative set to zero, gives the equation
\begin{equation}
\frac{d^2f}{dr^2} + \frac{(n-1)}{r}\,\frac{df}{dr} - \frac{(n-1)}{r^2}\,f
            - V'(f) = 0\ ,
\label{EQN:VECPROF}
\end{equation}
with boundary conditions $f(0)=0$, $f(\infty) = 1$. Of special interest
is the approach to saturation at large $r$. Putting $f(r) = 1-\epsilon(r)$
in (\ref{EQN:VECPROF}), and expanding to first order in $\epsilon$,
yields
\begin{equation}
 \epsilon(r) \simeq \frac{(n-1)}{V''(1)}\,\frac{1}{r^2}\ ,
\ \ \ \ \ \ \ r \to \infty\ .
\label{EQN:VECPROFTAIL}
\end{equation}
This should be contrasted with the exponential approach to saturation
(\ref{EQN:PROFILETAIL})
for scalar fields. A convenient definition of the `core size' $\xi$ is
through $f \simeq 1 - \xi^2/r^2$ for large $r$. This gives
$\xi = [(n-1)/V''(1)]^{1/2}$ for $n>1$.

\subsection{DEFECT ENERGETICS}
\label{SEC:DEF-ENERGETICS}
Consider an isolated, equilibrium defect of the $O(n)$ model in
$d$-dimensional space (with, of course, $n\le d$).
For a radially symmetric defect,
$\vec{\phi}({\bf r}) = f(r)\,\hat{\bf r}$, the energy per unit `core volume'
(e.g.\ per unit area for a wall, per unit length for a line, or per defect
for a point) is, from (\ref{EQN:VECTORF})
\begin{equation}
E = S_n \int dr\,r^{n-1}\,\left(\frac{(n-1)}{2r^2}\,f^2
                  + \frac{1}{2}\,(\nabla f)^2 + V(f)\right)\ ,
\label{EQN:DEF-ENERGY}
\end{equation}
where $S_n =2\pi^{n/2}/\Gamma(n/2)$ is the surface area of an
$n$-dimensional sphere.

For scalar fields ($n=1$), we have seen (section \ref{SEC:WALLS}) that the
terms in $(\nabla f)^2$ and $V(f)$ contribute equally to the wall energy.
For $n \ge 2$, the first term in (\ref{EQN:DEF-ENERGY}) dominates the
other two because, from (\ref{EQN:VECPROFTAIL}), the three terms in
the integrand fall off with distance as $r^{-2}$, $r^{-6}$ and
$V(f) \sim V''(1)(1-f)^2 \sim r^{-4}$ respectively as $r \to \infty$.
For $n \ge 2$, therefore, the first term gives a divergent integral
which has to be cut off as the system size $L_{sys}$, i.e.\
$E \sim \ln (L_{sys}/\xi)$ for $n=2$ and $E \sim L_{sys}^{n-2}$ for
$n >2$. Actually, the second and third terms give divergent integrals
for $n \ge 6$ and $n \ge 4$ respectively, but these are always
subdominant compared to the first term.

The above discussion concerns an {\em isolated} defect. In the phase
ordering system the natural cut-off is not $L_{sys}$ but $L(t)$, the
characteristic scale beyond which the field of a single defect will be
screened by the other defects. Of particular interest are
the dynamics of defect structures much smaller than $L(t)$.
These are the analogues of the small domains of the scalar system.
For $d=n=2$, these are vortex-antivortex pairs, for $d=3$, $n=2$ they are
vortex rings, while for $d=3=n$ they are monopole-antimonopole pairs.
For such a structure, the pair separation $r$ (for point defects) or
ring radius $r$ (for a vortex loop) provide the natural cut-off.
Including the factor $r^{d-n}$ for the volume of defect core,
the energy of such a structure is
\begin{eqnarray}
E & \sim & r^{d-2}\,\ln (r/\xi)\ ,\ \ \ \ \ d \ge n=2\ ,\nonumber \\
  & \sim & r^{d-2}\ ,\ \ \ \ \ \ \ \ \ \ \ \ d \ge n>2\ .
\label{EQN:DEF-ENERGY1}
\end{eqnarray}
The derivative with respect to $r$ of this energy provides the driving
force, $-dE/dr$, for the collapse of the structure.  Dividing by $r^{d-n}$
gives the force $F$ acting on a unit volume of core (i.e.\ per unit
length for strings, per point for points, etc.):
\begin{eqnarray}
F(r) & \sim & -r^{-1}\ ,\ \ \ \ \ \ \ \ \ \ \ \ \ \ \ d=n=2\ , \nonumber \\
  & \sim & -r^{n-3}\,\ln (r/\xi)\ ,\ \ \ \ \ d>n=2\ , \nonumber \\
  & \sim & -r^{n-3}\ ,\ \ \ \ \ \ \ \ \ \ \ \ \ \ \ d \ge n>2\ .
\label{EQN:DEF-FORCE}
\end{eqnarray}

In order to calculate the collapse time we need the analogue
of the `friction constant' $\eta$ (see section \ref{SEC:AC}) for vector
fields. This we calculate in the next section. Before doing so, we
compute the total energy density $\epsilon$
for vector fields. This can be obtained by putting $r \sim L(t)$ in
(\ref{EQN:DEF-ENERGY1}), and dividing by a characteristic volume $L(t)^d$
(since there will typically be of order one defect stucture, with size of
order $L(t)$, per scale volume $L(t)^d$),
\begin{eqnarray}
\epsilon & \sim & L(t)^{-2}\,\ln \left(L(t)/\xi\right)\ ,
\ \ \ \ \ d \ge n=2\ ,\nonumber \\
  & \sim & L(t)^{-2}\ ,\ \ \ \ \ \ \ \ \ \ \ \ \ \ d \ge n>2\ .
\label{EQN:DEF-ENERGY2}
\end{eqnarray}
For scalar systems, of course, $\epsilon \sim L(t)^{-1}$.

As a caveat to the above discussion, we note that we have explicitly
assumed that the individual defects possess an approximate radial
symmetry on scales small compared to $L(t)$. It has been known for some
time \cite{Ostlund}, however, that an isolated point defect for
$d >3$ can lower its energy by having the field uniform
(pointing `left', say) over most of space , with a narrow `flux tube'
of field in the opposite direction (i.e.\ pointing `right').
The energy is then linear in the size of the system,
$E \sim L_{sys}\xi^{d-3}$, which is smaller than the energy,
$\sim L_{sys}^{d-2}$, of the spherically symmetric defect, for $d>3$.
A defect-antidefect pair with separation $r$, connected by such a flux
tube, has an energy $E \sim r\xi^{d-3}$, which implies an $r$-independent
force for all $d \ge 3$, in contrast to (\ref{EQN:DEF-FORCE}).

How relevant are these considerations in the context of phase-ordering
dynamics? These single-defect and defect-pair calculations treat the
field as completely relaxed with respect to the defect cores. If this
were true we could estimate the energy density for typical defect spacing
$L(t)$ as $\xi^{d-3} L(t)^{1-d}$ for $d>3$. However, the smooth variation
(`spin waves') of the field between the defects gives a contribution
to the energy density of $(\nabla \phi)^2 \sim L(t)^{-2}$, which dominates
over the putative defect contribution for $d>3$. Under these circumstances,
we would not expect a strong driving force for point defects to adopt the
`flux tube' configuration, since the energy is dominated by spin waves.
Rather, our tentative picture is of the point defects `riding' on the
evolving spin wave structure for $d > 3$, although this clearly requires
further work. Note, however, that these concerns  are only relevant
for $d>3$: Eq.\  (\ref{EQN:DEF-FORCE}) is certainly correct for the
physically relevant cases $d \le 3$.

\subsection{DEFECT DYNAMICS}
\label{SEC:DEF-DYNAMICS}
Here we will consider only nonconserved fields.
Using the methods developed in section \ref{SEC:GROWTH}, however,
it is possible to generalise the results to conserved fields \cite{BRunpub}.
The caveats for $d>3$ discussed in the previous subsection also apply here.

The calculation of the friction constant $\eta$ proceeds as in section
\ref{SEC:AC}. Consider an isolated equilibrium defect, i.e.\ a vortex for
$d=n=2$, a monopole for $n=d=3$, a straight vortex line for
$n=2$, $d=3$ etc. Set up a Cartesian coordinate system $x_1, \ldots, x_d$.
For extended defects, let the defect occupy the (hyper)-plane defined by
the last $d-n$ Cartesian coordinates, and move with speed $v$ in the
$x_1$ direction. Then $\vec{\phi}$ only depends on coordinates
$x_1,\ldots,x_n$, and the rate of change of the system energy per unit
volume of defect core is
\begin{eqnarray}
dE/dt & = & \int dx_1 \ldots dx_n\,(\delta F/\delta \vec{\phi})\cdot
\partial\vec{\phi}/\partial t \nonumber \\
& = & -\int dx_1 \ldots dx_n\,(\partial\vec{\phi}/\partial t)^2\ .
\label{EQN:VECDISS}
\end{eqnarray}
The defect profile has the form
$\vec{\phi}(x_1,\ldots,x_n) = \vec{f}(x_1-vt,x_2,\ldots,x_n)$, where
the function $\vec{f}$ depends on $v$ in general. Putting this into
(\ref{EQN:VECDISS}) gives
\begin{eqnarray}
dE/dt & = & -v^2 \int dx_1 \ldots dx_n\,(\partial \vec{\phi}/\partial x_1)^2
\nonumber \\
      & = & -(v^2/n) \int d^nr (\nabla \vec{\phi})^2 = -\eta v^2\ ,
\end{eqnarray}
where the function $\vec{f}$ has been replace by its $v=0$ form to lowest
order in $v$, and $\eta$ is the friction constant per unit core volume.
The final expression follows from symmetry. It follows that $\eta$ is
(up to constants) equal to the defect energy per unit core volume.
In particular, it diverges with the system size for $n \ge 2$. For a small
defect structure of size $r$, we expect the divergence to be effectively
cut off at $r$. This gives a scale-dependent friction constant,
\begin{eqnarray}
\eta(r) & \sim & r^{n-2}\,\ln (r/\xi)\ ,\ \ \ \ \ d \ge n=2\ ,\nonumber \\
        & \sim & r^{n-2}\ ,\ \ \ \ \ \ \ \ \ \ \ \ d \ge n>2\ .
\label{EQN:eta}
\end{eqnarray}

Invoking the scaling hypothesis, we can now determine the growth laws for
non-conserved vector systems. Eqs.\ (\ref{EQN:DEF-FORCE}) and
(\ref{EQN:eta}) give the typical force and friction constant per unit
core volume as $F(L)$ and $\eta(L)$. Then a typical velocity is
$v \sim dL/dt \sim F(L)/\eta(L)$, which can be integrated to give,
asymptotically,
\begin{eqnarray}
L(t) & \sim & (t/\ln t)^{1/2}\ ,\ \ \ \ \ \ \ d = n =2\ ,\nonumber  \\
     & \sim & t^{1/2}\ ,\ \ \ \ \ \ \ {\rm otherwise}\ .
\label{EQN:VECGROWTH}
\end{eqnarray}
The result for $n=d=2$ was derived by Pargellis et al.\ \cite{BLOGGS},
and checked numerically by Yurke et al.\ \cite{Pargellis}.
The method used here follows their approach \cite{Turok}. The key concept
of a scale-dependent friction constant has been discussed by a number
of authors \cite{Musny}. A detailed analysis of monopole-antimonopole
annihilation, in the context of nematic liquid crystals, has been given by
Pismen and Rubinstein \cite{Pismen}.

A more general and powerful method to derive growth laws, valid for both
conserved and nonconserved systems, is the subject of section
\ref{SEC:GROWTH}. The results agree with the intuitive arguments presented
so far, with the possible exception of the case $n=d=2$, for which evidence
suggestive of scaling violations will be presented.

\subsection{POROD'S LAW}
\label{SEC:POROD}
The presence of topological defects, seeded by the initial conditions,
in the system undergoing phase ordering has an important effect on the
`short-distance' form of the pair correlation function $C({\bf r},t)$,
and therefore on the `large-momentum' form of the structure factor
$S({\bf k},t)$. To see why this is so, we note that, according to the
scaling hypothesis, we would expect a typical field gradient to be of
order $|\nabla\vec{\phi}| \sim 1/L$. At a distance $r$ from a defect
core, however, with $\xi \ll r \ll L$, the field gradient is much
larger, of order $1/r$ (for a vector field), because
$\vec{\phi} = \hat{r}$ implies $(\nabla\vec{\phi})^2 = (n-1)/r^2$.
Note that we require $r \gg \xi$ for the field to be saturated, and
$r \ll L$ for the defect field to be largely unaffected by other
defects (which are typically a distance $L$ away). This gives a meaning
to `short' distances ($\xi \ll r \ll L$), and `large momenta'
($L^{-1} \ll k \ll \xi^{-1}$). The large field gradients near
defects leads to a non-analytic behaviour at $x=0$ of the scaling
function $f(x)$ for pair correlations.

We start by considering scalar fields. Consider two points
${\bf x}$ and ${\bf x} + {\bf r}$, with $\xi \ll r \ll L$. The product
$\phi({\bf x})\,\phi({\bf x}+{\bf r})$ will be $-1$ if a wall passes
between them, and $+1$ if there is no wall. Since $r \ll L$, the
probability to find more than one wall can be neglected. The calculation
amounts to finding the probability that a randomly placed rod of length
$r$ cuts a domain wall. This probability is of order $r/L$, so we
estimate
\begin{eqnarray}
C({\bf r},t) & \simeq & (-1) \times (r/L) + (+1) \times (1-r/L) \nonumber \\
             & = & 1 - 2r/L\ ,\ \ \ \ \ \ \ r \ll L\ .
\label{EQN:SFSHORT}
\end{eqnarray}
The factor 2 in this result should not be taken seriously.

The important result is that (\ref{EQN:SFSHORT}) is non-analytic in
${\bf r}$ at ${\bf r} = 0$, since it is linear in $r \equiv |{\bf r}|$.
Technically, of course, this form breaks down inside the core region,
when $r < \xi$. We are interested, however, in
the scaling limit defined by $r \gg \xi$, $L \gg \xi$,
with $x=r/L$ arbitrary. The nonanalyticity is really in the scaling
variable $x$.

The nonanalytic form (\ref{EQN:SFSHORT}) implies a power-law tail in the
structure factor, which can be obtained from (\ref{EQN:SFSHORT})
by simple power-counting:
\begin{equation}
S({\bf k},t) \sim \frac{1}{L\,k^{d+1}}\ ,\ \ \ \ \ \ \ \ kL \gg 1\ ,
\label{EQN:POROD}
\end{equation}
a result known universally as `Porod's law'. It was first written
down in the general context of scattering from two-phase media \cite{Porod51}.
Again, one requires $k\xi \ll 1$ for the scaling regime. Although the
$k$-dependence of (\ref{EQN:POROD}) is what is usually referred to as
Porod's law, the $L$-dependence is equally interesting.
The factor $1/L$ is simply
(up to constants) the total area of domain wall per unit volume, a fact
appreciated by Porod, who proposed structure factor measurements as a
technique to determine the area of interface in a two-phase medium
\cite{Porod51}.
On reflection, the factor $1/L$ is not so surprising. For $kL \gg 1$,
the scattering function is probing structure on scales much shorter
than the typical interwall spacing or radius of curvature. In
this regime we would expect the structure factor to scale as the total
wall area, since each element of wall with linear dimension large
compared to $1/k$ contributes essentially independently to the structure
factor.

This observation provides the clue to how to generalize
(\ref{EQN:POROD}) to vector (and other) fields \cite{Bray93,BH93}.
The idea is that, for $kL \gg 1$, the
structure factor should scale as the total volume of defect core.
Since the dimension of the defects is $d-n$,
the amount of defect per unit volume scales as $L^{-n}$.
Extracting this factor from the general scaling form (\ref{EQN:STRUCT})
yields
\begin{equation}
\label{EQN:GENPOROD}
S({\bf k},t) \sim \frac{1}{L^n\,k^{d+n}}\ ,\ \ \ \ \ \ kL \gg 1\ ,
\end{equation}
for the $O(n)$ theory, a `generalized Porod's law'.

Equation (\ref{EQN:GENPOROD}) was first derived from approximate
treatments of the equation of motion (\ref{EQN:VECTORA})
for nonconserved fields \cite{BP,Toy92,LM,BH}. In these derivations,
however, the key role of topological defects was far from transparent.
The above heuristic derivation suggests that the result is in fact very general
(e.g., it should hold equally well for conserved fields), with extensions
beyond simple $O(n)$ models. The appropriate techniques, which also enable
the amplitude of the tail to be determined, were developed by Bray
and Humayun \cite{BH93}, and will be discussed in detail in section
\ref{SEC:SHORT}.

\subsection{NEMATIC LIQUID CRYSTALS}
\label{SEC:LIQUIDXTALS}
Liquid crystals have been a fertile area for recent experimental work
on the kinetics of phase ordering, largely due to the efforts of Yurke
and coworkers \cite{Mason93,BLOGGS,YurkeStrings,WWY,WWLY}.
Here we will concentrate on the simplest liquid crystal phase,
the nematic. In a simple picture which captures the
orientational degrees of freedom of the nematic, the liquid crystal can
be thought of as consisting of rod-like molecules \cite{DeGennes} which
have a preferential alignment with the `director' ${\bf n}$ in the
ordered phase. Because the molecules have a head-tail symmetry, however,
the free energy is invariant under the local transformation
${\bf n} \to -{\bf n}$. As a result, the nematic is usually described by
a tensor order parameter $Q$ that is invariant under this local
transformation, and has in the ordered phase the representation
\begin{equation}
Q_{ab} = S\,(n_an_b - \frac{1}{3}\,\delta_{ab})\ .
\label{EQN:Q}
\end{equation}
Note that $Q$ is a traceless, symmetric tensor. The simplest free energy
functional is one that is invariant under global rotations of the field
$n({\bf r})$. It has the form \cite{DeGennes}
\begin{equation}
F[Q] = \int d^3x\,\left(\frac{1}{2}\,{\rm Tr}\,(\nabla Q)^2
+\frac{r}{2}\,{\rm Tr}\,Q^2 - \frac{w}{3}\,{\rm Tr}\,Q^3
+\frac{u}{4}\,({\rm Tr}\,Q^2)^2\right)\ ,
\label{EQN:LCF}
\end{equation}
where we have retained only terms up to order $Q^4$, and the lowest
order term involving spatial gradients. (Note that ${\rm Tr}\,Q^4 =
(1/2)({\rm Tr}\,Q^2)^2$ for a $3 \times 3$ traceless, symmetric tensor,
so we don't include this term separately). The presence of the
cubic term, allowed by symmetry, leads to a first-order phase
transition in mean-field theory \cite{DeGennes}.
In experimental systems the transition is weakly first order.

The free energy functional (\ref{EQN:LCF}) is an idealisation of real
nematics, in the spirit of the Lebwohl-Lasher lattice Hamiltonian
$H_{LL} = -J \sum_{<i,j>} ({\bf n}_i\cdot {\bf n}_j)^2$, where ${\bf n}_i$
is the local director at site $i$. Both models are invariant under
global rotations of ${\bf n}$, as well local inversions,
${\bf n} \to -{\bf n}$. The gradient terms in (\ref{EQN:LCF}) can be
written, using (\ref{EQN:Q}), as (ignoring constants)
$\sum_{i,a}(\partial n^a/\partial x_i)^2$,
i.e.\ as $(\nabla \vec{n})^2$, which is a continuum version of $H_{LL}$.
The overarrow on $\vec{n}$ here indicates that, in these models, the spatial
and `internal' spaces can be considered as distinct (much as in the
$O(n)$ model, where $n$ and $d$ can be different). In real nematics,
however, these spaces are coupled. An appropriate gradient energy density
is the Frank energy
\begin{equation}
E_F = K_1\,(\nabla\cdot{\bf n})^2 + K_2\,({\bf n}\cdot\nabla\times{\bf n})^2
      +K_3\,[{\bf n}\times(\nabla\times{\bf n})]^2\ ,
\end{equation}
where the Frank constants $K_1$, $K_2$ and $K_3$ are associated with
`splay', `twist' and `bend' of the director \cite{DeGennes}. The isotropic
models discussed above correspond to the case $K_1=K_2=K_3$, the much-used
`equal-constant approximation'. We will limit our considerations
exclusively to this case.

In a similar spirit, we adopt the simplest possible dynamics, namely
the purely relaxational dynamics of model A. This captures correctly
the non-conserved nature of the dynamics, but ignores possible
complications due to hydrodynamic interactions. Recent work, comparing
experimental results with simulations based on relaxational dynamics,
provide some justification for this approach \cite{YurkePreprint}.
The equation of motion is
$\partial Q/\partial t = -\delta G[Q]/\delta Q$, where
$G[Q]=F[Q] -\int d^3x\,\lambda({\bf x}) Q({\bf x})$ and $\lambda$ is
a Lagrange multiplier introduced to maintain the condition
${\rm Tr}\,Q=0$. Imposing the constraint to eliminate $\lambda$ gives
\begin{equation}
\frac{\partial Q}{\partial t} = \nabla^2Q -rQ
+ w\left(Q^2 - \frac{1}{3} I\,{\rm Tr}\,Q^2\right) - uQ\,{\rm Tr}\,Q^2\ ,
\label{EQN:MODELN}
\end{equation}
where $I$ is the unit tensor. This equation will be discussed in more detail
in section \ref{SEC:NLC}.

Due to the extra local symmetry (compared to $O(3)$ models) under
${\bf n} \to -{\bf n}$, nematic liquid crystals support a number of
defect types \cite{Kleman}. In the present context, the most important
are string defects, or `disclinations', in which the director rotates
through  $\pm \pi$ on encircling the string, as sketched in Figure 11.
These $\pm 1/2$-strings are topologically stable, in contrast to
$\pm 1$ string configurations (in nematics and the $O(3)$ model) which
can be relaxed by smoothly canting the order parameter towards the string
axis (`escape in the third dimension'). The presence of string defects
(which have been observed by Yurke's group \cite{YurkeStrings}), makes
the nematic more akin to the $O(2)$ model than the $O(3)$ model as far
as its ordering kinetics are concerned.
In particular, if we make the natural assumption that the total string
length decreases as $L^{-2}$, a Porod law of the form (\ref{EQN:GENPOROD})
with $n=2$ is predicted. Scattering data were originally interpreted as
being consistent with $n=3$-like behaviour, i.e.\ an effective exponent
$d+n =6.0 \pm 0.3$ in (\ref{EQN:GENPOROD}) was observed \cite{WWY}, but
it is not clear whether the appropriate region of the structure factor tail
was fitted \cite{BPBS}. Numerical simulations of a `soft-spin' version
of the Lebwohl-Lasher model \cite{BB92} are fully consistent with a tail
exponent of 5 \cite{BPBS}. We will return to this point in section
\ref{SEC:NLC}.

\section{EXACTLY SOLUBLE MODELS}
\label{SEC:SOLUBLE}
There are few exactly solved models of phase ordering
dynamics and, unfortunately, these models are quite far from describing
systems of physical interest. However, the models are not without
interest, as some qualitative features survive in more physically
relevant models. In particular, such models are the only cases in
which the hypothesized scaling property has been explicitly established.

We begin by discussing phase ordering of a vector field in the limit
that the number of vector components of the field, $n$, tends to infinity.
This limit has been studied, mostly for nonconserved fields, by a large
number of authors \cite{CZ,NB90,Large-n,CRZ,NBM,Kiss93}.
In principle, the solution is the starting point for a systematic treatment
in powers of $1/n$. In practice, the calculation of the $O(1/n)$ terms is
technically difficult \cite{NB90,Kiss93}. Moreover, some important
physics is lost in this limit. In particular, there are no topological
defects, since clearly $n > d+1$ for any $d$ as $n \to \infty$.
As a consequence, Porod's law (\ref{EQN:GENPOROD}), for example, is not found.
It turns out, however, that similar techniques
can be applied for any $n$ after a preliminary transformation from the
physical order parameter field $\vec{\phi}$ to a suitably chosen
`auxiliary field' $\vec{m}$. This is discussed in section \ref{SEC:APPROXSF}.
The topological defects are incorporated through the functional dependence
of $\vec{\phi}$ on $\vec{m}$, and Porod's law is recovered.

\subsection{THE LARGE-$n$ LIMIT: NONCONSERVED FIELDS}
\label{SEC:LARGEnNC}
Although not strictly necessary, it is convenient to choose in
(\ref{EQN:VECTORA}) the familiar `$\phi^4$' potential, in the form
$V(\vec{\phi}) = (n - \vec{\phi}^2)^2/4n$, where the explicit $n$-dependence
is for later convenience in taking the limit $n \to \infty$. With this
potential, (\ref{EQN:VECTORA}) becomes
\begin{equation}
\partial\vec{\phi}/\partial t = \nabla^2 \vec{\phi} + \vec{\phi}
-\frac{1}{n}\,(\vec{\phi}^2)\,\vec{\phi}\ .
\end{equation}
The simplest way to take the limit is to recognize that, for $n \to \infty$,
$\vec{\phi}^2/n$ can be replaced by its average, to give
\begin{eqnarray}
\label{EQN:LARGEN}
\partial\phi/\partial t & = & \nabla^2 \phi + a(t)\,\phi \\
a(t) & = & 1 - \langle \phi^2 \rangle\ ,
\label{EQN:a(t)}
\end{eqnarray}
where $\phi$ now stands for (any) one of the components of $\vec{\phi}$.
Eq.\ (\ref{EQN:LARGEN}) can alternatively be derived by standard diagrammatic
techniques \cite{NB90}. Eq.\ (\ref{EQN:LARGEN}) can be solved exactly
for arbitrary time $t$ after the quench. However, we are mainly interested
in late times (i.e.\ the scaling regime), when the solution simplifies.
After Fourier transformation, the formal solution of (\ref{EQN:LARGEN}) is
\begin{eqnarray}
\label{EQN:FOURIER}
\phi_{\bf k}(t) & = & \phi_{\bf k}(0)\ \exp[-k^2t + b(t)]\ , \\
        b(t) & = & \int_0^t dt'\,a(t')\ ,
\label{EQN:FOURIERb}
\end{eqnarray}
giving
\begin{equation}
a(t) = db/dt = 1 - \Delta \sum_k \exp[-2k^2t + 2b(t)]\ ,
\label{EQN:b(t)}
\end{equation}
where (\ref{EQN:SRIC}) has been used to eliminate the initial condition.
Since we shall find {\em a posteriori} that $a(t) \ll 1$ at late times,
the left side of (\ref{EQN:b(t)}) is negligible for $t \to \infty$.
Using $\sum_{\bf k}\exp(-2k^2t) = (8\pi t)^{-d/2}$ gives
$b(t) \to (d/4)\ln(t/t_0)$, where
\begin{equation}
t_0 = \Delta^{2/d}/8\pi\ .
\label{EQN:Delta}
\end{equation}
Therefore, $a(t) \to d/4t$ for $t \to \infty$,
and the solution of (\ref{EQN:FOURIER}), valid at late times, is
\begin{equation}
\phi_{\bf k}(t) = \phi_{\bf k}(0)\,(t/t_0)^{d/4}\,\exp(-k^2t)\ .
\label{EQN:FOURIERSOLN}
\end{equation}
Using (\ref{EQN:SRIC}) once more, we obtain the structure factor, and
its Fourier transform, the pair correlation function as,
\begin{eqnarray}
\label{EQN:LARGENS}
S({\bf k},t) & = & (8\pi t)^{d/2}\,\exp(-2k^2t)\ , \\
C({\bf r},t) & = & \exp(-r^2/8t)\ .
\label{EQN:LARGENC}
\end{eqnarray}
These results exhibit the expected scaling forms (\ref{EQN:STRUCT}),
with length scale $L(t) \propto t^{1/2}$.
Note that the structure factor has a gaussian tail, in contrast to the
power-law tail (\ref{EQN:GENPOROD}) found in systems with $n \le d$.
It might be hoped, however, that the large-$n$ forms (\ref{EQN:LARGENS})
and (\ref{EQN:LARGENC}) would be qualitatively correct in systems with
no topological defects, i.e.\ for $n > d+1$. These cases will be discussed
in section \ref{SEC:SHORT}.

\subsection{TWO-TIME CORRELATIONS}
\label{SEC:TWOTIMES}
Within the large-$n$ solution, we can also calculate two-time correlations
to test the scaling form (\ref{EQN:TWOTIME}). It turns out (although
this becomes apparent only at $O(1/n)$) \cite{NB90,Kiss93}
that there is a new, non-trivial, exponent associated with the limit
when the two times are well separated \cite{Janssen}.

{}From (\ref{EQN:FOURIERSOLN}) it follows immediately that
\begin{eqnarray}
\label{EQN:TWOTIMES}
S({\bf k},t,t') & \equiv & \langle \phi_{\bf k}(t) \phi_{-{\bf k}}(t') \rangle
= [8\pi(tt')^{1/2}]^{d/2}\,\exp[-k^2(t+t')]\ , \\
C({\bf r},t,t') & \equiv & \langle \phi({\bf r},t) \phi({\bf 0},t') \rangle
= \left[\frac{4tt'}{(t+t')^2}\right]^{d/4}\,\exp\left[-\frac{r^2}{4(t+t')}
\right]\ .
\label{EQN:TWOTIMEC}
\end{eqnarray}
Eq.\ (\ref{EQN:TWOTIMEC}) indeed has the expected form (\ref{EQN:TWOTIME}).
In the limit $t \gg t'$, (\ref{EQN:TWOTIMEC}) becomes
\begin{eqnarray}
C({\bf r},t,t') & = & (4t'/t)^{d/4}\,\exp(-r^2/4t)\ , \\
                & = & (L'/L)^{\bar{\lambda}}\,h(r/L)\ ,
\label{EQN:TWO-TIME}
\end{eqnarray}
where the last equation defines the exponent $\bar{\lambda}$ through
the dependence on the {\em later} time $t$. Clearly,
$\bar{\lambda} = d/2$ for $n=\infty$. When the $O(1/n)$ correction is
included, however, an entirely non-trivial result is
obtained \cite{NB90,Kiss93}.

It is interesting to consider the special case where the earlier time
$t'$ is zero. Then $C({\bf r},t,0)$ is just the correlation with the
initial condition. This quantity is often studied in numerical
simulations as a convenient way to determine the exponent $\bar{\lambda}$.
Within the large-$n$ solution, Eqs.\ (\ref{EQN:FOURIERSOLN}) and
(\ref{EQN:Delta}) give, in Fourier and real space,
\begin{eqnarray}
\label{EQN:CORRlarge-n}
S({\bf k},t,0) & = & [8\pi(tt_0)^{1/2}]^{d/2}\,\exp(-k^2t)\ ,  \\
C({\bf r},t,0) & = & (4t_0/t)^{d/4}\,\exp(-r^2/4t)\ .
\end{eqnarray}
This is just what one gets by replacing $t'$ by $t_0$ in
(\ref{EQN:TWOTIMES}) and (\ref{EQN:TWOTIMEC}) (with $t_0$ playing the
role of a short-time cut-off), and then neglecting $t_0$ compared to $t$.

A related function is the {\em response} to the initial condition,
defined by
\begin{equation}
G({\bf k},t)=\langle\partial\phi_{\bf k}(t)/\partial\phi_{\bf k}(0)\rangle\ .
\label{EQN:RESPONSE}
\end{equation}
Within the large-$n$ solution, (\ref{EQN:FOURIERSOLN}) gives immediately
\begin{equation}
G({\bf k},t) = (t/t_0)^{d/4}\,\exp(-k^2t)\ .
\label{EQN:RESPONSElarge-n}
\end{equation}
Comparing (\ref{EQN:CORRlarge-n}) and (\ref{EQN:RESPONSElarge-n}), and
using (\ref{EQN:Delta}) once more gives the relation.
\begin{equation}
S({\bf k},t,0) = \Delta\,G({\bf k},t)\ .
\label{EQN:TWOTIMERELATION}
\end{equation}
In fact, this is an exact result, valid beyond the large-$n$ limit,
as may be proved easily using integration by parts on the gaussian
distribution for $\{\phi_{\bf k}(0)\}$. The general scaling form for
$G({\bf k},t)$,
\begin{equation}
G({\bf k},t) = L^{\lambda}\,g_R(kL)\ ,
\label{EQN:GSCALING}
\end{equation}
defines a new exponent $\lambda$, equal to $d/2$ for $n=\infty$.
Since, however, the correlation with the initial condition has the
scaling form $C({\bf r},t,0) = L^{-\bar{\lambda}}\,f(r/L)$, the
identity (\ref{EQN:TWOTIMERELATION}) gives immediately \cite{Note3}
\begin{equation}
\bar{\lambda} = d-\lambda\ .
\label{EQN:lambdas}
\end{equation}
(The symbol $\lambda$ is also used for the transport coefficient in systems
with conserved dynamics. This should not be a source of confusion, as the
meaning will be clear from the context).

Before leaving this section, it is interesting to consider to what extent
the results depend on the specific form (\ref{EQN:SRIC}) chosen for the
correlator of the initial conditions. Let us replace the right-hand side
of (\ref{EQN:SRIC}) by a function $\Delta(|{\bf x}-{\bf x}'|)$, with
Fourier transform $\Delta(k)$. Then $\Delta(k)$ will appear inside
the sum over ${\bf k}$ in Eq.\ (\ref{EQN:b(t)}). The dominant $k$ values
in the sum, however, are of order $t^{-1/2}$, so for late times we can
replace $\Delta(k)$ by $\Delta(0)$, provided the latter exists. This
means that universal results are obtained when only {\em short-range}
spatial correlations are present at $t=0$. For sufficiently {\em long-range}
correlations however, such that $\Delta(k)$ diverges for $k \to 0$, new
universality classes are obtained. We shall return to the role of initial
conditions, from a more general perspective, in section \ref{SEC:RG}.

\subsection{THE LARGE-$n$ LIMIT: CONSERVED FIELDS}
\label{SEC:LARGEnC}
For conserved fields, the calculation proceeds as before, but with an extra
$(-\nabla^2)$ on the right-hand side of the equation of motion. Making as
before, the replacement $\vec{\phi}^2/n \to \langle \phi^2 \rangle$ for
$n \to \infty$, where $\phi$ is (any) one component of $\vec{\phi}$, one
obtains
\begin{equation}
\partial{\phi}/\partial t = -\nabla^4 \phi - a(t)\,\nabla^2\phi\ ,
\end{equation}
with $a(t)$ still given by (\ref{EQN:a(t)}). Transforming to Fourier space,
the solution is
\begin{equation}
\phi_{\bf k}(t) = \phi_{\bf k}(0)\,\exp[-k^4t + k^2 b(t)]\ ,
\label{EQN:FOURIERCOP}
\end{equation}
The function $b(t)$, defined as in (\ref{EQN:FOURIERb}), satisfies the equation
\begin{equation}
a(t) = db/dt = 1 - \Delta \sum_{\bf k} \exp[-2k^4t + 2k^2b(t)]\ .
\label{EQN:b(t)COP}
\end{equation}
This equation was solved by Coniglio and Zannetti \cite{CZ}, by
first expressing the sum over ${\bf k}$ as a parabolic cylinder function,
then taking the large-$t$ limit. Here we will take the large-$t$ limit
from the outset, and recognize that the sum can then be evaluated using
steepest descents. Just as for the nonconserved case, we can show {\em a
posteriori} that $db/dt \ll 1$ at late times, so that this term can be
dropped from (\ref{EQN:b(t)COP}). After the change of variable
${\bf k} = [b(t)/t]^{1/2}{\bf x}$, we obtain
\begin{equation}
1 = \Delta C_d (b/t)^{d/2} \int_0^\infty x^{d-1}dx\,\exp[2\beta(x^2-x^4)]\ ,
\label{EQN:x}
\end{equation}
where $C_d$ is an uninteresting constant, and
\begin{equation}
\beta(t) = b^2(t)/t\ .
\end{equation}
Provided $\beta(t) \to \infty$ for $t \to \infty$ (which can be verified
{\em a posteriori}), the integral on the right of (\ref{EQN:x}) can be
evaluated by steepest descents. Including the gaussian fluctuations
around the maximum of the integrand at $x = 1/\surd{2}$ gives
\begin{equation}
1 = {\rm const}\,\Delta\,\beta^{-1/2}\,(\beta/t)^{d/4}\,\exp(\beta/2)\ ,
\end{equation}
with asymptotic solution
\begin{equation}
\beta \simeq (d/2) \ln t\ ,\ \ \ \ \ \ \ \ t \to \infty\ ,
\label{EQN:beta}
\end{equation}
justifying the use of the steepest descents method for large $t$.
Putting this result into (\ref{EQN:FOURIERCOP}) gives the final result
for the structure factor \cite{CZ}
\begin{eqnarray}
\label{EQN:MULTI}
S({\bf k},t) & \simeq & t^{(d/4)\phi(k/k_m)} \\
         k_m & \simeq & \left(\frac{d}{8}\,\frac{\ln t}{t}\right)^{1/4} \\
     \phi(x) &   =    & 1 - (1-x^2)^2\ .
\end{eqnarray}
Here $k_m(t)$ is the position of the maximum in $S({\bf k},t)$.
A slightly more careful treatment (retaining the leading subdominant term
in (\ref{EQN:beta})), gives an additional logarithmic prefactor, of order
$(\ln t)^{(2-d)/4}$, in (\ref{EQN:MULTI}), such that (asymptotically in time)
$\sum_{\bf k} S({\bf k},t) = 1$.

Eq.\ (\ref{EQN:MULTI}) is interesting because, in contrast to the nonconserved
result (\ref{EQN:LARGENS}), it does not have the conventional scaling form.
Rather it exhibits `multiscaling' \cite{CZ}. In particular there are two,
logarithmically different, length scales, $k_m^{-1}$ and $L=t^{1/4}$.
For simple scaling, these two scales would be the same. Furthermore,
for fixed scaling variable, which can be written as $k/k_m$, the structure
factor would vary as $L(t)^d$, with a {\em prefactor} depending on the
scaling variable. In the multiscaling form (\ref{EQN:MULTI}), for fixed
scaling variable, $S({\bf k},t) \sim L^{d\phi(k/k_m)}$, i.e. the
{\em exponent} depends continuously on the scaling variable.

After the discovery of multiscaling in the $n \to \infty$ limit, some effort
was devoted to looking for similar phenomena at finite $n$, notably for
scalar systems \cite{CZOS}, but also for $n=2$ \cite{MG,SR} and $n=3$
\cite{CR}. However, no evidence was found for any departure from simple
scaling for any finite $n$. At the same time, Bray and Humayun showed,
within the context of an approximate calculation based on an idea of
Mazenko, that simple scaling is recovered asymptotically for any finite
$n$ \cite{BH92}. This result is discussed in detail in section
\ref{SEC:APPROXSF}.

\subsection{THE ONE-DIMENSIONAL ISING MODEL}
\label{SEC:GLAUBER}
An exceptionally simple system that can be solved exactly \cite{Glauber}
is the Ising model in one dimension with Glauber dynamics. It is defined
by the Glauber equation for the spin probability weight:
\begin{eqnarray}
\frac{d}{dt}\,P(S_1, \ldots, S_N;t)  & = & - P(S_1, \ldots, S_N;t)
\sum_i\left(\frac{1-S_i \tanh \beta h_i}{2}\right) \nonumber \\
& & + \sum_i P(S_1,\ldots,-S_i,\ldots,S_N;t)\,
\left(\frac{1+S_i \tanh \beta h_i}{2}\right),
\label{EQN:GLAUBER}
\end{eqnarray}
where $\beta=1/T$, $h_i = J(S_{i-1} + S_{i+1})$ is the local field at
site $i$, and periodic boundary conditions, $S_{i+N}=S_i$, have been
adopted.

{}From (\ref{EQN:GLAUBER}), it is straightforward to derive the equation of
motion for the pair correlation function,
$C_{ij}(t) = \langle S_i(t) S_j(t)\rangle$, where the brackets indicate
an average over the distribution $P$. After averaging also over the initial
conditions, $C_{ij}$ depends only on the difference $r=|i-j|$ if the
ensemble of initial conditions is invariant under translations. Then one
obtains
\begin{equation}
(d/dt) C(r,t) = C(r+1,t) - 2\,C(r,t) + C(r-1,t)\ ,\ \ \ \ \ r \ne 0\ .
\label{EQN:GLAUBER1}
\end{equation}
For $r=0$ one has trivially $C(0,t)=1$ for all $t$. To solve for $C$ in
the scaling limit, it is simplest to take the continuum limit, when
(\ref{EQN:GLAUBER1}) reduces to the diffusion equation,
$\partial C/\partial t = \partial^2 C/\partial r^2$, with constraint
$C(0,t)=1$. A scaling solution obviously requires $L(t) =t^{1/2}$.
Inserting $C(r,t) = f(r/t^{1/2})$ in the diffusion equation gives
$f'' = -(x/2)f'$, which can be integrated with boundary conditions
$f(0)=1$, $f(\infty)=0$ to give $f(x) = {\rm erfc}\,(x/2)$, where
${\rm erfc}$ is the complementary error function. Thus the scaling
solution is
\begin{equation}
C(r,t) = {\rm erfc}\,(r/2t^{1/2})\ .
\label{EQN:erfc}
\end{equation}
In particular, the solution exhibits the expected Porod regime,
$C = 1 - r/(\pi t)^{1/2} + O(r^3/t^{3/2})$ at short distance. A more
complete discussion can be found in \cite{Glauber}.

The scalar TDGL equation (\ref{EQN:MODELA}) is also soluble in one
dimension, in the sense that the scaling functions can be exactly
calculated \cite{Nagai}. When $L(t) \gg \xi$, neighboring domain
walls interact only weakly, with a force of order $\exp(-L/\xi)$,
leading to a logarithmic growth law, $L \sim \xi\ln t$.
Moreover, in the limit $L/\xi \to \infty$, the closest pair of
domain walls interact strongly compared to other pairs, so
that the other walls can be treated as stationary while the closest
pair annihilate. This leads to a simple recursion for the domain
size distribution, with a scaling solution \cite{Nagai}. It is
interesting that the fraction of the line which has never been
traversed by a wall decays with a non-trivial power of the mean
domain size \cite{BDG}. A similar phenomenon (but with a different
power) occurs for Glauber dynamics \cite{DBG}.

\subsection{THE ONE-DIMENSIONAL XY MODEL}
\label{SEC:XY}
As our final example of a soluble model, we consider the case $d=1$, $n=2$
with non-conserved order parameter. The solution, first given by Newman
et al \cite{NBM}, is interesting for the `anomalous' growth law obtained,
$L(t) \sim t^{1/4}$. Here we shall give a more detailed discussion than
appears in \cite{NBM}, emphasizing the scaling violations exhibited
by, in particular, the two-time correlation function.
In section \ref{SEC:GROWTH}, we shall present a general technique,
developed with A. D. Rutenberg \cite{BR}, for determining growth laws
for phase ordering systems. The scaling form (\ref{EQN:TWOTIME}) plays
an important role in the derivation. For the $d=1$, $n=2$ model, however,
our method fails to predict the correct $t^{1/4}$ growth. The reason is
precisely the unconventional form (i.e.\ different from (\ref{EQN:TWOTIME}))
of the two-time correlation function for this system.

It is simplest to work with `fixed length' fields, i.e.\ $\vec{\phi^2}=1$,
with Hamiltonian $F = (1/2)\int dx (\partial\vec{\phi}/\partial x)^2$.
The constraint can be eliminated by the representation
$\vec{\phi} = (\cos\theta,\sin\theta)$, where $\theta$ is the phase angle,
to give $F = (1/2) \int dx (\partial\theta/\partial x)^2$.
The `model A' equation of motion,
$\partial \theta/\partial t = -\delta F/\delta\theta$,
becomes
\begin{equation}
\partial \theta/\partial t = \partial^2 \theta/\partial x^2\ ,
\label{EQN:DIFF}
\end{equation}
i.e.\ a simple diffusion equation for the phase. In general dimensions, it
is difficult to include vortices, which are singularities in the
phase field, in a simple way. Such singularities, however, are absent for
$d=1$.

Eq.\ (\ref{EQN:DIFF}) has to be supplemented by suitable initial conditions.
It is convenient to choose the probability distribution for $\theta(r,0)$
to be gaussian: in Fourier space
\begin{equation}
P(\{\theta_k(0)\}) \propto \exp[-\sum_k(\beta_k/2)\theta_k(0)\theta_{-k}(0)]\ .
\label{EQN:1DXYIC}
\end{equation}
Then the real-space correlation function at $t=0$ is readily evaluated using
the gaussian property of the $\{\theta_k(0)\}$:
\begin{eqnarray}
C(r,0) & = & \langle \cos[\theta(r,0)-\theta(0,0)] \rangle \nonumber \\
       & = & \exp\{-(1/2)\langle [\theta(r,0)-\theta(0,0)]^2 \rangle\}
\nonumber \\
       & = & \exp\{-\sum_k (1-\cos kr)/\beta_k\}\ .
\label{EQN:t=0}
\end{eqnarray}
The choice $\beta_k = (\xi/2) k^2$ yields $C(r,0) = \exp(-|r|/\xi)$,
appropriate to a quench from a disordered state with correlation length $\xi$.

The general two-time correlation function can be calculated in the same
way \cite{Kay}. Using $\beta_k = (\xi/2)k^2$, and the solution
$\theta_k(t) = \theta_k(0) \exp(-k^2t)$ of (\ref{EQN:DIFF}), gives
\begin{eqnarray}
C(r,t_1,t_2) & = & \langle \cos[\theta(r,t_1) - \theta(0,t_2)] \rangle
\nonumber \\
 & = & \exp\{-(1/2)\langle[\theta(r,t_1) - \theta(0,t_2)]^2\rangle\}
\nonumber \\
 & = & \exp\left(-\sum_k\frac{1}{\xi k^2}
\left\{[\exp(-k^2t_1)-\exp(-k^2t_2)]^2 \right.\right.\nonumber \\
 &  & \ \ \ \ \ \ \ \ \left.\left.+2(1-\cos kr) \exp\{-k^2(t_1+t_2)\}
\right\}\right)\ .
\end{eqnarray}
Since we shall find that $r$ is scaled by $(t_1+t_2)^{1/4}$, we can take
$kr \ll 1$ in the summand for the $r$-values of interest, i.e.\ we can
replace $(1-\cos kr)$ by $(kr)^2/2$. Evaluation of the sums then gives
\begin{equation}
C(r,t_1,t_2) = \exp\left\{-\frac{1}{\xi\surd{\pi}}
\left[\frac{r^2}{2(t_1+t_2)^{1/2}} + 2(t_1+t_2)^{1/2}
- (2t_1)^{1/2} - (2t_2)^{1/2}\right]\right\}\ .
\label{EQN:1dXY}
\end{equation}

For the special case $t_1=t_2=t$, (\ref{EQN:1dXY}) reduces to
$C(r,t,t) = \exp(-r^2/2\xi(2\pi t)^{1/2})$, which has the standard
scaling form (\ref{EQN:STRUCT}), with growth law $L(t) \sim t^{1/4}$.
This growth law is unusual: we shall show in section \ref{SEC:GROWTH}
that the generic form for nonconserved fields is $L(t) \sim t^{1/2}$,
just as in the large-$n$ result (\ref{EQN:LARGENC}). Another, related,
feature of (\ref{EQN:1dXY}) is the explicit appearance of $\xi$, the
correlation length for the initial condition. The large-$n$ solution,
for example, becomes independent of $\xi$ for $L(t) \gg \xi$. The
most striking feature of (\ref{EQN:1dXY}), however, is the breakdown of
the scaling form (\ref{EQN:TWOTIME}) for the two-time correlations.
It is this feature that invalidates the derivation of the result
$L(t) \sim t^{1/2}$ given in section \ref{SEC:GROWTH}. It is possible
that a similar anomalous scaling is present in the {\em conserved}
$d=1$ XY model, for which simulation results \cite{MG,CR} suggest
$L(t) \sim t^{1/6}$, instead of the $t^{1/4}$ growth derived in
section \ref{SEC:GROWTH} assuming simple scaling for two-time
correlations. Unfortunately, an exact solution for the conserved
case is non-trivial.

The explicit dependence of (\ref{EQN:1dXY}) on $\xi$ suggests an unusual
sensitivity to the initial conditions in this system. A striking
manifestation of this is obtained by choosing initial conditions with a
non-exponential decay of correlations. For example, choosing
$\beta_k \sim |k|^\alpha$ for small $|k|$ in (\ref{EQN:1DXYIC}) gives,
via (\ref{EQN:t=0}), $C(r,0) \sim \exp(- const\, |r|^{\alpha-1})$ for
large $|r|$, provided $1<\alpha<3$. The calculation of the pair correlation
function is again straightforward. For example, the equal-time function
has a scaling form given by
$C(r,t) = \exp(- const\,r^2/t^{(3-\alpha)/2})$, implying a growth
law $L(t) \sim t^{(3-\alpha)/4}$, but the two-time correlation still does
not scale properly.

An especially interesting case is $\alpha=1$, which generates power-law
spatial correlations in the initial condition. Thus we choose
$\beta_k = |k|/\gamma$ for $|k| \le \Lambda$, and $\beta_k =\infty$ for
$|k| > \Lambda$, where $\Lambda$ is an ultraviolet cut-off. Then the
initial-condition correlator has the form
$C(r,0) \sim (\Lambda r)^{-\gamma/\pi}$ for $\Lambda r \gg 1$.
The general two-time correlation function now has the conventional scaling
form (\ref{EQN:TWOTIME}), with $L(t) \sim t^{1/2}$. Its form is
\begin{equation}
C(r,t_1,t_2) = f\left(\frac{r}{(t_1+t_2)^{1/2}}\right)\,
\left(\frac{4t_1 t_2}{(t_1 + t_2)^2}\right)^{\gamma/4\pi}\ ,
\end{equation}
where $f(x)$ is the equal-time correlation function. In particular, for
$t_2 \gg t_1$ this gives $C(r,t_1,t_2) \simeq f(r/\surd{t_2})\,
(4t_1/t_2)^{\gamma/4\pi}$, so the exponent $\bar{\lambda}$ defined by
(\ref{EQN:lambdabar}) is $\gamma/2\pi$ for this model. Also the large
distance behaviour of the equal-time correlation function is
$f(x) \sim x^{-\gamma/\pi}$, exhibiting the same power-law decay as the
initial condition. These two results are in complete agreement with the
general treatment \cite{IC} of initial conditions with power-law
correlations given in section \ref{SEC:IC}.

\section{APPROXIMATE THEORIES FOR SCALING FUNCTIONS}
\label{SEC:APPROXSF}
While the determination of growth laws (i.e.\ the form of $L(t)$)
has proved possible using fairly simple arguments (as in section
\ref{SEC:L(t)}), which can be made precise by the use of exact relations
between correlation functions (section \ref{SEC:GROWTH}), or
Renormalization Group methods (section \ref{SEC:RG}), the calculation
of scaling functions, e.g. the pair correlation
scaling function $f(x)$ (see Eq.\ (\ref{EQN:STRUCT})), has been a
long-standing challenge. In the previous section we have shown
that this function can be calculated exactly in a number of
soluble models. With the exception of the $1-d$ Glauber model,
however, these models lack the topological defects that play
such an important role in realistic models. In particular, these
defects are responsible for the power-law tail (\ref{EQN:GENPOROD})
in the structure factor.

In this section we will review some of the approximate theories
that have been put forward for the scaling function $f(x)$ of the
pair correlation function. The most successful by far are theories
for nonconserved fields. Even these, however, are not quite as
good as has been believed, as we shall show. We shall propose a
new approach which can in principle lead to systematically
improvable calculations of scaling functions for nonconserved fields.
For conserved fields the theory is in a less satisfactory state.
We shall try to give some indication of why this is so. Finally
we emphasize that the discussion is limited throughout to the late-stage
scaling regime.

\subsection{NONCONSERVED FIELDS}
A number of approximate scaling functions have been
proposed for non-conserved fields, but in my view none of them is
completely satisfactory. The most physically appealing approach for scalar
fields is that of Ohta, Jasnow and Kawasaki (OJK) \cite{OJK}, which starts
from the Allen-Cahn equation (\ref{EQN:AC}) for the interfaces. Below
we will review the OJK method, as well as an earlier approach by
Kawasaki, Yalabik and Gunton (KYG) \cite{KYG}, and more recent work by
Mazenko \cite{Maz89,Maz90,Maz91}.
Finally we discuss in detail a new approach \cite{BHsys} which has the virtue
that it can, in principle, be systematically improved.

\subsubsection{The OJK Theory}
\label{SEC:OJK}
A common theme, introduced by Ohta, Jasnow, and Kawasaki \cite{OJK} (OJK),
in the approximate theories of scaling functions is the replacement
of the physical field $\phi({\bf x},t)$, which is $\pm 1$ everywhere
except at domain walls, where it varies rapidly,  by an  auxiliary
field $m({\bf x},t)$, which varies smoothly in space. This is achieved
by using a non-linear function $\phi(m)$ with a `sigmoid' shape
(such as $\tanh m$).
In the OJK theory, the dynamics of the domain walls themselves,
defined by the zeros of $m$, are considered. The normal velocity of a point
on the interface is given by the Allen-Cahn equation (9),
$v = -K=-\nabla \cdot {\bf n}$, where $K$ is the
curvature, and ${\bf n} = \nabla m/|\nabla m|$ is a unit vector normal
to the wall. This gives
\begin{equation}
v = \{-\nabla^2 m + n_a n_b \nabla_a \nabla_b m\}
/|\nabla m|\ .
\label{EQN:OJK1}
\end{equation}
In a frame of reference comoving with the interface,
\begin{equation}
dm/dt = 0 = \partial m/\partial t + {\bf v}\cdot\nabla m\ .
\label{EQN:COMOVE}
\end{equation}
But since ${\bf v}$ is parallel to $\nabla m$ (and defined in the same
direction), ${\bf v}.\nabla m = v|\nabla m|$ so
\begin{equation}
v = - \frac{1}{|\nabla m|}\,\frac{\partial m}{\partial t}\ .
\label{EQN:OJK2}
\end{equation}
Eliminating $v$ between (\ref{EQN:OJK1}) and (\ref{EQN:OJK2}) gives the
OJK equation
\begin{equation}
\partial m/\partial t = \nabla^2 m - n_a n_b \nabla_a \nabla_b m\ .
\label{EQN:OJK3}
\end{equation}
Since ${\bf n} = \nabla m/|\nabla m|$, this equation is non-linear.
To make further progress, OJK made the simplifying approximation of
replacing $n_a n_b$ by its spherical average $\delta_{ab}/d$,
obtaining the simple diffusion equation
\begin{equation}
\partial m/\partial t = D\,\nabla^2 m\ ,
\label{EQN:OJK4}
\end{equation}
with diffusion constant $D=(d-1)/d$.

Providing there are no long-range correlations present, we do not expect
the form of the random initial conditions to play an important role
in the late-stage scaling. A convenient choice is a gaussian distribution
for the field $m({\bf x},0)$, with mean zero and correlator
\begin{equation}
\langle m({\bf x},0) m({\bf x'},0) \rangle
= \Delta \delta({\bf x}-{\bf x'})\ .
\label{EQN:OJKIC}
\end{equation}
Then the linearity of (\ref{EQN:OJK4}) ensures that the field $m({\bf x},t)$
has a gaussian distribution at all times. Solving (\ref{EQN:OJK4}),
and averaging over initial conditions using (\ref{EQN:OJKIC}) gives the
equal-time correlation function
\begin{equation}
\langle m(1) m(2) \rangle = \frac{\Delta}{(8\pi Dt)^{d/2}}\,
\exp\left(-\frac{r^2}{8Dt}\right)\ ,
\end{equation}
where `1' and `2' represent space points separated by $r$. Of special
relevance in what follows is the normalized correlator
\begin{equation}
\gamma(12) \equiv \frac{\langle m(1) m(2) \rangle}
             {\langle m(1)^2 \rangle^{1/2}\langle m(2)^2 \rangle^{1/2}}
           = \exp\left(-\frac{r^2}{8Dt}\right)\ .
\label{EQN:gamma}
\end{equation}
The generalization to different times is straightforward \cite{YeungJasnow}
and will be given explicitly below.

To calculate the pair correlation function of the original field $\phi$,
we need to know the joint probability distribution for $m(1)$ and $m(2)$.
For a gaussian field this can be expressed in terms of the second
moments of $m$:
\begin{equation}
P\left(m(1),m(2)\right) = N\,\exp\left(-\frac{1}{2(1-\gamma^2)}\,
\left[\frac{m(1)^2}{S_0(1)}+\frac{m(2)^2}{S_0(2)} -2\gamma\frac{m(1)m(2)}
{\sqrt{S_0(1)S_0(2)}}\right]\right)\ ,
\label{EQN:JOINTPROB}
\end{equation}
where $\gamma=\gamma(12)$, and
\begin{equation}
S_0(1)=\langle m(1)^2 \rangle,\ \ S_0(2)=\langle m(2)^2 \rangle,\ \
N = (2\pi)^{-1}[(1-\gamma^2)S_0(1)S_0(2)]^{-1/2}\ .
\label{EQN:ESSES}
\end{equation}
Note that (\ref{EQN:JOINTPROB}) is a general expression for the joint
probability distribution of a gaussian field, with $\gamma$ defined by the
first part of (\ref{EQN:gamma}). Now `1' and `2'  represent arbitrary
space-time points. For the special case where $m$ obeys the diffusion
equation (\ref{EQN:OJK4}), $\gamma$ is given by
\begin{equation}
\gamma = \left(\frac{4t_1t_2}{(t_1+t_2)^2}\right)^{d/4}\,
\exp\left(-\frac{r^2}{4D(t_1+t_2)}\right)\ ,
\label{EQN:TWOTIMEgamma}
\end{equation}
a simple generalization of (\ref{EQN:gamma}).

The pair correlation function is given by
$C({\bf r},t) = \langle \phi\left(m(1)\right) \phi\left(m(2)\right) \rangle$.
In the scaling regime, one can replace the function $\phi(m)$ by
${\rm sgn}\,(m)$, because the walls occupy a negligible volume fraction.
In a compact notation,
\begin{equation}
C(12) =  \langle {\rm sgn}\,m(1)\,{\rm sgn}\,m(2) \rangle
= (2/\pi)\,\sin^{-1}(\gamma)\ .
\label{EQN:ARCSIN}
\end{equation}
The gaussian average over the field $m$ required in (\ref{EQN:ARCSIN}) is
standard (see, e.g., \cite{Oono88}). Eqs.\ (\ref{EQN:gamma}) and
(\ref{EQN:ARCSIN}) define the `OJK scaling function' for equal-time pair
correlations. Note that (apart from the trivial dependence through $D$)
it is independent of the spatial dimension $d$. We will
present arguments that it becomes exact in the large-$d$ limit.
The OJK function fits experiment and simulation data very well.
As an example, we show the function $f(x)$ for the $d=2$ scalar theory
in Figure 12.

The general two-time correlation function is especially interesting in the
limit $t_1 \gg t_2$ that defines (see, e.g., (\ref{EQN:TWO-TIME})) the
exponent $\bar{\lambda}$. Since $\gamma \ll 1$ in this limit,
(\ref{EQN:ARCSIN}) can be linearised in $\gamma$ to give
$C({\bf r},t_1,t_2) \sim (4t_1/t_2)^{d/4}\,\exp(-r^2/4Dt_2)$, i.e.\
$\bar{\lambda}=d/2$ within the OJK approximation.

\subsubsection{The KYG Method}
\label{SEC:KYG}

An earlier approach, due to Kawasaki, Yalabik and Gunton (KYG)
\cite{KYG}, building on still earlier work of Suzuki \cite{Suzuki},
was based on an approximate resummation of the direct perturbation
series in the non-linearity, for the quartic potential
$V(\phi) = (1/4)(1-\phi^2)^2$. The equation of motion (\ref{EQN:MODELA})
for this potential is
\begin{equation}
\partial \phi/\partial t = \nabla^2 \phi + \phi - g \phi^3\ ,
\label{EQN:KYG1}
\end{equation}
with $g=1$. The basic idea is treat $g$ as small, expand
in powers of $g$, extract the leading asymptotic (in $t$) behaviour
of each term in the series, and set $g=1$ at the end.
However, an uncontrolled approximation is made in simplifying the
momentum dependence of each term (the expansion is performed in Fourier
space). After setting $g=1$, the final result
can be expressed in terms of the mapping
\begin{equation}
\phi(m) = m/(1+m^2)^{1/2}\ .
\label{EQN:KYG2}
\end{equation}
It is found that $m$ obeys the equation
\begin{equation}
\partial m/\partial t = \nabla^2 m + m\ ,
\label{EQN:KYG3}
\end{equation}
instead of (\ref{EQN:OJK4}), which gives an exponential growth superimposed on
the diffusion. After the replacement $\phi(m) \to {\rm sgn}(m)$, however,
this drops out: the OJK scaling function (\ref{EQN:ARCSIN}) is recovered,
with $\gamma$ given by (\ref{EQN:gamma}) (but with $D=1$).

The nature of the approximation involved can be clarified by putting
(\ref{EQN:KYG2}) into (\ref{EQN:KYG1}) (with $g=1$) to derive the exact
equation satisfied by $m$:
\begin{equation}
\frac{\partial m}{\partial t} = \nabla^2 m + m
                                   - 3\,\frac{m(\nabla m)^2}{1+m^2}\ .
\end{equation}
In contrast to a claim made in \cite{KYG}, there is no reason to neglect
the final term. On a physical level, the fact that this approach gives
the correct growth law, $L(t) \sim t^{1/2}$, seems to be fortuitous
(see the discussion in section \ref{SEC:GROWTH}). In particular, the
crucial role of the interfacial curvature in driving the growth is
not readily apparent in this method. By contrast the OJK approach,
while giving the same final result, clearly contains the correct
physics.

Despite its shortcomings, the KYG method has the virtue that it can be
readily extended to vector fields \cite{BP,Puri90}.
Eq.\ (\ref{EQN:KYG3}) is again obtained,
but with $m$ replaced by a vector auxiliary field $\vec{m}$, with
$\vec{\phi} = \vec{m}/(1+\vec{m}^2)^{1/2}$.
At late times, $\vec{\phi} \to \hat{m}$, a unit vector, almost everywhere
and $C(12) = \langle \hat{m}(1)\cdot\hat{m}(2) \rangle$. Taking gaussian
initial conditions for $\vec{m}$, the resulting scaling function is
\cite{BP}, with $\gamma$ again given by (\ref{EQN:gamma}) (but with $D=1$),
\begin{equation}
C(12) = \frac{n\gamma}{2\pi}\,\left[B\left(\frac{n+1}{2},\frac{1}{2}\right)
\right]^2\,F\left(\frac{1}{2},\frac{1}{2};\frac{n+2}{2};\gamma^2\right)\ ,
\label{EQN:BPT}
\end{equation}
where $B(x,y)$ is the beta function and $F(a,b;c;z)$ the hypergeometric
function $_2F_1$. The same scaling function was obtained independently by
Toyoki \cite{Toy92}. We will call it the `BPT scaling function'. The
result (\ref{EQN:gamma}) for $\gamma$ implies $L(t) \sim t^{1/2}$ for
all $n$ within this approximation.

Both (\ref{EQN:OJK4}) and (\ref{EQN:KYG3}) suffer from the weakness that
(for scalar fields) the width of the interface changes systematically
with time.
Since $\phi(m)$ is linear in $m$ for small $m$, and $|\nabla \phi|$ is
fixed (by the interface profile function) in the interface,
we expect $\langle (\nabla m)^2 \rangle = {\rm const}$.
Eqs.\ (\ref{EQN:OJK4}) and (\ref{EQN:KYG3}), however, give
$\sim t^{-(d+2)/2}$ and $\sim \exp(2t)/t^{(d+2)/2}$ for this quantity,
corresponding to increasing and decreasing interface widths respectively.
Oono and Puri \cite{Oono88} showed that this unphysical feature can be
eliminated by introducing an extra term $h(t)\,m$ in (\ref{EQN:OJK4}).
Since this term vanishes at the interfaces, where $m=0$, it's inclusion
does not change the underlying physics. Fixing $h(t)$ by the requirement
$\langle (\nabla m)^2 \rangle = {\rm const.}$ gives $h(t) \simeq (d+2)/4t$
at late times. The scaling function (\ref{EQN:ARCSIN}), however, is
unaffected by the presence of the extra term. In section \ref{SEC:SYSTEMATIC}
we shall find that the Oono-Puri result arises naturally within a systematic
treatment of the problem.

\subsubsection{Mazenko's Method}
\label{SEC:MAZENKO}
In an interesting series of papers, Mazenko \cite{Maz89,Maz90,Maz91} has
introduced a new approach that deals with the interface in a natural way.
This approach combines a clever choice for
the function $\phi(m)$ with the minimal assumption that the field $m$
is gaussian. Specifically $\phi(m)$ is chosen to be the {\em equilibrium
interface profile function}, defined by (compare Eq.\ (\ref{EQN:PROFILE}))
\begin{equation}
\phi''(m) = V'(\phi)\ ,
\label{EQN:MAZENKO1}
\end{equation}
with boundary conditions $\phi(\pm \infty)=\pm 1$, $\phi(0)=0$. The field
$m$ then has a physical interpretation, near walls, as a coordinate
normal to the wall. Note that this mapping transforms a problem with {\em two}
length scales, the domain scale $L(t)$ and the interface width $\xi$, into
one with only a {\em single} length scale, namely $L(t)$ (see Fig.\ 13).
With the choice (\ref{EQN:MAZENKO1}) for $\phi(m)$, the TDGL equation
(\ref{EQN:MODELA}) becomes
\begin{equation}
\partial_t \phi = \nabla^2 \phi - \phi''(m)\ .
\label{EQN:MAZENKO2}
\end{equation}
Multiplying by $\phi$ at a different space point and averaging over
initial conditions gives
\begin{equation}
(1/2)\partial_t C(12) = \nabla^2 C(12) - \langle \phi''(m(1))\,\phi(m(2))
\rangle\ .
\label{EQN:MAZENKO3}
\end{equation}

So far this is exact. In order to simplify the final term in
(\ref{EQN:MAZENKO3}), Mazenko assumes that $m$ can be treated as
a gaussian field.  Then the final term can be expressed in terms of
$C(12)$ itself as follows, exploiting the Fourier decomposition
of $\phi(m)$ and the gaussian property of $m$ \cite{Maz90}:
\begin{eqnarray}
\langle \phi''(m(1))\,\phi(m(2))\rangle\ & = &
\sum_{k_1,k_2} \phi_{k_1}\phi_{k_2}
(-k_1^2)\langle \exp[ik_1m(1)+ik_2m(2)]\rangle \nonumber \\
 & = & \sum_{k_1,k_2} \phi_{k_1}\phi_{k_2} (-k_1^2)
\exp[ - k_1^2 S_0(1)/2 - k_2^2 S_0(2)/2 \nonumber \\
    &  & \hspace{5cm} - k_1k_2 C_0(12)] \nonumber \\
 & = & 2\,\partial C(12)/\partial S_0(1)\ .
\label{EQN:MAZENKO4}
\end{eqnarray}
where $S_0(1)$, $S_0(2)$, are given by (\ref{EQN:ESSES}) and
$C_0(12) = \langle m(1) m(2) \rangle$. The derivative in
(\ref{EQN:MAZENKO4}) is taken holding $S_0(2)$ and $C_0(12)$ fixed.
Since, from the definition (\ref{EQN:gamma}),
$\gamma(12) = C_0(12)/\sqrt{S_0(1)S_0(2)}$, the general result
(\ref{EQN:ARCSIN}) for gaussian fields implies
\begin{eqnarray}
2\frac{\partial C(12)}{\partial S_0(1)} & = & 2\frac{dC(12)}{d\gamma(12)}\,
\frac{\partial\gamma(12)}{\partial S_0(1)} \nonumber \\
 & = & a(t)\,\gamma(12)\,\frac{dC(12)}{d\gamma(12)}\ .
\label{EQN:MAZENKO5}
\end{eqnarray}
where
\begin{equation}
a(t) = 1/S_0(1) = \langle m(1)^2 \rangle^{-1}\ .
\label{EQN:MAZENKO6}
\end{equation}
Putting it all together, and suppressing the arguments, the final equation
for $C$ is
\begin{equation}
(1/2)\,\partial_t C = \nabla^2 C + a(t)\,\gamma dC/d\gamma\ .
\label{EQN:MAZENKO7}
\end{equation}

Using (\ref{EQN:ARCSIN}) for $C(\gamma)$ gives
$\gamma dC/d\gamma = (2/\pi) \tan [(\pi/2) C]$. Then (\ref{EQN:MAZENKO7})
becomes a closed non-linear equation for $C$. For a scaling solution,
one requires $L(t) \sim t^{1/2}$ and $a(t)=\lambda/2t$ for large $t$ in
(\ref{EQN:MAZENKO7}), so that each of the terms scales as $1/t$ times
a function of the scaling variable $r/t^{1/2}$. Setting $C({\bf r},t)
=f(r/t^{1/2})$ gives the equation
\begin{equation}
0 = f'' + \left(\frac{d-1}{x} + \frac{x}{4}\right)f'
     + \frac{\lambda}{\pi}\,\tan\left(\frac{\pi}{2}f\right)
\label{EQN:MAZENKO8}
\end{equation}
for the scaling function $f(x)$. The constant $\lambda$ is fixed by
the requirement that the large-distance behaviour of $C$ be
physically reasonable \cite{Maz90}. Linearization of
(\ref{EQN:MAZENKO8}) (valid for large $x$) leads to two linearly
independent large-$x$ solutions with gaussian and power-law tails.
The constant $\lambda$ is chosen to eliminate the `unphysical'
power-law term.

It is straightforward to adapt this approach to nonconserved
vector fields \cite{LM,BH}. A significant simplification is that for
gaussian fields, the joint probability distribution for $\vec{m}(1)$
and $\vec{m}(2)$ factors into a product of separate distributions
of the form (\ref{EQN:JOINTPROB}) for each component. This results
is an equation of form (\ref{EQN:MAZENKO7}) for any $n$, but with the
function $C(\gamma)$ given by (\ref{EQN:BPT}) for general $n$ instead
of (\ref{EQN:ARCSIN}). Again, $a(t)=\lambda/2t$, with $\lambda$ chosen
to eliminate the power-law tail in the scaling function $f(x)$.
The values $\lambda$ for various $n$ and $d$ are given in table 1.

It is interesting that the `unphysical' power-law tails in real space
become physical when sufficiently long-range spatial correlations
are present in the initial state. This will be shown using Renormalisation
Group arguments \cite{IC} in section \ref{SEC:IC}. It also emerges
within the Mazenko treatment \cite{BHLR}.

The general two-time correlation function $C({\bf r},t_1,t_2)$ can also be
evaluated within this scheme \cite{LM,BH}. It is given by a simple
generalisation of (\ref{EQN:MAZENKO7}), namely
\begin{equation}
\partial_{t_1} C = \nabla^2 C + a(t_1)\,\gamma dC/d\gamma\ ,
\end{equation}
with $a(t_1) = \lambda/2t_1$. This equation simplifies for $t_1 \gg t_2$,
because $C$ is then small and the linear relation between $C$ and $\gamma$
for small $C$ (see Eq.\ (\ref{EQN:BPT})) implies $\gamma dC/d\gamma = C$,
i.e.\
\begin{equation}
\partial_{t_1} C = \nabla^2 C + (\lambda/2t_1)\,C\ ,\ \ \ \ \ t_1 \gg t_2\ .
\label{EQN:LINEAR}
\end{equation}
This linear equation can be solved by spatial Fourier transform. Choosing
an initial condition at $t_1 = \alpha t_2$, with $\alpha \gg 1$ to justify
the use of (\ref{EQN:LINEAR}) for all $t_1 \ge \alpha t_2$, gives
\begin{equation}
S({\bf k},t_1,t_2) = \left(\frac{t_1}{\alpha t_2}\right)^{\lambda/2}
  \exp\{-k^2(t_1-\alpha t_2)\}\,S({\bf k},\alpha t_2,t_2)\ .
\end{equation}
Imposing the scaling form $S({\bf k},\alpha t_2,t_2) = t_2^{d/2}g(k^2t_2)$,
with $g(0) = {\rm constant}$, and Fourier transforming back to real space
gives, for $t_1 \gg \alpha t_2$,
\begin{equation}
C(12) = {\rm constant}\,\left(\frac{t_2}{t_1}\right)^{(d-\lambda)/2}\,
\exp\left(-\frac{r^2}{4t_1}\right)\ .
\label{EQN:MAZTWO-TIME}
\end{equation}
The constant cannot be determined from the linear equation alone: it is,
of course, independent of $\alpha$.

Comparison of (\ref{EQN:MAZTWO-TIME}) with the general form
(\ref{EQN:TWO-TIME}), shows that $\bar{\lambda} = d - \lambda$, i.e.\
the parameter $\lambda$ of the Mazenko theory is precisely the exponent
$\lambda$ associated with the response function $G({\bf k},t)$
(Eq.\ (\ref{EQN:GSCALING})), related to $\bar{\lambda}$ by
(\ref{EQN:lambdas}). This connection was first pointed out by Liu and
Mazenko \cite{LM91}. The values of $\lambda$ obtained (table 1) are
in reasonable agreement with those extracted from simulations
\cite{FH88,NBM,BHlambda,BHvector}. For example, for the scalar theory
in $d=2$ simulations \cite{FH88,BHlambda,LM91} give $\lambda \simeq 0.75$
(argued to be 3/4 exactly in \cite{FH88}), compared to $0.711$ from table 1.
\begin{center}
\begin{tabular}{ccccc}\hline
  &     &     &     & \\
d & n=1 & n=2 & n=3 & n=4 \\
  &     &     &     &    \\ \hline
  &     &     &     &    \\
1 &  0  & 0.301 & 0.378 & 0.414 \\
  &     &     &     &    \\
2 & 0.711 & 0.829 & 0.883 & 0.912 \\
  &     &     &     &    \\
3 & 1.327 & 1.382 & 1.413 & 1.432 \\
  &     &     &     &    \\ \hline
\end{tabular}

\medskip

\underline{Table 1} Exponent $\lambda$ within the Mazenko theory.
\end{center}

To summarise, the virtues of Mazenko's approach are (i) only the
assumption that the field $m$ is gaussian is required, (ii) the scaling
function has a non-trivial dependence on $d$ (whereas, apart from the
trivial dependence through the diffusion constant $D$,
(\ref{EQN:gamma}), (\ref{EQN:ARCSIN}) and (\ref{EQN:BPT}) are
independent of $d$), and (iii) the non-trivial behaviour of
{\em different-time} correlation functions \cite{NB90} emerges in a
natural way \cite{LM91}. In addition, the OJK result (\ref{EQN:ARCSIN}),
and its generalisation (\ref{EQN:BPT}), are reproduced for  $d \to \infty$,
while the exact scaling function (\ref{EQN:erfc}) of the $1-d$ Glauber
model is recovered from (\ref{EQN:MAZENKO8}) in the
limit $d \to 1$ \cite{LMHighLow}.
In practice, however, for $d \ge 2$ the shape of the scaling
function $f(x)$ differs very little from that of the OJK function given
by (\ref{EQN:ARCSIN}) and (\ref{EQN:gamma}), or its generalization
(\ref{EQN:BPT}) for vector fields \cite{BH}.  All these functions are in
good agreement with  numerical simulations. The Mazenko function for $n=1$,
$d=2$ is included in Figure 12, while the BPT results for vector fields
are compared with simulations in Figures 14 and 15. The Mazenko approach
can also be used, with some modifications, for conserved scalar \cite{Maz91}
and vector \cite{BH92} fields.

To conclude this section we note that the crucial gaussian approximation,
used in all of these theories, has recently been critically discussed
by Yeung et al.\ \cite{YSO}. By explicit simulation they find that the
distribution $P(m)$ for the field $m$ at a single point is flatter than a
gaussian at small $m$. In section \ref{SEC:SHORT} we shall show that the
joint distribution $P(m(1),m(2))$ can be calculated analytically when
$|m(1)|$, $|m(2)|$ and $r$ are all small compared to $L(t)$. The result
is non-gaussian, but is consistent with the gaussian form
(\ref{EQN:JOINTPROB}) in the limit $d \to \infty$. Below, we present
evidence that the gaussian approximation becomes exact for
$d \to \infty$. Finally we note that very recent work by Mazenko presents
a first attempt to go beyond the gaussian approximation \cite{Mazenko94}.

\subsection{A SYSTEMATIC APPROACH}
\label{SEC:SYSTEMATIC}
All of the treatments discussed above suffer from the disadvantage that
they invoke an uncontrolled approximation at some stage. Very recently,
however, a new approach has been developed \cite{BHsys} which recovers
the OJK and BPT scaling functions in leading order, but has the
advantage that it can, in principle, be systematically improved.

\subsubsection{Scalar Fields}
For simplicity of presentation, we will begin with scalar fields.
The TDGL equation for a non-conserved scalar field $\phi({\bf x},t)$
is given by Eq.\ (\ref{EQN:MODELA}). We recall that, according to the
Allen-Cahn equation (\ref{EQN:AC}), the interface motion is determined
solely by the local curvature. It follows that the detailed form of the
potential $V(\phi)$ is not important, a fact that we can
usefully exploit: the principal role of the double-well potential
is to establish and maintain well-defined interfaces.

Following Mazenko \cite{Maz90} we define the function $\phi(m)$ by
Eq.\ (\ref{EQN:MAZENKO1}) with boundary conditions
$\phi(\pm\infty) = \pm 1$. We have noted that $\phi(m)$
is just the equilibrium domain-wall profile function, with $m$
playing the role of the distance from the wall. Therefore, the spatial
variation of $m$ near a domain wall is completely smooth (in fact, linear).
The additional condition $\phi(0)=0$ locates the center of the wall at $m=0$.
Figure 13 illustrates the difference between $\phi$ and $m$ for a cut
through the system. Note that, while $\phi$ saturates in the interior
of domains, $m$ is typically of order $L(t)$, the domain scale.
Rewriting (\ref{EQN:MODELA}) in terms of $m$, and using
(\ref{EQN:MAZENKO1}) to eliminate $V'$, gives
\begin{equation}
\partial_t m = \nabla^2 m - \frac{\phi''(m)}{\phi'(m)}\,(1-(\nabla m)^2)\ .
\label{EQN:SYS1}
\end{equation}

For general potentials $V(\phi)$, Eq.\ (\ref{EQN:SYS1}) is a complicated
non-linear equation, not obviously simpler than the original TDGL equation
(\ref{EQN:MODELA}).
For reasons discussed in section \ref{SEC:AC}, however, we expect the scaling
function $f(x)$ to be {\em independent} both of the detailed form of the
potential and of the particular choice for the distribution of initial
conditions. Physically, the motion of the interfaces
is determined by their {\em curvature}. The potential
$V(\phi)$ determines the domain wall {\em profile}, which is irrelevant
to the large-scale structure.

Similarly, the initial conditions determine the early-time locations
of the walls, which should again be irrelevant for late-stage scaling
properties. For example, in Mazenko's approximate theory, both the
potential and the initial conditions drop out from the equation for
the scaling function $f(x)$.

The key step in the present approach is to exploit the notion that the
scaling function should be independent of the potential (or, equivalently,
independent of the wall profile) by choosing a particular $V(\phi)$
such that Eq.\ (\ref{EQN:SYS1}) takes a much simpler form
(Eq.\ (\ref{EQN:SYS4})).
Specifically we choose the domain-wall profile function $\phi(m)$ to satisfy
\begin{equation}
\phi''(m) = -m\,\phi'(m)\ .
\label{EQN:SYS2}
\end{equation}
This is equivalent, via (\ref{EQN:MAZENKO1}), to a particular choice of
potential, as discussed below. First we observe that (\ref{EQN:SYS2})
can be integrated, with boundary conditions
$\phi(\pm \infty) = \pm 1$ and $\phi(0)=0$ to give the wall profile function
\begin{equation}
\phi(m) = (2/\pi)^{1/2}\int_0^mdx\,\exp(-x^2/2) = {\rm erf}\,(m/\surd{2})\ ,
\label{EQN:SYS3}
\end{equation}
where ${\rm erf}\,(x)$ is the error function. Also, (\ref{EQN:MAZENKO1})
can be integrated once, with the zero of potential defined by
$V(\pm 1)=0$, to give
\begin{equation}
V(\phi) = (1/2)\,(\phi')^2 = (1/\pi)\,\exp(-m^2)
= (1/\pi)\,\exp(-2[{\rm erf}^{-1}\,(\phi)]^2)\ ,
\label{EQN:SYSPOT}
\end{equation}
where ${\rm erf}^{-1}\,(x)$ is the inverse function of ${\rm erf}\,(x)$.
In particular, $V(\phi) \simeq 1/\pi - \phi^2/2$ for $\phi^2 \ll 1$,
while $V(\phi) \simeq (1/4)(1-\phi^2)^2|\ln(1-\phi^2)|$ for
$(1-\phi^2) \ll 1$ \cite{Note2}.

With the choice (\ref{EQN:SYS2}), Eq.\ (\ref{EQN:SYS3}) reduces to the
much simpler equation
\begin{equation}
\partial_t m = \nabla^2 m + (1-(\nabla m)^2)\,m\ .
\label{EQN:SYS4}
\end{equation}
This equation, though still non-linear, represents a significant
simplification of the original TDGL equation. It is clear, however,
on the basis of the physical arguments discussed above, that it
retains all the ingredients necessary to describe the universal
scaling properties.

We now proceed to show that the usual OJK result is recovered
by simply replacing $(\nabla m)^2$ by its average (over the ensemble of
initial conditions) in (\ref{EQN:SYS4}), and choosing a gaussian
distribution for the initial conditions. In order to make this
replacement in a controlled way, however, and to facilitate the
eventual computation of corrections to the
leading order results, we systematize the treatment by attaching
to the field $m$ an internal `colour' index $\alpha$ which runs from
1 to $N$, and generalize (\ref{EQN:SYS4}) to
\begin{equation}
\partial_t m_{\alpha} = \nabla^2 m_{\alpha} +
(1-N^{-1}\sum_{\beta=1}^N(\nabla m_{\beta})^2)\,m_{\alpha}\ .
\end{equation}
Eq.\ (\ref{EQN:SYS4}) is the case $N=1$. The OJK result is obtained, however,
by taking the limit $N \to \infty$, when $N^{-1}\sum_{\beta=1}^N
(\nabla m_{\beta})^2$ may be replaced by its average. In this limit
(\ref{EQN:SYS4}) becomes (where $m$ now stands for one of the $m_\alpha$)
\begin{eqnarray}
\label{EQN:SYS5}
\partial_t m & = & \nabla^2 m + a(t)\,m \\
        a(t) & = & 1 - \langle (\nabla m)^2 \rangle\ ,
\label{EQN:SYS6}
\end{eqnarray}
a self-consistent {\em linear} equation for $m({\bf x},t)$.

It is interesting that the replacement of $(\nabla m)^2$ by its average in
(\ref{EQN:SYS4}) is also justified in the limit $d \to \infty$,
because $(\nabla m)^2 = \sum_{i=1}^d (\partial m/\partial x_i)^2$. If m is a
gaussian random field (and the self-consistency of this assumption follows
from (\ref{EQN:SYS5}) -- see below) then the different derivatives
$\partial m/\partial x_i$ at a given point $x$ are independent random
variables, and the central limit theorem gives, for $d \to \infty$,
$(\nabla m)^2 \to d \langle (\partial m/\partial x_i)^2 \rangle
= \langle (\nabla m)^2 \rangle$, with fluctuations of relative order
$1/\surd{d}$. While this approach is not so simple to systematize as that
adopted above, it seems clear that the leading order results become
exact for large $d$.

As discussed above, we will take the initial conditions for $m$ to be
gaussian, with mean zero and correlator (in Fourier space)
\begin{equation}
\langle m_{\bf k}(0)\,m_{-\bf k'}(0) \rangle = \Delta\,\delta_{\bf k,k'}\ ,
\label{EQN:FOURIERIC}
\end{equation}
representing short-range spatial correlations at $t=0$. Then $m$ is a
gaussian field at all times. The solution of (\ref{EQN:SYS5}) is
\begin{eqnarray}
m_{\bf k}(t) & = & m_{\bf k}(0)\exp(-k^2t + b(t))\ , \\
b(t) & = & \int_0^t dt'\,a(t')\ .
\end{eqnarray}
Inserting this into (\ref{EQN:SYS6}) yields
\begin{equation}
a(t) \equiv db/dt = 1 - \Delta \sum_{\bf k}k^2\,\exp(-2k^2t +2b)\ .
\end{equation}
After evaluating the sum one obtains, for large $t$ (where
the $db/dt$ term can be neglected), $\exp(2b) \simeq (4t/\Delta d)\,
(8\pi t)^{d/2}$, and hence $a(t) \simeq (d+2)/4t$. This form for $a(t)$
in (\ref{EQN:SYS5}), arising completely naturally in this scheme,
reproduces exactly the Oono-Puri modification of the OJK theory \cite{Oono88},
designed to keep the wall-width finite as $t \to \infty$, which was discussed
in section \ref{SEC:KYG}.

The explicit result for $m_{\bf k}(t)$, valid for large $t$, is
\begin{equation}
m_{\bf k}(t) = m_{\bf k}(0)\,(4t/\Delta d)^{1/2}\,(8\pi t)^{d/4}\,\exp(-k^2t)\
,
\end{equation}
from which the equal-time two-point correlation functions in Fourier and
real space follow immediately:
\begin{eqnarray}
\langle m_{\bf k}(t)\,m_{\bf -k}(t)\rangle & = & (4t/d)\,(8\pi t)^{d/2}\,
\exp(-2k^2t)\ , \\
\langle m(1)\,m(2)\rangle & = & (4t/d)\,\exp(-r^2/8t)\ ,
\label{EQN:SYS7}
\end{eqnarray}
where `1', `2', are the usual shorthand for space-time points $({\bf r}_1,t)$,
$({\bf r}_2,t)$, and $r=|{\bf r}_1 - {\bf r}_2|$.

We turn now to the evaluation of the correlation function of the original
fields $\phi$. Since, from (\ref{EQN:SYS7}), $m$ is typically of order
$\surd{t}$ at late times it follows from (\ref{EQN:SYS3}) that the field
$\phi$ is saturated (i.e.\ $\phi = \pm 1$) almost everywhere at late times.
As a consequence, the relation (\ref{EQN:SYS3}) between $\phi$ and $m$
may, as usual, be simplified to $\phi={\rm sgn}\,(m)$ as far as
the late-time scaling behavior is concerned.
Thus $C(12) = \langle {\rm sgn}\,(m(1))\,{\rm sgn}\,(m(2)) \rangle$.
The calculation of this average for a gaussian field $m$ proceeds
just as in the OJK calculation. The OJK result, given by (\ref{EQN:ARCSIN})
and (\ref{EQN:gamma}), (with $D=1$) is recovered.
The present approach, however, makes possible a systematic treatment in
powers of $1/N$. The work involved in calculating the next term is
comparable to that required to obtain the $O(1/n)$ correction to the
$n=\infty$ result for the $O(n)$ model \cite{NB90,Kiss93}.

\subsubsection{Vector Fields}
For vector fields, the TDGL equation is given by (\ref{EQN:VECTORA}),
where $V(\vec{\phi})$ is the usual `mexican hat' potential with ground-state
manifold $\vec{\phi}^2 = 1$.  This time we introduce a {\em vector} field
$\vec{m}({\bf x},t)$, related to $\vec{\phi}$ by the vector analog of
(\ref{EQN:MAZENKO1}), namely \cite{LM,BH}
\begin{equation}
\nabla_m^2 \vec{\phi} = \partial V/\partial \vec{\phi}\ ,
\label{EQN:SYSVEC1}
\end{equation}
where $\nabla_m^2$ means $\sum_{a=1}^n \partial^2/\partial m_a^2$ for an
$n$-component field. We look for a radially symmetric solution of
(\ref{EQN:SYSVEC1}),
$\vec{\phi}(\vec{m}) = \hat{m}\,g(\rho)$, with boundary conditions
$g(0)=0$, $g(\infty)=1$, where $\rho=|\vec{m}|$ and $\hat{m}=\vec{m}/\rho$.
Then the function $g(\rho)$ is the defect profile function for a topological
defect in the $n$-component field, with $\rho$ representing
the distance from the defect core \cite{LM,BH}.
In terms of $\vec{m}$, the TDGL equation for a vector field reads
\begin{equation}
\sum_b \frac{\partial \phi_a}{\partial m_b}\,\frac{\partial m_b}{\partial t}
= \sum_b \frac{\partial \phi_a}{\partial m_b}\,\nabla^2 m_b
+ \sum_{bc} \frac{\partial^2 \phi_a}{\partial m_b \partial m_c}\,
\nabla m_b \cdot \nabla m_c  - \nabla_m^2 \phi_a\ .
\label{EQN:SYSVEC2}
\end{equation}
Just as in the scalar theory, we can attach an additional `colour' index
$\alpha$ ($= 1, \ldots,N$) to the vector field $\vec{m}$, such that
the theory in the limit $N \to \infty$ is equivalent to replacing
$\nabla m_b \cdot \nabla m_c$ by its mean,
$\langle (\nabla m_b)^2 \rangle\, \delta_{bc}$ in (\ref{EQN:SYSVEC2}).
Noting also that $\langle (\nabla m_b)^2 \rangle$ is independent of $b$ from
global isotropy, (\ref{EQN:SYSVEC2}) simplifies in this limit to
\begin{equation}
\sum_b \frac{\partial \phi_a}{\partial m_b}\,\frac{\partial m_b}{\partial t}
= \sum_b \frac{\partial \phi_a}{\partial m_b}\,\nabla^2 m_b
- \nabla_m^2 \phi_a\,(1 - \langle (\nabla m_1)^2 \rangle)\ ,
\end{equation}
where $m_1$ is any component of $\vec{m}$. Finally, this equation can be
reduced to the linear form (\ref{EQN:SYS5}),
with $m$ replaced by $\vec{m}$, through the choice
$\nabla_m^2 \phi_a = -\sum_b (\partial \phi_a/\partial m_b)\,m_b$
or, more compactly,
$\nabla_m^2 \vec{\phi} = -(\vec{m} \cdot \nabla_m)\,\vec{\phi}$, to determine
the function $\vec{\phi}(\vec{m})$. Substituting the radially symmetric
form $\vec{\phi} = \hat{m}g(\rho)$ gives the equation
\begin{equation}
g'' + \left(\frac{n-1}{\rho} + \rho\right)\,g' - \frac{n-1}{\rho^2}\,g = 0\ ,
\end{equation}
a generalization of (\ref{EQN:SYS2}),
for the profile function $g(\rho)$, with boundary
conditions $g(0)=0$, $g(\infty)=1$. The solution is linear in $\rho$ for
$\rho \to 0$, while $g(\rho) \simeq 1-(n-1)/2\rho^2$ for $\rho \to \infty$.
The potential $V(\vec{\phi})$ corresponding to this profile function can be
deduced from (\ref{EQN:SYSVEC1}),
though we have been unable to derive a closed form
expression for it. Note that we are making here the natural assumption that
scaling functions are independent of the details of the potential for
vector fields, as well as for scalar fields.

For the vector theory, Eqs.\ (\ref{EQN:SYS5}) and (\ref{EQN:SYS6})
hold separately for each component
of the field. Taking gaussian initial conditions, with correlator
(\ref{EQN:OJKIC}), yields $a(t) \simeq (d+2)/4t$ again,
giving (\ref{EQN:SYS7}) for each component.
The final step, the evaluation of the two-point function
$C(12) = \langle \vec{\phi}(1) \cdot \vec{\phi}(2) \rangle$, proceeds
exactly as in the KYG treatment of section \ref{SEC:KYG}:
since $|\vec{m}|$ scales as $\surd{t}$, we can replace the function
$\vec{\phi}(\vec{m})$ by $\hat{m}$ at late times. Then
$C(12) = \langle \hat{m}(1) \cdot \hat{m}(2) \rangle$ in the scaling regime.
The required gaussian average over the fields $\vec{m}(1)$, $\vec{m}(2)$
yields the BPT scaling function (\ref{EQN:BPT}).
Again, it can be systematically improved by expanding in $1/N$.

\subsubsection{The Porod Tail}

It is easy to show \cite{BP,Toy92,LM,BH} that (\ref{EQN:BPT}) contains
the singular term of order $r^n$ (with an additional logarithm for
even $n$) that generates the Porod tail (\ref{EQN:GENPOROD})
in the structure factor. This feature was effectively
built into the theory through the mapping $\vec{\phi}(\vec{m})$.
Specifically, the singular part of (\ref{EQN:BPT}) for $\gamma \to 1$
is \cite{Abramowitz}
\begin{equation}
C_{sing} = \frac{n\gamma}{2\pi}\left[B\left(\frac{n+2}{2},\frac{1}{2}\right)
\right]^2\,\frac{\Gamma((n+2)/2)\Gamma(-n/2)}
{\Gamma^2(1/2)}\,(1-\gamma^2)^{n/2}\ .
\end{equation}
Using $\gamma = \exp(-r^2/8t) = 1 - r^2/8t + \cdots$ for $r \ll t^{1/2}$,
and simplifying the beta and gamma functions, gives
\begin{equation}
C_{sing} = \frac{1}{\pi}\,\frac{\Gamma^2((n+1)/2)\Gamma(-n/2)}{\Gamma(n/2)}\,
\frac{r^n}{(4t)^{n/2}}\ .
\label{EQN:SHORTGAUSSIAN}
\end{equation}
It will be interesting to compare this result with the exact short-distance
singularity derived in section \ref{SEC:SHORT}.

In Figures 14 and 15, we compare the BPT scaling function with numerical
simulation results \cite{BBunpub,BSB}, both for the pair correlation
function $C({\bf r},t)$ and the structure factor $S({\bf k},t)$. Since the
defect density $\rho$ scales as $L^{-n}$, a natural choice for the scaling
length $L$ is $\rho^{-1/n}$. Note that $\rho$ can be measured independently
in the simulation, so using $r\rho^{1/n}$ as scaling variable provides a
direct, zero parameter test of the scaling hypothesis itself. For the
scalar systems, the scaling variable $r\langle 1-\phi^2 \rangle$ was
employed \cite{BSB}: because  $1-\phi^2$ is non-zero only near domain
walls, $\langle 1-\phi^2 \rangle$ is equal to $\rho$, up to a
time-independent constant.

The resulting scaling plots (Figure 14) provide very good evidence for
scaling, except for $d=2=n$ where clear scaling violations are apparent:
the data drift to the right with increasing time, i.e.\ they are
`undercollapsed'. In this case we can apparently make the data scale,
however, by plotting against $r/L(t)$ with $L(t)$  chosen independently
at each time $t$ to provide the best data collapse. The collapse is then
as good as for any of the other systems.

The theoretical curves in Figure 14 are the BPT function (\ref{EQN:BPT}),
which reduces to the OJK scaling function for $n=1$. In making the fits,
$\gamma$ was replaced by $\exp(-\alpha r^2/L(t)^2)$ with the scale factor
$\alpha(n,d)$ adjusted to give the best fit by eye.

The structure factor plots of Figure 15, on a log-log scale, confirm
the existence of the Porod tail (\ref{EQN:GENPOROD}) in the data.
On the logarithmic scale, the poor scaling of the $d=2=n$ data against
$r\rho^{1/2}$ is reflected in a slight spreading of the data at small
$k$ in the corresponding structure factor plot. We do not show the
structure factor plots for $n=1$: the existence of the Porod tail
for scalar systems is implicit in the linear short-distance regime
in the real-space plots.

It should be noted that the real-space correlation function
(\ref{EQN:BPT}) is {\em independent of the space dimensionality d}.
The $d$-dependence of the structure factor enters only through the
process of Fourier transformation. Within the BPT theory, therefore,
the Porod tail is obtained for {\em any} $n$ and $d$. The same
feature is present in the structure factor computed using Mazenko's
method \cite{LM,BH}. In section \ref{SEC:SHORT}, however, we shall
show that the Porod tail is a direct consequence of the presence
of stable topological defects in the system and, furthermore, that
the {\em amplitude} of the tail can be evaluated exactly in terms of
density of defect core. Since stable defects are only possible for
$n \le d$, the Porod tail obtained from the BPT function (\ref{EQN:BPT})
for $n>d$ is an artefact of the approximations invoked. This scenario
is consistent with our claim that the BPT function actually represents
an exact solution in the limit $d \to \infty$. In this limit, of course,
the condition $n \le d$ is always satisfied!

\subsubsection{External Fields, Thermal Noise, Quenched Disorder \ldots}
\label{SEC:EXTERNAL-FIELDS}
Remarkably, the systematic approach can be readily extended to treat
the situation where a general (space- and/or time-dependent) external
field is present and/or the initial conditions contain a bias. This
also allows the effects of thermal fluctuations to be incorporated
to a limited extent. For simplicity we will only treat scalar fields.

Consider the following equation of motion:
\begin{equation}
\partial\phi/\partial t = \nabla^2\phi -V'_0(\phi) + h({\bf x},t)V'_1(\phi)\ .
\label{EQN:MODELAH}
\end{equation}
Here $V_0(\phi)$ is the usual symmetric, double-well potential sketched in
Figure 3, while $V_1(\phi)$ has the sigmoid form sketched in Figure 16(a).
The full potential, $V(\phi;{\bf x},t) = V_0(\phi) - h({\bf x},t)V_1(\phi)$,
has (for given ${\bf x}$ and $t$) the asymmetric double-well form shown in
Figure 16(b), with the right-hand minimum lower for $h>0$.

As in our treatment of the case $h=0$, we can exploit the insensitivity of
the domain growth to specific details of the potential to choose
especially convenient forms for $V_0$ and $V_1$. This rests on the physical
truth that the motion of an interface depends only on the local curvature
$K$ and the local field. To see this, consider again Eq.\ (\ref{EQN:AC1})
for the interface motion, this time for a general potential $V(\phi)$.
Multiplying through by $(\partial\phi/\partial g)_t$, integrating over $g$
through the interface, and using (\ref{EQN:SIGMA}), gives the local
velocity of the interface as
\begin{equation}
v = -K + \Delta V/\sigma\ ,
\end{equation}
instead of (\ref{EQN:AC}), where $\Delta V$ is the potential difference
across the interface. The essential point is that the interface motion
depends on the external field only through $\Delta V$. This gives us
a great deal of flexibility in the choice of $V_1(\phi)$, since all that
matters is the potential difference between the minima of $V(\phi)$.
For example, we can choose the minima to remain at $\phi=\pm 1$, as in
Figure (16b).

With these insights, we now change variables to the auxiliary field
$m$, with $\phi=\phi(m)$. Then (\ref{EQN:MODELAH}) becomes
\begin{equation}
\phi'\,\partial m/\partial t = \phi'\nabla^2 m + \phi''(\nabla m)^2
                     -V'_0(\phi) + h({\bf x},t)V'_1(\phi)\ ,
\end{equation}
where $\phi'\equiv d\phi/dm$ etc. Simplifications analogous to those that
led to (\ref{EQN:SYS4}) are achieved through the choices
\begin{eqnarray}
\label{EQN:V0}
V'_0(\phi) & = & \phi'' = -m\phi' \\
V'_1(\phi) & = & \phi' \ ,
\label{EQN:V1}
\end{eqnarray}
which give immediately
\begin{equation}
\partial_t m = \nabla^2 m + (1-(\nabla m)^2)\,m + h({\bf x},t)\ ,
\label{EQN:SYSH}
\end{equation}
a simple extension of (\ref{EQN:SYS4}).

The right part of (\ref{EQN:V0}) gives (with the appropriate boundary
conditions) the usual error function profile (\ref{EQN:SYS3}), while
the left part leads to the previous form (\ref{EQN:SYSPOT}) for
$V_0(\phi)$. Integrating (\ref{EQN:V1}) gives, with the boundary
condition $V_1(0)=0$, the result
\begin{equation}
V_1(\phi) = \frac{2}{\pi}\int_0^m dt\,\exp(-t^2)
= \frac{1}{\surd{\pi}}\,{\rm erf}\,(m)
= \frac{1}{\surd{\pi}}\,{\rm erf}\,[\sqrt{2}\,{\rm erf}^{-1}\,(\phi)]\ .
\end{equation}
Again, this only defines $V_1(\phi)$ for $-1 \le \phi \le 1$, but this is
the only region we require for the $T=0$ dynamics.

The difference $V_1(1)-V_1(-1)$ is $2/\surd{\pi}$, so the difference
between the minima of the full potential, $V = V_0 -hV_1$, is $-2h/\surd{\pi}$,
corresponding to an effective magnetic field $h_{eff} = h/\surd{\pi}$ as
far as the interface dynamics are concerned.

\noindent{\underline{External Fields/Initial Bias} \\
As a simple application of (\ref{EQN:SYSH}), consider the case
$h({\bf x},t)=h$, representing a uniform, time-independent magnetic field.
In order to solve the equation, we take the same limit ($d \to \infty$, or
number of `colours', $N$, large) as in section \ref{SEC:SYSTEMATIC},
enabling the replacement of $(\nabla m)^2$ by its mean. Additionally,
we allow for a bias, $\langle m(0) \rangle = m_0(0)$, in the (gaussian)
initial conditions, while the other Fourier components (${\bf k} \ne 0$)
of $m$ still satisfy (\ref{EQN:FOURIERIC}). Then the equations for
the ${\bf k} \ne 0$ components of $m$, and the self-consistency condition,
are unchanged by the field: $a(t) = 1 - (\nabla m)^2$, and
$b(t) = \int_0^t dt'\,a(t')$ are the same as for $h=0$. The equation for
the ${\bf k}=0$ component is $dm_0/dt+a(t)m_0 = h$, with
solution
\begin{equation}
m_0(t) = m_0(0)\,\exp\{b(t)\} + h\int_0^t dt'\,\exp\{b(t)-b(t')\}\ .
\end{equation}
Inserting the result $\exp(b) \simeq (4t/\Delta d)^{1/2}(8\pi t)^{d/4}$,
valid for large $t$, from section \ref{SEC:SYSTEMATIC} gives, for large
$t$,
\begin{equation}
m_0(t) = m_0(0)\,\left(\frac{4t}{\Delta d}\right)^{1/2}(8\pi t)^{d/4}
          + h \int_{t_0}^t dt' \left(\frac{t}{t'}\right)^{(d+2)/4}\ ,
\label{EQN:BIAS}
\end{equation}
where $t_0 \sim (\Delta d)^{2/(d+2)}$ is a short-time cut-off (to allow
for the breakdown, at short times, of the form used for $b(t)$).

Exploiting the gaussian property of $m$ (which now has a non-zero mean
given by (\ref{EQN:BIAS})), we can calculate the expectation value of
the original field $\phi$:
\begin{equation}
\langle \phi \rangle = \langle {\rm sign}\,(m) \rangle
= {\rm erf}\,\left(\frac{\langle m \rangle}{\sqrt{2 \langle m^2 \rangle_c}}
\right)\ ,
\end{equation}
where $\langle m^2 \rangle_c \equiv \langle m^2 \rangle - \langle m \rangle^2$
is the second cumulant of $m$. It is given by the same expression,
Eq.\ (\ref{EQN:SYS7}) with $1=2$, as for $h=0$:
$\langle m^2 \rangle_c = 4t/d$. So,
\begin{equation}
\langle \phi \rangle = {\rm erf}\,\left(\frac{m_0(0)}{\sqrt{2\Delta}}
(8\pi t)^{d/4} + h\,\left(\frac{d}{8t}\right)^{1/2} \int_{t_0}^t dt'
\left(\frac{t}{t'}\right)^{(d+2)/4}\right)\ .
\label{EQN:PHIBIAS}
\end{equation}

The time-dependence of the mean order parameter $\langle \phi \rangle$
depends on $d$. Consider the argument of the error function.
The initial bias $m_0(0)$ gives a contribution of order $t^{d/4}$
for any $d$, but the contribution from the external field $h$ scales
as $t^{1/2}$ for $d<2$ (when times $t'$ of order $t$ dominate the
integral in (\ref{EQN:PHIBIAS}), as $t^{1/2}\ln(t/t_0)$ for $d=2$,
and as $t^{d/4}$ for $d>2$ (when the integral is dominated by times
near the lower cut-off). Thus for $t$ large the external field dominates
over the initial bias for $d \le 2$, whereas for $d>2$ both terms
are of the same order. For an arbitrary time-dependent field $h(t)$,
the final term in the argument of the error function is simply
$(d/8t)^{1/2} \int_{t_0}^t dt'\,h(t')(t/t')^{(d+2)/4}$. This shows
that the field becomes less important at late times and, for $d>2$,
a constant field has all its effect at early times of order $t_0$.
For $d \le 2$, a constant field continues to have an effect at late times.

In fact one can make exact statements \cite{Kissner92} about the
`initial growth' regime, where $\langle \phi \rangle \ll 1$ .
The main modification is that
$t^{d/4}$ ($=L^{d/2}$) gets replaced by $L^\lambda$, where $\lambda$
is the exponent in the scaling form (\ref{EQN:GSCALING}) for the
response to the initial condition. This result is essentially obvious
from the definition (\ref{EQN:RESPONSE}) of the response function.
The crossover (in $d$) between the two regimes no longer occurs at
$d=2$, but at the dimension where $\lambda=1$ \cite{Kissner92}.
The virtue of (\ref{EQN:PHIBIAS}) is that it gives the complete
time dependence, from the initial regime to final saturation
($\langle \phi \rangle = 1$).

\noindent\underline{Thermal Fluctuations} \\
Thermal fluctuations can be included, to some extent, within the
present formalism by choosing $h({\bf x},t)$ to be a gaussian
white noise, with mean zero and correlator
$\langle h({\bf x},t)h({\bf x}',t') \rangle =
2D\delta({\bf x}-{\bf x}')\delta(t-t')$. The original equation of motion
(\ref{EQN:MODELAH}) may be recast using (\ref{EQN:V0}) and (\ref{EQN:V1})
as $\partial\phi/\partial t = \nabla^2\phi - V'_0(\phi)
+ \sqrt{2V_0(\phi)}\,h({\bf x},t)$. Recall that $V_0(\phi)$ vanishes is the
bulk phases, $V_0(\pm 1)=0$. The noise in the $\phi$ equation, therefore,
also vanishes in the bulk phases, differing from zero only in the
interfaces. Consequently, this noise will be effective in thermally
roughening the interfaces, but will be incapable of nucleating bubbles
of stable phase from a metastable state, or thermally exciting reversed
regions within a domain.

\noindent\underline{Quenched Disorder}\\
Quenched random fields are generated by a time independent, spatially
random field $h({\bf x})$. Again, in the original $\phi$ variable the
field is multiplied by $\sqrt{2V_0(\phi)}$, and so is active only at
the interfaces. Since driving forces due to the field only act
at the interfaces, this way of including a random field is perfectly
adequate. Unfortunately, however, our leading order approximation
of replacing $(\nabla m)^2$ by its average misses the important
interface pinning effects induced by the disorder, so this term
has to be kept in full.
A detailed discussion of quenched disorder, using Renormalisation
Group concepts, is given in section \ref{SEC:RGRANDOM}.

\subsection{HIGHER-ORDER CORRELATION FUNCTIONS}
\label{SEC:C_4}
Until now we have focussed exclusively on the pair correlation function
$C({\bf r},t)$ and its the Fourier transform, the structure factor
$S({\bf k},t)$. These primarily probe the spatial correlations in the
{\em sign}, or {\em direction} (for vector fields), of the order parameter.
However, one can also study the spatial correlations in the {\em amplitude}
of the order parameter \cite{Bray93}.
This is worthwhile for two reasons. In certain
systems, such as superconductors and superfluids, the (complex scalar)
order parameter $\psi$ does not directly couple to experimental probes.
Rather, such probes couple to $|\psi|^2$, and any scattering experiment,
for example, measures the Fourier transform of
$\langle |\psi(1)|^2 |\psi(2)|^2\rangle$. The second reason to study these
correlation functions is that the simultaneous calculation of two
different correlation functions provides an exacting test of theory.
This is because plotting one correlation function against another
provides an `absolute' (i.e.\ free of adjustable parameters)
prediction \cite{BSB}.
Tested this way, the predictions of the gaussian theories of the
`OJK' and `BPT' (or `Mazenko') type are not quite as impressive as
they at first seem.

In this section we will be concerned specifically with the normalised
correlation function
\begin{equation}
C_4(12) = \frac{\langle \{1-\vec{\phi}(1)^2\}\{1-\vec{\phi}(2)^2\}\rangle}
{\langle 1-\vec{\phi}(1)^2\rangle\,\langle 1-\vec{\phi}(2)^2\rangle}\ ,
\label{EQN:C_4}
\end{equation}
where the `1' in each bracket represents the saturated (i.e.\ equilibrium)
value of $\vec{\phi}^2$. The function $C_4(12)$ can be evaluated using
any of the gaussian field methods discussed above \cite{Bray93}.
For definiteness, we
adopt the `systematic approach' of section \ref{SEC:SYSTEMATIC}.
The details of the calculation are qualitatively different for scalar
and vector fields.

\subsubsection{Scalar Fields}

In terms of the gaussian auxiliary field $m$ the numerator in
(\ref{EQN:C_4}) is given by
\begin{equation}
C_4^N = \int dm(1) \int dm(2)\,P(m(1),m(2))\,
         \{1-\phi(m(1))^2\}\{1-\phi(m(2))^2\}\ ,
\label{EQN:C_4^N}
\end{equation}
where $P$ is the probability distribution (\ref{EQN:JOINTPROB}).
Since $(1-\phi^2(m))$ approaches zero exponentially fast for scalar
fields, the integrals are dominated by values of $m(1)$ and $m(2)$
close to zero (i.e.\ within an interfacial width of zero).
The variation of $P$ with $m(1)$ and $m(2)$, on the other hand,
is set by the length scales $r$ and $L(t)$, which are both large
in the scaling limit. Defining the interfacial width $\xi$ by
$\xi =\int dm (1-\phi(m)^2)$ gives, in the scaling limit,
\begin{equation}
C_4^N = \xi^2 P(0,0) = \frac{\xi^2}{\sqrt{2\pi (1-\gamma^2)}}\ ,
\end{equation}
while the normalised correlator $C_4$ is
\begin{equation}
C_4 = (1-\gamma^2)^{-1/2}\ .
\end{equation}
Here we recall that $\gamma \equiv \gamma(12)$ is the normalised correlator
(\ref{EQN:gamma}) of the field $m$. In particular, $\gamma(0)=1$ and
$\gamma(\infty)=0$. Using $\gamma = 1-const\,r^2/t$ for $r \ll t^{1/2}$,
we see that $C_4 \sim L/r$ for $r \ll L \sim t^{1/2}$. This result
will be derived using elementary arguments in section \ref{SEC:HOCFs}.
Note that the $1/r$ dependence at small $r$ implies a power-law tail,
$S_4({\bf k}) \sim Lk^{-(d-1)}$, in the Fourier transform of $C_4$.

By eliminating $\gamma$ between $C_4$ and the pair correlation function
$C = (2/\pi) \sin^{-1}\gamma$ (see (\ref{EQN:ARCSIN})), we obtain the
`absolute' relation
\begin{equation}
1/C_4 = \cos(\pi C/2)
\label{EQN:ABSOLUTE}
\end{equation}
between the two correlation functions, with no adjustable parameters.
We emphasize that (\ref{EQN:ABSOLUTE}) is a prediction of all
gaussian theories, which differ only in the relation between $\gamma$
and the scaling variable $x \equiv r/t^{1/2}$. Thus a test of
(\ref{EQN:ABSOLUTE}) is a test of the gaussian assumption itself.

In Figure 17 we show $1/C_4$ plotted against $C$, where $C_4$
and $C$ were measured simultaneously in `cell dynamics' simulations
\cite{CDS} in $d=2$ and $d=3$ \cite{BBunpub}. Also shown is the
prediction (\ref{EQN:ABSOLUTE}). It is clear that the agreement
is much poorer than one obtains by fitting $C$ alone (see Figure (12)).
The agreement is significantly better, however, for $d=3$ than for
$d=2$, consistent with our claim in section \ref{SEC:SYSTEMATIC} that
the gaussian assumption becomes exact for $d \to \infty$.

\subsubsection{Vector Fields}
The first step is a simple generalisation of (\ref{EQN:C_4^N}) to vector
fields:
\begin{equation}
C_4^N = \int d\vec{m}(1) \int d\vec{m}(2)\,P(\vec{m}(1),\vec{m}(2))\,
         \{1-\vec{\phi}(\vec{m}(1))^2\}\{1-\vec{\phi}(\vec{m}(2))^2\}\ ,
\label{EQN:VECC_4^N}
\end{equation}
where $P$ is a product of separate factors (\ref{EQN:JOINTPROB}) for
each component (since $\vec{m}$ is assumed to be gaussian).
The subsequent analysis is different from the scalar case,
however, because for vector fields
$\vec{\phi}(\vec{m})^2$ approaches its saturated value of unity for
$|\vec{m}| \to \infty$ only as a power law. To see this we recall that
the function $\vec{\phi}(\vec{m})$ is defined as the equilibrium
profile function for a radially symmetric topological defect. The
amplitude equation satisfies (\ref{EQN:VECPROF}) with $f \to |\vec{\phi}|$
and $r \to |\vec{m}|$.  From (\ref{EQN:VECPROFTAIL}) we obtain directly
\begin{equation}
1 - \vec{\phi}(\vec{m})^2 \to \xi^2/|\vec{m}|^2\ ,\ \ \ \ \
|\vec{m}| \to \infty\ ,
\label{EQN:VECTAIL}
\end{equation}
where $\xi^2 = 2(n-1)/V''(1)$. We will use this to define the `core size'
$\xi$ for topological defects in vector fields.

Inserting (\ref{EQN:VECTAIL}) in (\ref{EQN:VECC_4^N}), we see that for
$n>2$ the factors $(1-\vec{\phi}(\vec{m})^2)$ do not, in contrast to scalar
fields, converge the integral at small $|\vec{m}|$
(i.e.\ at $|\vec{m}|\sim \xi$). Instead, the integrals are converged
in this case by the probability distribution $P$, which sets a
typical scale $L(t)$ for $|\vec{m}|$. This justifies the use of the
asymptotic form (\ref{EQN:VECTAIL}) in the scaling limit:
\begin{equation}
C_4^N = \xi^4 \int \frac{d\vec{m}(1)}{|\vec{m(1)}|^2}
\int \frac{d\vec{m}(2)}{|\vec{m(2)}|^2}\,P(\vec{m}(1),\vec{m}(2))\ .
\label{EQN:C_4VECSCALING}
\end{equation}
It is now a straightforward matter to evaluate the $\vec{m}$ integrals
\cite{Bray93}. Dividing by the large-distance limit (corresponding to
$\gamma=0$), gives the normalised correlator (\ref{EQN:C_4}) as
\begin{equation}
C_4 = F(1,1;n/2;\gamma^2)\ ,
\label{EQN:C_4FINAL}
\end{equation}
where $F$ is again the hypergeometric function $_2F_1$.

For $\gamma \to 1$, $C_4$ has a short-distance singularity proportional to
$(1-\gamma^2)^{(n-4)/2} \sim (L/r)^{4-n}$ (with logarithmic corrections
for even $n$). It follows that the Fourier transform has the power-law tail
$S_4 \sim L^{4-n}k^{-(d+n-4)}$ \cite{Bray93}, for $n>2$.

For the special case $n=2$ one has to be more careful, as the integral
(\ref{EQN:C_4VECSCALING}) is formally logarithmically divergent at small
$|\vec{m(1)}|$, $|\vec{m(2)}|$, and has to be cut off at
$|\vec{m}| \sim \xi$. A careful analysis \cite{Bray93} shows that
$C_4^N$ exhibits logarithmic scaling violations in this case. However,
in the scaling limit $r \to \infty$, $L(t) \to \infty$ with
$r/L(t)$ fixed, the extra logarithm cancels in the {\em normalised}
correlator $C_4$, and (\ref{EQN:C_4FINAL}) is recovered, but with
logarithmic corrections to scaling \cite{Bray93,BBunpub}.

In fact (\ref{EQN:C_4FINAL}) simplifies for physical (i.e.\ integer)
values of $n$, giving $(1-\gamma^2)^{-1}$ for $n=2$ and
$\sin^{-1}(\gamma)/\gamma(1-\gamma^2)^{1/2}$ for $n=3$.
As for the scalar theory, one can eliminate $\gamma$ between
(\ref{EQN:C_4FINAL}) and (\ref{EQN:BPT}) to obtain a parameter-free
relation between $C_4$ and $C$ that may be used as an absolute test
of the gaussian assumption. Figure 18 shows data for $1/C_4$
plotted against $1-C$, from cell-dynamics simulations \cite{BBunpub},
and the corresponding predictions of the gaussian theory.
It can be seen that the gaussian theory is again rather poor but, as for
the scalar theory, it improves with increasing $d$, once more in accord
with our argument that it becomes exact for large $d$.

\subsubsection{Defect-Defect Correlations}
\label{SEC:DEFECT-DEFECT}
As a final example, we consider the correlation functions of the
defect density itself. In terms of the auxiliary field $\vec{m}$,
the defect density is $\rho({\bf x}) = \delta(m({\bf x}))\,J$, where
$J$ is the Jacobian between the field $\vec{m}$ and the spatial
coordinate ${\bf x}$, for example $J=|\nabla m|$ for scalar fields.
A significant simplification is achieved by choosing Mazenko's
definition of $\vec{m}$, near defects, as a coordinate normal to the
defect. Then $J=1$ holds identically at defects, giving simply
$\rho = \delta(\vec{m})$. Making {\em now} the gaussian approximation
for $m$, the one-point distribution function is
\begin{equation}
P(\vec{m}) = (2\pi S_0)^{-n/2}\,\exp(-\vec{m}^2/2S_0)\ ,
\end{equation}
where $S_0 = \langle m^2 \rangle$ is mean-square value of {\em one
component} of $\vec{m}$. Eq.\ (\ref{EQN:SYS7}) gives $S_0=4t/d$.
The mean defect density is, therefore,
\begin{equation}
\rho_{def}^{gauss1} = \langle \delta(m) \rangle = P(0) = (d/8\pi t)^{n/2}\ .
\label{EQN:GAUSS1}
\end{equation}
The superscript `gauss1' indicates that this is {\em one} way to calculate
$\rho_{def}$ within the gaussian approximation. An alternative approach,
pursued by Liu and Mazenko \cite{Defect-Defect}, is to retain the
Jacobian explicitly. This gives a different result, because $J=1$ at
defects is true for the {\em exact} $\vec{m}$, but not for the gaussian
approximation. With the Jacobian retained, the calculation has not been
completed for general $d$ and $n$.

With $J=1$, the pair correlation function is also trivially evaluated:
\begin{equation}
\langle \rho(1) \rho(2) \rangle = P(0,0)
= (d/8\pi t)^n\,(1-\gamma^2)^{-n/2}\ ,
\label{EQN:rhorho}
\end{equation}
where we used (\ref{EQN:JOINTPROB}) for $P(0,0)$. In the short-distance
limit, $r \ll t^{1/2}$, this becomes,
\begin{equation}
\langle \rho(1) \rho(2) \rangle \to \frac{\rho_{def}}{r^n}
\left(\frac{d}{2\pi}\right)^{n/2}\ ,\ \ \ \ \ \ r \ll L(t)\ .
\label{EQN:DEF-DEF1}
\end{equation}

Again, the short-distance behaviour can be evaluated exactly (see section
\ref{SEC:HOCFs}), and the $r^{-n}$ short-distance behaviour recovered from
simple geometrical arguments, which exclude, however, point defects (i.e.\
$n=d$). This failure to capture the correct short-distance behaviour for
point defects is another weakness of the gaussian approximation.

\subsection{NEMATIC LIQUID CRYSTALS}
\label{SEC:NLC}
We have not succeeded in applying the systematic approach to the
equation of motion (\ref{EQN:MODELN}) for nematics. Application of
the KYG method (see section \ref{SEC:KYG}), however, is relatively
straightforward \cite{BPBS}. For orientation purposes, we first
recall the use of the KYG method for vector fields \cite{BP}.
Recall that, in the scaling regime, the relation between the order
parameter field $\vec{\phi}$ and the auxiliary field $\vec{m}$ can
be simplified to $\vec{\phi}=\hat{m}$, a unit vector, and that $\vec{m}$
may be taken to satisfy the diffusion equation
$\partial_t \vec{m} = \nabla^2 \vec{m}$. As was stressed in section
\ref{SEC:KYG}, this approach is somewhat ad hoc, and is not even guaranteed
to yield the correct time-dependence for $L(t)$. In practice, however,
it gives good results for scaling functions since it builds in, through
the zeros of $\vec{m}$, the correct topological defects. Therefore,
we adopt this as a reasonable first attempt. It turns out that for
nematics, we do in fact recover the correct growth, $L \sim t^{1/2}$,
as shown in section \ref{SEC:GROWTH}.

The first step is to introduce the (traceless, symmetric) {\em tensor}
auxiliary field $m$, satisfying the diffusion equation. The only tricky
part is to determine the mapping $Q(m)$, between the auxiliary field
and the order parameter, analogous to $\phi(m)={\rm sgn}\,(m)$  for
scalar fields and $\vec{\phi}(\vec{m}) = \hat{m}$ for vector fields.
The key observation is that these latter results simply represent
the mapping from an initial value of $m$ to the nearest minimum of
the potential or, equivalently, they describe the attractors of the
dynamics (\ref{EQN:MODELN}) for a spatially uniform initial state.
It is easy to show \cite{BPBS} that for a nematic, an equivalent
procedure is the following. The director {\bf n} at a given space-time
point `1' is obtained as the eigenvector with largest eigenvalue of
the tensor $m(1)$ obtained by evolving the diffusion equation
$\partial_t m = \nabla^2 m$ forward from a random initial condition.
The physical tensor $Q(1)$ then has elements $Q_{ab}(1) = S\,[n_a(1)n_b(1)
- \delta_{ab}/3]$, where $S$ is an arbitrary amplitude that has the
value 3/2 for the particular coefficients in the equation of motion
(\ref{EQN:MODELN}). The pair correlation function is then obtained as
\begin{equation}
C(12) = (2/3)\,\langle {\rm Tr}\,Q(1) Q(2) \rangle\ ,
\label{EQN:NLC1}
\end{equation}
where the factor 2/3 normalises (for $S=3/2$) the correlation function to
unity when points `1' and `2' are the same. The average in (\ref{EQN:NLC1})
is over the (gaussian) joint probability distribution for $m(1)$ and $m(2)$,
which can be deduced from the diffusion equation for $m$ and the assumed
gaussian initial conditions.

The results for the pair correlation function and scaled structure factor
are shown in Figure 19, along with the simulation data of Blundell and
Bray \cite{BB92}, and the experimental structure factor data of Wong et
al.\ \cite{WWLY}. The inset in Figure 19(a) shows that the real-space
scaling function $f(x)$ has the short-distance behaviour
$f(x) = 1 + a\,x^2 \ln x -b\,x^2 + \cdots$. This is the same short-distance
form as the $O(2)$ model and leads to the same $k^{-5}$ tail in the structure
factor, reflecting the presence of line defects (disclinations). The fit to
the simulation data (with the length scale $L(t)$ adjusted at each time)  is
good. Remarkably, the BPT function (\ref{EQN:BPT}) for $n=2$ fits just as
well, and indeed the simulation data for the two systems are essentially
indistinguishable. This provides a dramatic illustration of the central
role played by the topological defects: the nematic might naively be
regarded as more like an $n=3$ than an $n=2$ system.

The theoretical curve in Figure 19(b) represents the $O(2)$ theory, as this
was simpler to obtain, by numerical Fourier transform of the analytic result
for the real-space scaling function, than the Fourier transform of the
nematic correlation function, which had to be generated numerically
\cite{BPBS}. Again the agreement is quite good. The data of Wong et al.\
can be shifted by an arbitrary amount, both horizontally and vertically,
but we were unable to collapse it precisely on to the analytic result or
simulation data. In addition, the experimental data have not yet reached
the asymptotic $k^{-5}$ regime expected on the basis of the string
defects present. A line of slope -5 is included as a guide to the eye.

\subsection{CONSERVED FIELDS}
We have seen that the OJK scaling function (\ref{EQN:ARCSIN}) and its
generalisation (\ref{EQN:BPT}) to vector fields provide a very good
description of the pair correlation function for {\em nonconserved}
fields, subject to the caveat that the scale length $L(t)$ is fitted
when comparing with data.
Furthermore, we have argued that the gaussian
approximation for the auxiliary field $\vec{m}$ is exact in certain
limits, and provides a starting point for a systematic treatment.

For conserved fields, the theory is less well developed. The most naive
approach, for example, does not even give the correct growth law,
$L(t) \sim t^{1/3}$ (for scalar fields). One can still attempt to
to make progress by introducing an auxiliary field $m$, but in contrast
to nonconserved fields, there is no evidence for any simple limit
in which the theory for $m$ becomes tractable.

To put the difficulties into context, we start with scalar fields, and
recall that the chemical potential $\mu$ satisfies the Laplace equation
(\ref{EQN:LAPLACE}), with boundary conditions (\ref{EQN:GT2}) imposed
at the interfaces. In fact the interfaces act as sources of the field
$\mu$. To see this we integrate $\nabla^2 \mu$ over a volume element
$dV$ enclosing an interface surface element $dS$.
Using (\ref{EQN:VELOCITY}) gives a source density
$-2v({\bf r})\delta(m({\bf r}))|\nabla m({\bf r})|$, where $v$ is the
interface velocity (measured in the direction of increasing $\phi$) and
$m({\bf r})$ is an auxiliary field whose zeros define the interfaces
(so that $\delta(m({\bf r}))|\nabla m({\bf r})|$ gives the volume
density of interfacial area). Our usual choice for the field $m$,
defined by (\ref{EQN:MAZENKO1}), gives $|\nabla m| = 1$ at interfaces.
This gives the Poisson equation
\begin{equation}
\nabla^2 \mu = -2\,v({\bf r})\,\delta(m({\bf r}))\ ,
\end{equation}
with solution (for $d=3$)
\begin{equation}
\mu({\bf r}) = \frac{1}{2\pi}\int \frac{d{\bf r}'}{|{\bf r}-{\bf r}'|}\,
v({\bf r}')\,\delta(m({\bf r}'))\ .
\label{EQN:POISSONSOLN}
\end{equation}
The Gibbs-Thomson boundary condition gives $\mu=-\sigma K/2$ at an
interface, where $K=\nabla \cdot {\bf n} = \nabla^2 m$ is the
interface curvature. Using (\ref{EQN:OJK2}) (with $|\nabla m|=1$) for
the interface velocity in (\ref{EQN:POISSONSOLN}), and the Gibbs-Thomson
boundary condition for $\mu$, gives
\begin{eqnarray}
\frac{\sigma}{2}\nabla^2 m
 & = & \frac{1}{2\pi} \int \frac{d{\bf r}'}{|{\bf r}-{\bf r}'|}\,
\frac{\partial m({\bf r}')}{\partial t}\,\delta\left(m({\bf r}')\right)
\nonumber \\
 & = & \frac{1}{4\pi} \int \frac{d{\bf r}'}{|{\bf r}-{\bf r}'|}\,
\frac{\partial}{\partial t}\,{\rm sgn}\left(m({\bf r}')\right)
\label{EQN:CONS1}
\end{eqnarray}
at interfaces. The same result could, in fact, be obtained directly from
the Cahn-Hilliard equation, $\partial_t \phi = \nabla^2 \mu$, by operating
on both sides by the inverse Laplacian and setting $\mu = -\sigma K/2
= -(\sigma/2) \nabla^2 m$ on the interfaces.

For {\em nonconserved fields}, the extension of the interface equation away
from the interfaces (as in, for example, the OJK theory) is a mathematical
convenience which does not change the underlying physics of interfaces moving
under their local curvature. For conserved fields, however, the interface
dynamics are nonlocal and the extension of (\ref{EQN:CONS1}) away from
interfaces is non-trivial. Any extension should satisfy the following
criteria: (i) the equation reduces to (\ref{EQN:CONS1}) at interfaces,
(ii) the chemical potential satifies $\nabla^2 \mu=0$ {\em except} at
interfaces, (iii) $\nabla\mu$ is discontinuous at interfaces, the normal
component of the discontinuity generating the interface velocity as in
(\ref{EQN:VELOCITY}), (iv) the conservation of the order parameter should be
preserved. Unfortunately, it is very difficult to construct an approximate
theory that satisfies all these criteria, and I am not aware of any
successful attempts. In addition, a good theory would ideally incorporate
two further features: (v) the structure factor should vanish as $k^4$ for
$k \to 0$ (see section \ref{SEC:SMALLk}), and (vi) the short-distance
expansion of the real-space scaling function should contain (after the
leading `1') only odd powers of $r$, the so-called `Tomita sum rule'
\cite{Tomita}, deriving from the smoothness of the interfaces.

A number of approximate theories have been proposed which satisfy a
subset of these requirements. The theories of Ohta and Nozaki
\cite{Nozaki}, Tomita \cite{Tomita93} and Yeung et al.\ \cite{YSO} all
involve a gaussian approximation for the correlator of the auxiliary
field $m$. The correct $t^{1/3}$ growth is obtained, and the scaling
functions describe the real-space
simulation data \cite{ShinOono} very well out to reasonable values of
the scaling variable, but violate the conservation law. A recent attempt
by Kramer and Mazenko \cite{Kramer,Mazenko94a} corresponds to an off-interface
extension of (\ref{EQN:CONS1}) in which the left-hand side replaced by
$-\mu = (\sigma/2)(\nabla^2 m + u\phi/L)$, with $u$ a constant.
The real-space scaling function is obtained by multipling through by
$\phi$ at a different space point, and averaging both sides with the
usual gaussian assumption for $m$. The resulting scaling function has
the desired $k^4$ small-$k$ form in Fourier space, although this behaviour
does not have the same origin as in the derivation of this form in
section \ref{SEC:SMALLk}. The real-space fit is not as good as earlier
theories. In addition, the chemical potential (rather than
its gradient) is discontinuous at interfaces.

Yeung et al.\ \cite{YSO} have critically analysed approximation schemes
based on a gaussian assumption for $m$. Using data of Shinozaki and Oono
\cite{ShinOono} for the pair correlation function $C({\bf r},t)$ in $d=3$
to infer a value
for the normalised correlator of $m$, namely $\gamma = \sin(\pi C/2)$
(see (\ref{EQN:ARCSIN})), they found that the Fourier transform
$\gamma_k$ would have to be negative at small $k$ to fit the data. However,
this is impossible since $\gamma_k \ge 0$ by definition. Yeung et al.\
concluded that no gaussian theory could adequately describe the data,
at least for scalar fields in $d=3$.  Nevertheless, we conclude this
subsection by considering the approach of Mazenko for conserved fields,
which is explicitly built on a gaussian assumption for $m$. This is
especially interesting for vector fields, since it allows us to make
contact with the large-$n$ calculation of section \ref{SEC:LARGEnC}.
The reader may recall that the exact solution of Coniglio and Zannetti
\cite{CZ} for conserved fields with $n=\infty$ exhibits a novel
`multiscaling' behaviour. A natural question is whether this behaviour
survives at finite $n$. Employing the Mazenko approach, we find
multiscaling behaviour for $n$ strictly infinite, but conventional scaling
for any finite $n$ \cite{BH92}.

\subsubsection{The Mazenko Method for Conserved Vector Fields}
\label{SEC:BH92}
A naive application of Mazenko's technique to the Cahn-Hilliard
equation (\ref{EQN:MODELB}), and its generalization to vector fields, yields,
in complete analogy to (\ref{EQN:MAZENKO7}),
\begin{equation}
(1/2)\,\partial C/\partial t =
-\nabla^2(\nabla^2 C + a(t)\,\gamma dC/d\gamma)\ ,
\label{EQN:MAZCON1}
\end{equation}
with $a(t)$ still defined by (\ref{EQN:MAZENKO6}). The only difference
between (\ref{EQN:MAZENKO7}) and (\ref{EQN:MAZCON1}) is the extra
$(-\nabla^2)$ on the right-hand side. The form (\ref{EQN:MAZCON1}) is
obtained for any $n$, with the function $C(\gamma)$ given by
(\ref{EQN:BPT}).

If we seek a scaling solution of (\ref{EQN:MAZCON1}), of the form
$C({\bf r},t) = f(r/L(t))$, it is immediately clear that consistency
requires $a(t) \sim 1/L(t)^2$ and $L(t) \sim t^{1/4}$. We shall show
in section \ref{SEC:GROWTH} that this in fact is the correct growth law for
$n > 2$, while there is a logarithmic correction,
$L(t) \sim (t\ln t)^{1/4}$, for $n=2$ (and $d>2$). For $n=1$, however,
(\ref{EQN:MAZCON1}) fails to give the correct $t^{1/3}$ growth. The reason
is clear. Taking $\phi$ to be a sigmoid function of a gaussian field
from the outset overlooks the vital role of the bulk diffusion field
in transferring material between interfaces. Recognizing this fact, Mazenko
\cite{Maz91} writes $\phi$ as a superposition of `ordering' and `diffusing'
components, $\phi = \psi(m) + \tilde{\phi}$, with $m$ a gaussian field.
It is then possible to construct a consistent theory with $t^{1/3}$
growth \cite{Maz91}, although the results do not agree well with
simulations \cite{Shin}.

Here we will concentrate on vector fields with $n>2$, for which
(\ref{EQN:MAZCON1}) does give the correct $t^{1/4}$ growth. For
general $n$ this equation can be solved numerically for the scaling
function $f(x)$ \cite{Rojas}. Somewhat surprisingly, the solution
for $d=3$, $n=2$ is very close to the simulation data of Siegert and
Rao \cite{SR}, despite the (logarithmically) wrong growth law.
For large $n$, however, we can make analytic progress, and contribute
to the debate on the possibility of multiscaling for finite $n$.

For large $n$, (\ref{EQN:MAZCON1}) is simplified as follows. Expanding
(\ref{EQN:BPT}) to first order in $1/n$ gives
$C = \gamma -\gamma(1-\gamma^2)/2n + O(n^{-2})$, and so
$\gamma dC/d\gamma = C + C^3/n + O(n^{-2})$. Putting this in
(\ref{EQN:MAZCON1}) gives
\begin{equation}
(1/2)\,\partial C/\partial t = -\nabla^4C - a(t)\,\nabla^2 (C+C^3/n)
\label{EQN:MAZCON2}
\end{equation}
correct to $O(1/n)$. The solution of this equation is very different
for $n$ strictly infinite than for $n$ large but finite.

For $n=\infty$, the $C^3/n$ term can be dropped and the resulting
linear equation solved by Fourier transformation to give
\begin{equation}
S({\bf k},t) = S({\bf k},0)\,\exp[-2k^4t + 2k^2 b(t)]
\end{equation}
for the structure factor, where $b(t)=\int_0^t dt'\,a(t')$. This result
is {\em identical} to the exact result obtained by Coniglio and Zannetti
\cite{CZ} in the same limit (compare for example, equation
(\ref{EQN:FOURIERCOP})), and leads to same multiscaling form
(\ref{EQN:MULTI}).

For $n$ finite, we try scaling forms consistent with the expected $t^{1/4}$
growth, namely $S({\bf k},t)=t^{d/4}g(kt^{1/4})$,
$a(t) = q_m^2/t^{1/2}$, and $C({\bf r},t) = f(r/t^{1/4})$, where $f(x)$ is
the Fourier transform of $g(q)$ and $q_m$ is a constant. Using these
in (\ref{EQN:MAZCON2}) gives
\begin{eqnarray}
\label{EQN:BH1}
dg/dq & = & -(d/q + 8q^3 - 8q_m^2q)g + qB(q)\ , \\
B(q) & = & (8q_m^2/n)\,(f^3)_{\bf q}\ ,
\label{EQN:BH2}
\end{eqnarray}
where $(f^3)_{\bf q}$ indicates the Fourier transform of $f(x)^3$.
Note that $g(0)=0$ must hold for a
conserved order parameter, otherwise $S({\bf k},0)$ would grow as
$L(t)^d$, violating the conservation law. Integrating
(\ref{EQN:BH1}) with initial condition $g(0)=0$ gives
\begin{equation}
g(q) = q^{-d}\,\exp(-2q^4 + 4q_m^2 q^2)\int_0^q dq'\,q'^{d+1}
       B(q')\,\exp(2q'^4 -4q_m^2 q'^2)\ .
\label{EQN:BH3}
\end{equation}

The constant $q_m$ is fixed by the condition $f(0)=1$,
i.e.\ $\sum_{\bf q} g(q)=1$.
For very large $n$ we find {\em a posteriori} that $q_m$ is large.
Then $g(q)$ is strongly peaked near $q=q_m$, justifying a steepest
descent evaluation of the sum over ${\bf q}$. For $q$ near $q_m$,
the integral in (\ref{EQN:BH3}) is dominated by $q'$ values of
order $q_m^{-1}$, giving \cite{BH92}
\begin{equation}
g(q) \simeq 2^{-(d+3)} \Gamma (1+d/2) B(0) q_m^{-(d+2)} q^{-d}\,
\exp(-2q^4 + 4q_m^2q^2)
\end{equation}
for $q$ near $q_m$. Using this form in $\sum_{\bf q}g(q)=1$, and evaluating
the sum by steepest descents, gives
\begin{equation}
1 = 2^{-(d+5)} \sqrt{2\pi} K_d \Gamma (1+d/2) q_m^{-(d+4)}\,\exp(2q_m^4)B(0)\ ,
\label{EQN:BH4}
\end{equation}
where $K_d = 2/(4\pi)^{d/2}\Gamma (d/2)$. Using this to eliminate $B(0)$ from
(\ref{EQN:BH3}) gives the desired scaling solution, valid for
$q_m^{-1} \ll q \stackrel{<}{\sim} q_m$,
\begin{equation}
g(q) = (4/K_d\sqrt{2\pi})\,q_m^2\,q^{-d}\,\exp[-2(q^2-q_m^2)^2]\ .
\label{EQN:BH5}
\end{equation}
In the limit $q_m \to \infty$, the width of the peak at $q=q_m$ vanishes
as $q_m^{-1}$, so in this limit we can write
\begin{equation}
g(q) \to K_d^{-1}\,q_m^{1-d}\,\delta (q-q_m)\ ,\ \ \ \ \ q_m \to \infty\ .
\label{EQN:BH6}
\end{equation}
The final step is to use (\ref{EQN:BH2}) with $q=0$ to obtain a second
relation (in addition to (\ref{EQN:BH4}) between $B(0)$ and $q_m$. From
(\ref{EQN:BH2}), $B(0) = (8q_m^2/n)\,\int d^dx\,f(x)^3$. Using
(\ref{EQN:BH6}) for $g(q)$ gives $f(x) = A(q_m x)$, where
$A(y) = {\rm const.}\,J_\nu(y)/y^\nu$, with $\nu=(d-2)/2$, and hence
\begin{equation}
B(0) = (8q_m^2/n) \int d^dx\,A(q_mx)^3 = const.\,q_m^{2-d}/n\ .
\end{equation}
Putting this in (\ref{EQN:BH4}) gives $1 = {\rm const.}\,q_m^{-2(d+1)}\,
\exp(2q_m^4)$, giving $q_m \simeq [(\ln n)/2]^{1/4}$ for large $n$. This
in turn implies a characteristic length scale $L(t) \sim (t/\ln n)^{1/4}$.

To summarise, we have shown that, within the Mazenko approximation, scaling
solutions are obtained for any finite $n$. Only for $n$ strictly infinite is
the multiscaling form (\ref{EQN:MULTI}) of Coniglio and Zannetti recovered.
Note that the amplitude of the $t^{1/4}$ growth depends in a singular
way on $n$ (as $[\ln n]^{-1/4}$) for $n \to \infty$, i.e.\ our result is
non-perturbative in $1/n$ and could not be obtained by expanding around the
large-$n$ solution. The scaling solution is obtained when the limit $t \to
\infty$ is taken at fixed $n$.

\subsubsection{The Small-$k$ Behaviour of the Structure Factor}
\label{SEC:SMALLk}
One slightly unsatisfactory aspect of the above treatment is that is does
not recover the correct small-$q$ behaviour (indeed, the same shortcoming
afflicts the scalar version of the calculation \cite{Maz91}). For $q \to 0$,
Eq.\ (\ref{EQN:BH3}) gives $g(q) \to B(0)q^2/(d+2)$. There are compelling
arguments, however, for a $q^4$ behaviour at small $q$
\cite{Yeung88,Furukawa89,Furukawa89a,Furukawa89b,Tomita91,Fratzl91},
strongly supported by numerical simulations \cite{ShinOono}, as well as
experiment \cite{Fratzl89}, for a scalar order parameter. Here we discuss
both scalar and vector fields. We begin by deriving an inequality for
the small-$q$ beaviour, using an approach based on that of Yeung
\cite{Yeung88}.

We recall that the equation of motion for conserved fields takes the form
$\partial_t \phi = \nabla^2 \mu$, where $\mu$ is the chemical potential.
Multiplying through by $\phi$ at a different space point, averaging, and
Fourier transforming, gives
\begin{eqnarray}
\frac{1}{2}\,\frac{\partial S({\bf k})}{\partial t} & = &
-k^2\,\langle \mu_{\bf k}\,\phi_{-{\bf k}} \rangle\ \nonumber \\
& \le & k^2\,[S({\bf k})]^{1/2}\,
\langle \mu_{\bf k}\,\mu_{-{\bf k}} \rangle^{1/2}\ ,
\label{EQN:CS1}
\end{eqnarray}
where the final line follows from the Cauchy-Schwartz inequality.
Now impose scaling: $S({\bf k}) = L^d\,g(kL)$. Since $\mu \sim 1/L$ for
scalar fields, the analogous scaling form is
$\langle \mu_{\bf k}\,\mu_{-{\bf k}} \rangle = L^{d-2}g_{\mu}(kL)$.
Putting these into (\ref{EQN:CS1}) and using $L \sim t^{1/3}$ for scalar
fields, gives
\begin{equation}
d\,g(q) + q\,g'(q) \leq const.\,q^2\,[g(q)\,g_{\mu}(q)]^{1/2}\ .
\label{EQN:CS2}
\end{equation}
For $q \to 0$ one expects $g(q) \sim q^\delta$ and $g_\mu(q) \to const.$,
because $\mu$ is not a conserved field. Then (\ref{EQN:CS2}) gives
the inequality $\delta \ge 4$. An approximate treatment which gives the
expected $q^4$ small-$q$ behaviour has been proposed by Kramer and
Mazenko \cite{Kramer}.

For vector fields, it is shown in sections \ref{SEC:GROWTH} and
\ref{SEC:RG} that $L(t) \sim t^{1/4}$ provided $n>2$ (see section
\ref{SEC:GROWTH} for a discussion of $n=2$).  This suggests that the
(vector) chemical potential scales as $\vec{\mu} \sim 1/L^2$ for $n>2$,
which gives (\ref{EQN:CS2}) again and $\delta \ge 4$ as before.

It is easy to show, using an argument of Furukawa
\cite{Furukawa89,Furukawa89a,Furukawa89b}, that the
lower-bound for $\delta$ implied by Yeung's argument is in fact realized,
i.e.\ $\delta=4$. Integrating the equation of motion $\partial_t \phi_{\bf k}
= -k^2 \mu_{\bf k}$ gives $\phi_{\bf k}(t) = \phi_{\bf k}(0)
- k^2 \int_0^t dt_1\,\mu_{\bf k}(t_1)$. This yields
\begin{equation}
S({\bf k},t) = - S({\bf k},0) +
                 2\langle\phi_{\bf k}(t)\phi_{-\bf k}(0)\rangle +
k^4\int_0^t dt_1 \int_0^t dt_2\,
\langle \mu_{\bf k}(t_1)\mu_{-\bf k}(t_2) \rangle\ .
\label{EQN:Furuq4}
\end{equation}
It is clear that the first two terms on the right must be neglible in the
scaling regime, otherwise Yeung's inequality would be violated!
It is simple, however, to show this explicitly.
In the scaling limit ($k \to 0$, $L \to \infty$ with $kL$ fixed),
$S({\bf k},t)$ increases as $L^d$. Therefore the time-independent first
term on the right of (\ref{EQN:Furuq4}) is certainly negligible.
Now consider the second term. The autocorrelation function,
$A(t) \equiv \langle \phi({\bf x},t) \phi({\bf x},0) \rangle$, decreases
as $A(t) \sim L(t)^{-\bar{\lambda}}$. This implies that
$\langle\phi_{\bf k}(t)\phi_{-\bf k}(0)\rangle = L^{d-\bar{\lambda}} a(kL)$,
where $a(q)$ is a scaling function. Since this term grows less rapidly
than $L^d$, it also is a negligible contribution to $S({\bf k},t)$ for
large $L$. It follows that the structure-factor scaling function $g(q)$ is
obtained entirely from the final term in (\ref{EQN:Furuq4}). It vanishes
as $q^4$ because $\mu$ is not a conserved field, so $\langle\mu_{\bf k}
\mu_{-\bf k}\rangle$ is non-zero at $k=0$, as discussed above.
Furukawa has gone slightly further, and shown explicitly that this final
term in indeed of order $L^d$. Inserting the two-time scaling form (for
scalar fields) $\langle \mu_{\bf k}(t_1)\mu_{-\bf k}(t_2) \rangle
= L_1^{d-2} g_\mu(kL_1,L_2/L_1)$
(a natural generalization of the equal-time scaling form given above),
using $L \sim t^{1/3}$, and evaluating the double time integral by power
counting, gives $S({\bf k},t) \sim k^4 t^{2+(d-2)/3} \sim k^4L^{d+4}$ as
required.

Finally we note that Tomita  has given a rather general argument for the $k^4$
behaviour based on the isotropy of the scaling functions \cite{Tomita91}.

\subsection{BINARY LIQUIDS}
\label{SEC:BINARY-LIQUIDS}
The equation of motion appropriate to binary liquids,
Eq.\ ({\ref{EQN:BINLIQS}), was derived in section \ref{SEC:BINLIQS}
for the case where the inertial terms in the Navier-Stokes equation can
be neglected. Eq.\ ({\ref{EQN:BINLIQS}) leads to an asymptotic linear
growth, $L(t) \sim t$. Here we will discuss how one might attempt to
calculate an approximate scaling function for pair correlations in this
regime.

In the regime where $L(t) \sim t$, the `advective term' in
({\ref{EQN:BINLIQS}) dominates the `diffusive term', $\lambda \nabla^2 \mu$,
on the right hand side, so we will discard the latter. In the spirit of
the `systematic approach' of section \ref{SEC:SYSTEMATIC}, we introduce
an auxiliary field $m$ defined by (\ref{EQN:MAZENKO1}), with the additional
choice (\ref{EQN:SYS2}), corresponding to a convenient choice of the
potential $V(\phi)$. Then the chemical potential $\mu$ can be expressed
as
\begin{eqnarray}
\mu \equiv \delta F/\delta \phi & = & V'(\phi) - \nabla^2 \phi \nonumber \\
 & = & -\phi'(m)\{\nabla^2 m + [1-(\nabla m)^2]m\}\ .
\label{EQN:muBINLIQS}
\end{eqnarray}
Expressing the equation of motion ({\ref{EQN:BINLIQS}) in term of $m$, and
using (\ref{EQN:muBINLIQS}) for $\mu$ gives
\begin{eqnarray}
\partial m({\bf r})/\partial t & = & \int d{\bf r}'\,
[\nabla m({\bf r})\cdot T({\bf r}-{\bf r}')\cdot\nabla' m({\bf r}')]\,
\{\phi'(m({\bf r}'))\}^2\,\{\nabla'^2 m({\bf r}') \nonumber \\
 & & \hspace{6cm} + [1-(\nabla' m({\bf r}'))^2] m({\bf r'})\}\ .
\label{EQN:BINLIQS1}
\end{eqnarray}
Now we recall that $\{\phi'(m)\}^2$ acts very much like a delta function on
the interfaces. In fact, the result (compare Eq.\ (\ref{EQN:SIGMA}))
$\sigma = \int dm (d\phi/dm)^2$ for the surface tension leads to the
identification $(d\phi/dm)^2 = \sigma\delta(m)$. Using this in
(\ref{EQN:BINLIQS1}) gives
\begin{equation}
\partial m({\bf r})/\partial t = \sigma\int d{\bf r}'\,
[\nabla m({\bf r})\cdot T({\bf r}-{\bf r}')\cdot\nabla' m({\bf r}')]\,
\delta(m({\bf r}'))\,\nabla'^2 m({\bf r}')\ .
\end{equation}

With the identification $\phi({\bf r}) = {\rm sgn}\,m({\bf r})$, this
equation can be rewritten as an equation for $\phi$, by multiplying
both sides by $\delta(m({\bf r}))$:
\begin{equation}
\frac{\partial{\phi({\bf r})}}{\partial t} = \frac{\sigma}{2}
\int d{\bf r}'\,[\nabla\phi({\bf r})\cdot T({\bf r}-{\bf r}')
\cdot\nabla'\phi({\bf r}')]\,\nabla'^2 m({\bf r}')\ .
\label{EQN:CBINLIQS}
\end{equation}
This equation could serve as a convenient starting point for approximate
treatments of the pair correlation function $C({\bf r},t)$.  Note, however,
that  Eq.\ (\ref{EQN:CBINLIQS}) is {\em fundamentally} nonlinear, and
approximate scaling functions that capture the correct physics
are difficult to construct. The correct growth law is, nevertheless, built
in to (\ref{EQN:CBINLIQS}): simple power-counting (remembering that $m$
scales as $L(t)$) gives immediately $L(t) \sim \sigma t/\eta$, as required.

In the above discussion, the field $m$ was taken to have zero mean,
corresponding to a critical quench. An off-critical quench can be
handled in similar fashion by allowing $m$ to have a non-zero mean
\cite{LMOffCritical}. An outstanding challenge is to devise an
approximate treatment which includes the `switching off' of the linear
growth at small volume fractions, when the minority phase is no longer
continuous, and to properly incorporate thermal fluctuations
in this regime (see the discussion in section \ref{SEC:BINLIQS}).

\section{SHORT-DISTANCE SINGULARITIES AND POROD TAILS}
\label{SEC:SHORT}
In the previous section various approximate treatments of correlation
functions were discussed. In this section we show that exact statements
can be made about the short distance behaviour or, more precisely, the
short-distance singularities, of these functions. In particular,
the qualitative arguments of section \ref{SEC:POROD} can be
made precise \cite{BH93}, and the {\em amplitude} of the $k^{-(d+n)}$ Porod
tail obtained in terms of the density of defect core, $\rho_{def}$, which
scales as $L^{-n}$. The basic result is a generalization of
(\ref{EQN:SFSHORT}),
in which the leading singular contribution to $C({\bf r},t)$ is a term
of order $|{\bf r}|^n$ for $n$ odd (or $n$ real, in a continuation
of the theory to real $n$), and $|{\bf r}|^n\ln |{\bf r}|$ for n even.
This in turn implies a power law tail $k^{-(d+n)}$ in $S({\bf k},t)$.

We first illustrate the method for the case of point defects ($n=d$).
The extension to the general case $n \le d$ is relatively straightforward.
Next we discuss the short-distance singularities of some higher-order
correlation functions, namely the function $C_4$ defined by equation
(\ref{EQN:C_4}), and the defect-defect correlation function.
We also calculate the `Porod tails' in the corresponding
structure factors. Finally, we compute for scalar fields the joint
probability distribution $P(m(1),m(2))$ of the auxiliary field $m$
of section \ref{SEC:APPROXSF}, in the limit where $|m(1)|$, $|m(2)|$ and
the distance $r=|{\bf r}_1-{\bf r}_2|$ are all small compared to $L(t)$.
We find that the distribution is {\em not}
gaussian, except perhaps for $d=\infty$.

\subsection{POINT DEFECTS ($n = d$)}
Consider the field $\vec{\phi}$ at points
${\bf x}$ and ${\bf x} +{\bf r}$ in the presence of a point defect
at the origin. We consider the case where $|{\bf x}|$,
$|{\bf x}+{\bf r}|$ and $|{\bf r}|$ are all small compared to a typical
inter-defect distance $L$, but large compared to the defect core size
$\xi$. Then the field at the points ${\bf x}$ and
${\bf x}+{\bf r}$ is saturated in length (i.e.\ of unit length) and not
significantly distorted by the presence of other defects. Moreover, the
field can be taken, up to a global rotation, to be directed radially
outward from the origin, as illustrated in Figure 10. Thus
\begin{equation}
\vec{\phi}({\bf x}).\vec{\phi}({\bf x}+{\bf r}) =
\frac{{\bf x}.({\bf x}+{\bf r})}{|{\bf x}|\,|{\bf x}+{\bf r}|}\ .
\label{EQN:PHI.PHI}
\end{equation}
With ${\bf r}$ held fixed we average (\ref{EQN:PHI.PHI}) over all possible
relative positions of the point defect, i.e.\ over all values of ${\bf x}$
within a volume of order $L^n$ around the pair of points, with the
appropriate probability density $\rho_{def}$. Focussing on the
{\em singular} part of the correlation function we obtain
\begin{equation}
C_{sing}({\bf r},t)  =  \rho_{def}\,\int^{L}\, d^{n}x\,
\left(\frac{{\bf x}.({\bf x}+{\bf r})}{|{\bf x}|\,|{\bf x}+{\bf r}|}
 - {\rm analytic\ terms}\,\right)\ .
\label{EQN:CSING1}
\end{equation}
The `analytic terms' in (\ref{EQN:CSING1}) serve to converge the
${\bf x}$-integral at large $|{\bf x}|$, and allow us to extend the
integral over all space. We include as many terms in the expansion of
(\ref{EQN:PHI.PHI})) in powers of ${\bf r}$ as
are necessary to ensure the convergence of the integral. When $n$ is
even, there is a residual logarithmic singularity. This case can be
retrieved from the general $n$ result by taking a suitable limit (see below).

At this point two comments are in order. Firstly, by taking the field to be
directed radially outward from the origin, we seem to be limiting ourselves
to `defects', and excluding `antidefects'. The antidefect of a point
defect, however, can be generated (up to arbitrary rotations), by
`inverting' ($\phi_i \to -\phi_i$) an odd number of Cartesian components
of the vector $\vec{\phi}$. Reference to Figure 10(a) shows immediately that,
for $n=2=d$ for example, the `antivortex' can be generated from the `vortex'
by this construction. Clearly, however, the scalar product
$\vec{\phi}(1)\cdot\vec{\phi}(2)$ required for the evaluation of
$C_{\rm sing}$ is invariant under this operation.
Secondly, we seemed to need the assumption that the field near a given
defect is not significantly distorted by the presence of other defects.
Actually, this is not strictly necessary. Any distortion generated by
the other defects provides an analytic background field that does not
affect the contribution of the given defect to the singular part of $C$
\cite{MZ}.

The integral (\ref{EQN:CSING1}) (extended over all space) is evaluated in
\cite{BH93}. The result is
\begin{equation}
C_{sing} = n\,\pi^{n/2-1}\,B\left(\frac{n+1}{2},\frac{n+1}{2}\right)\,
\Gamma\left(-\frac{n}{2}\right)\,\rho_{def}\,|{\bf r}|^n\ ,
\label{EQN:CSING2}
\end{equation}
where $\Gamma(x)$ is the gamma function, and
$B(x,y) = \Gamma(x)\Gamma(y)/\Gamma(x+y)$ is the beta function. Note
that the dependence on $|{\bf r}|$ can be extracted simply by a
change of variable in (\ref{EQN:CSING1}).

The pole in the $\Gamma(-n/2)$ factor for even values of $n$ signals a
contribution of the form $|{\bf r}|^n\,\ln (|{\bf r}|/L)$ to $C_{sing}$
for those cases. We shall discuss these cases explicitly when we have
the result for general $n \le d$.

\subsection{THE GENERAL CASE $n \le d$}
For $n<d$ the defects are spatially extended, but the analysis is only
slightly more complicated. The defect defines a surface, or subspace,
of dimension $d-n$ in the $d$-dimensional space.
On the length scales of interest (small compared to $L(t)$),
the defect is effectively `flat' (walls) or `straight' (lines), etc.,
and the vector $\vec{\phi}$ can be taken  to lie in the
$n$-dimensional subspace orthogonal to the defect (the `orthogonal
subspace'). The vector
${\bf r}$ can be resolved into components ${\bf r}_{\perp}$ and
${\bf r}_{||}$ lying in this subspace and in the $d-n$ dimensional
subspace of the defect respectively. Now consider the points ${\bf x}$
and ${\bf x} + {\bf r}$, where ${\bf x}$ lies in the orthogonal subspace,
with the origin of ${\bf x}$ lying on the defect (see the illustration
in Figure 20 for the case $d=3$, $n=2$). Then (\ref{EQN:PHI.PHI}) has the
same form, but with ${\bf r}$ replaced by ${\bf r}_{\perp}$ on the
right-hand side. Proceeding as for point defects, the integration over
the $n$-dimensional vector ${\bf x}$, with ${\bf r}$ fixed, gives
\begin{equation}
C_{sing} = n\,\pi^{n/2-1}\,B\left(\frac{n+1}{2},\frac{n+1}{2}\right)\,
\Gamma\left(-\frac{n}{2}\right)\,\rho_{def}\,|{\bf r}_{\perp}|^n\ ,
\label{EQN:CSING3}
\end{equation}
where $\rho_{def}$ is, as usual, the density of defect core.

The final step is to take the isotropic average of (\ref{EQN:CSING3})
over the orientations of ${\bf r}$, i.e. to compute
$\langle |{\bf r}_{\perp}|^n \rangle$ where the brackets indicate an
isotropic average. For generality, and because we will need it later,
we in fact compute $\langle |{\bf r}_{\perp}|^\alpha \rangle$
for general $\alpha$. To do this we set up generalized polar
coordinates with the first $d-n$ polar axes in the subspace of the defect.
Then
\begin{eqnarray}
|{\bf r}_{\perp}| & = & r\prod_{i=1}^{d-n}|\sin\theta_i|\ , \nonumber \\
\langle |{\bf r}_{\perp}|^\alpha \rangle & = & r^\alpha \prod_{i=1}^{d-n}
\frac{\int_0^{\pi/2} d\theta_i\,(\sin\theta_i)^{\alpha + d-1-i}}
{\int_0^{\pi/2} d\theta_i\,(\sin\theta_i)^{d-1-i}} \nonumber \\
& = & r^{\alpha} \frac{\Gamma(d/2)\Gamma((\alpha+n)/2)}
{\Gamma(n/2)\Gamma((\alpha+d)/2)}\ .
\label{EQN:PERPAVE}
\end{eqnarray}

Using (\ref{EQN:PERPAVE}) with $\alpha=n$ to perform the isotropic average
of (\ref{EQN:CSING3}) gives the final result for the singular part of
the correlation function, valid for all $n \le d$:
\begin{equation}
C_{sing} = \pi^{n/2-1}\,\frac{\Gamma(-n/2)\Gamma(d/2)\Gamma^2((n+1)/2)}
{\Gamma((d+n)/2) \Gamma(n/2)}\,\rho_{def}\,|{\bf r}|^n\ ,
\label{EQN:CSING4}
\end{equation}
which reduces to (\ref{EQN:CSING2}) for $n=d$.

We remarked in the previous section that for even $n$ the leading singularity
is of the form $r^{n}\ln r$. The precise result can be extracted by
setting $n=2m+\epsilon$, with $m$ an integer, letting $\epsilon$ go to
zero, and picking up the term of order unity. The leading pole
contribution, proportional to $\epsilon^{-1}\,(r^2)^m$, is
analytic in $|{\bf r}|$ and therefore does not contribute to $C_{sing}$.
The $O(1)$ term (in the expansion in powers of $\epsilon$) generates the
logarithmic correction from the expansion of $|{\bf r}|^{2m + \epsilon}$.
This gives, for even n,
\begin{equation}
C_{sing} = -(4/n)\,\pi^{n/2-1}(-1)^{n/2}\frac{\Gamma(d/2)\Gamma^2((n+1)/2)}
{\Gamma((d+n)/2)\Gamma^2(n/2)}\,\rho_{def}r^n\ln r\ .
\end{equation}
It turns out that the Fourier transform $S({\bf k},t)$ of
$C_{sing}({\bf r},t)$ has the same form for even, odd and real
$n$, so we will not need to consider the even $n$ case separately.

We can now compare the exact result (\ref{EQN:CSING4}) with the
equivalent result (\ref{EQN:SHORTGAUSSIAN}) obtained within the gaussian
theory for nonconserved fields. Equating these provides another way of
estimating the defect density within the gaussian theory, namely
\begin{equation}
\rho_{def}^{gauss2} = \frac{1}{(4\pi t)^{n/2}}\,
\frac{\Gamma((d+n)/2)}{\Gamma(d/2)}\ .
\label{EQN:GAUSS2}
\end{equation}
Comparing this with (\ref{EQN:GAUSS1}), we see that the two estimates
differ for general $d$, but agree for $d \to \infty$. This is another
indication that the gaussian approximation becomes exact in this limit.

\subsection{THE STRUCTURE FACTOR TAIL}
It remains to Fourier transform (\ref{EQN:CSING4}) to obtain the tail of the
structure factor. Although the Fourier transform of (\ref{EQN:CSING4}) by
itself does not technically exist (because the required integral does not
converge), the following method gives the large-momentum tail correctly,
as may be checked by back-transforming the result. Of course, the Fourier
transform of the complete correlation function $C({\bf r},t)$ does exist,
since $C({\bf r},t)$ vanishes at infinity.

Simple power counting on (\ref{EQN:CSING4}) gives immediately the power-law
tail $S({\bf k},t) \sim k^{-(d+n)}$.
To derive the {\em amplitude} we employ the integral representation
\begin{equation}
\Gamma(-n/2) |{\bf r}|^n = \int_0^\infty du\,u^{-n/2-1}\,\{\exp(-ur^2)
                                         -{\rm analytic\ terms}\}\ ,
\label{EQN:CSING5}
\end{equation}
where `analytic terms' indicates, once more, as many terms in the expansion
of $\exp(-ur^2)$ as are necessary to converge the integral. These terms will
not contribute to the tail of the Fourier transform, and can be dropped
once the transform has been taken. The Fourier transform of
(\ref{EQN:CSING5}) is, therefore,
\begin{eqnarray}
&&\int_0^\infty du\,u^{-n/2-1}\,\int d^dr \exp(-ur^2 -i{\bf k}\cdot{\bf r})
\nonumber \\
\ \ \ \ \ \ \ \ &=& \pi^{d/2}\,\int_0^\infty du\,u^{-(d+n)/2-1}\,\exp(-k^2/4u)
\nonumber \\
\ \ \ \ \ \ \ \ &=& \pi^{d/2}\,\Gamma\left(\frac{d+n}{2}\right)\,
\left(\frac{2}{k}\right)^{d+n}\ .
\end{eqnarray}
Inserting the remaining factors from (\ref{EQN:CSING4}) gives the final result,
\begin{equation}
S({\bf k},t) = \frac{1}{\pi}\,(4\pi)^{(d+n)/2}\,\frac{\Gamma^2((n+1)/2)
\Gamma(d/2)}{\Gamma(n/2)}\,\frac{\rho_{def}}{k^{d+n}}\ .
\label{EQN:EXACTTAIL}
\end{equation}
We note that this expression is smooth as $n$ passes through the even
integers. The generality of the result should be noted: it is
independent of any details of the dynamics, e.g.\ whether the order
parameter is conserved or non-conserved, and holds independently of
whether the scaling hypothesis is valid. We note that, as well as
providing an exact result against which to test approximate theories,
Eq.\ (\ref{EQN:EXACTTAIL}) can also be used to {\em determine} the defect
density experimentally from scattering data.

Measurements of the tail amplitude in numerical simulations are in good
agreement with (\ref{EQN:EXACTTAIL}) for both scalar \cite{BH93} and
vector \cite{BBunpub} fields. As an example, we show in Figure 21 the
results for the cases $n=d=2$, $n=d=3$ and $n=2$, $d=3$ \cite{BBunpub}.
It is interesting that in all cases the asymptotic behaviour is approached
{\em from above}. For a scalar order parameter ($n=1$), the leading
correction to (\ref{EQN:EXACTTAIL}) has been calculated by Tomita
\cite{Tomita}. It is a term of order $k^{-(d+3)}$, associated with the
curvature of the interfaces and obtained from a short-distance expansion
of the form $C({\bf r} = 1 - ar + br^3 - \cdots$. The absence of an $r^2$
term leads to the `Tomita sum rule' for the structure factor \cite{Tomita},
$\int_0^\infty dk\,[k^{d+1}S(k) - A] = 0$, where $A$ is the amplitude of the
Porod tail.

The discussion has so far been restricted to the cases $n \le d$, where
singular topological defects exist. What can be said about the structure
factor tail for $n>d$? The case $n=d+1$ may be complicated by the presence
of topological textures \cite{Textures}. For $n > d+1$, preliminary
numerical studies for nonconserved dynamics \cite{HumayunUnpub} suggest
that the structure-factor scaling function $g(y)$ has a `stretched-exponential'
tail,  $g(y) \sim \exp(-y^\delta)$, with $\delta \simeq 1$ for $d=1,n=3$ and
$\delta \simeq 1/2$ for $d=2,n=4$. Similar studies by Toyoki were analysed
in terms of a power-law tail, $g(y) \sim y^{-\chi}$, with $\chi > d+n$, but
are also consistent with stretched exponential decay \cite{Toyoki94}.
In contrast to $n \le d$, the tail behaviour for $n>d$ may be different
for conserved and nonconserved dynamics. Recent results for conserved
dynamics \cite{CR} suggest a stretched exponential form but with
$\delta \simeq 1.7$ for both $d=2,n=4$ and $d=1,n=3$, while
$\delta \sim 2.7$ for $d=1,n=2$.

\subsection{HIGHER-ORDER CORRELATION FUNCTIONS}
\label{SEC:HOCFs}
We consider first the correlation function $C_4$, defined by (\ref{EQN:C_4}),
of the square of the order-parameter. We concentrate here on the numerator
$C_4^N$, defined by,
\begin{equation}
C_4^N = \langle \{1-\vec{\phi}(1)^2\}\{1-\vec{\phi}(2)^2\}\rangle\ .
\label{EQN:C_4^N-1}
\end{equation}
For $n \le d$, the presence of topological defects leads to a
singular short-distance behaviour that can be evaluated in direct analogy
to that of the usual pair correlation function $C$. As in the approximate
calculation of $C_4$, using gaussian auxiliary field methods, in section
\ref{SEC:C_4}, we have to distinguish between scalar and vector fields.

\subsubsection{Scalar Fields}
For scalar fields, $1-\phi^2$ is sharply peaked near domain walls. It is
convenient to introduce the auxiliary field $m$ defined as in Mazenko's
approximate theory, but {\em not} assumed to be gaussian! We recall that
the function $\phi(m)$ represent the equilibrium domain wall profile,
with $m$ the coordinate normal to the wall. Since $\phi^2$ saturates to
unity with a width of order $\xi$ of the wall, we can use, for the
calculation of scaling functions,
\begin{equation}
1-\phi^2(m) = \xi\delta(m)\ ,
\label{EQN:C_4^N-2}
\end{equation}
where we have used our usual definition of the `wall width',
$\xi = \int_{-\infty}^{\infty} dm\,(1-\phi^2(m))$.
Putting (\ref{EQN:C_4^N-2}) in (\ref{EQN:C_4^N-1}) gives
\begin{eqnarray}
C_4^N(1,2) & = & \xi^2 \langle \delta(m(1))\,\delta(m(2)) \rangle\ \\
           & = & \xi^2 P(1,2) \\
           & = & \xi^2 P(1|2)\,P(2)\ .
\label{EQN:C_4^N-3}
\end{eqnarray}
Here we have used the fact that the wall area density (area per unit volume),
$\delta(m)|\nabla m|$, can be simplified to $\delta(m)$ since $|\nabla m|=1$
at interfaces from the definition of $m$ as a coordinate normal to the
interface. In (\ref{EQN:C_4^N-3}), therefore, $P(1,2)$ indicates the
joint probability density to find both `1' and `2' in an interface. In the
final equality,  $P(1|2)$ is the probability density to find `1' in an
interface given that `2' is in an interface. Clearly
$P(2) = \rho$, the average wall density.  For $r \ll L(t)$, $P(1|2)$ is
dominated by cases where `1' and `2' lie in the {\em same} wall, which can
be regarded as flat on this scale, as illustrated in Figure 22. For
general $d$, $P(1|2) = S_{d-1}/S_d r$, where $S_d = 2\pi^{d/2}/\Gamma(d/2)$
is the surface area of a $d$-dimensional sphere of unit radius.
Assembling everything in (\ref{EQN:C_4^N-3}),
\begin{equation}
C_4^N \to \xi^2\,\rho\,\frac{S_{d-1}}{S_d r}\ ,\ \ \ \ \
\xi \ll r \ll L(t)\ .
\label{EQN:C_4^N-4}
\end{equation}
This result breaks down when $r$ becomes comparable with $\xi$, since it
is no longer adequate to neglect the thickness of the walls. The small-$r$
behaviour (\ref{EQN:C_4^N-4}) implies the power-law tail
$S_4({\bf k},t) \sim k^{-(d-1)}$
for the corresponding structure factor, valid when $k\xi \ll 1 \ll kL(t)$.

To obtain the normalised correlation function $C_4$, we divide
(\ref{EQN:C_4^N-4}) by its large-$r$ limit $\langle 1-\phi^2 \rangle^2$.
But
\begin{equation}
\langle 1-\phi^2 \rangle = \xi \langle \delta(m) \rangle = \xi\rho\ ,
\end{equation}
giving
\begin{equation}
C_4 \to \frac{S_{d-1}}{S_d\,\rho r}\ ,\ \ \ \ \ \xi \ll r \ll L(t)\ ,
\label{EQN:C_4SHORT}
\end{equation}
from which the wall width $\xi$ has dropped out. The exact result for
the tail of the corresponding structure factor is obtained by Fourier
transformation. Inserting the expression for $S_d$ gives
\begin{equation}
S_4({\bf k},t) \to (2/\rho)\,(4\pi)^{(d-2)/2}\,\Gamma(d/2)\,k^{-(d-1)}\ ,
\ \ \ \ \ k\xi \ll 1 \ll kL(t)\ .
\end{equation}
This could be measured by small-angle scattering in a situation where
all the scattering was from the interfaces. The $d=3$ result has been
derived by Onuki, together with the leading correction, of relative
order $1/(kL)^2$ \cite{Onuki}.

A heuristic derivation of the $k^{-(d-1)}$ tail, based purely on scaling,
proceeds as follows \cite{Bray93}.
Since $\langle 1-\phi^2 \rangle \sim \rho \sim 1/L$,
$C_4^N$ has the scaling form $C_4^N = L^{-2} f(r/L)$, giving
$S_4^N = L^{d-2} g(kL)$. But for $kL \gg 1$, the scattering intensity
should scale as the defect density, i.e. as $1/L$. This requires
$g(y) \sim y^{-(d-1)}$ for $y \gg 1$. This argument can be
generalised to vector fields, as we shall see.

It is interesting that (\ref{EQN:C_4SHORT}) can be combined with
(\ref{EQN:CSING4}) (with $n=1$) to obtain an exact relation,
valid at short distances, between the two correlation functions.
For $n=1$, (\ref{EQN:CSING4}) implies the short-distance behaviour
\begin{equation}
C({\bf r},t) = 1 - \frac{2}{\surd{\pi}}\,\frac{\Gamma(d/2)}{\Gamma((d+1)/2)}\,
               \rho r\ .
\label{EQN:CSING6}
\end{equation}
Eliminating $\rho r$ between (\ref{EQN:C_4SHORT}) and (\ref{EQN:CSING6})
yields
\begin{equation}
C_4^{-1} = \frac{\pi}{2}\,\frac{\Gamma((d-1)/2)\Gamma((d+1)/2)}
            {\Gamma^2(d/2)}\,(1-C)\ ,
\label{EQN:ABSOLUTE-SHORT}
\end{equation}
valid at `short' distances (i.e.\ $\xi \ll r \ll L(t)$). Let us compare
this exact result with (\ref{EQN:ABSOLUTE}), obtained using gaussian auxiliary
field methods. For short distances, when $C$ is close to unity,
(\ref{EQN:ABSOLUTE}) becomes $C_4^{-1} = (\pi/2)\,(1-C)$. This has the same
form as (\ref{EQN:ABSOLUTE-SHORT}), but with a different coefficient of
$(1-C)$. However, the exact coefficient approaches the `gaussian' value
of $\pi/2$ in the limit $d\to \infty$, consistent with our argument that the
auxiliary field $m$ is indeed gaussian in this limit.

\subsubsection{Vector Fields}
For vector fields, $\vec{\phi}^2$ saturates to unity only as an inverse
power of the distance from a defect, and the representation
(\ref{EQN:C_4^N-2}) is no longer appropriate. Instead,
(\ref{EQN:VECPROFTAIL}) (see also (\ref{EQN:VECTAIL})) implies
\begin{equation}
1 - \vec{\phi}^2 \to \xi^2/r^2
\end{equation}
when the distance $r$ from the core satisfies $\xi \ll r \ll L(t)$.
The calculation of the singular part of $C_4^N$, due to the presence
of defects, then follows that of the usual pair correlation function
$C$:
\begin{equation}
C_{4\,sing}^N = \xi^4 \rho_{def} \int d^nx\,\left\langle
        \frac{1}{{\bf x}^2({\bf x}+{\bf r}_\perp)^2} \right\rangle\ ,
\end{equation}
where ${\bf r}_\perp$ is the usual component of ${\bf r}$ in the plane
perpendicular to the defect, and the angled brackets indicate an
isotropic average over the orientations of ${\bf r}$. Evaluating the
${\bf x}$-integral first gives
\begin{equation}
C_{4\,sing}^N = \xi^4\rho_{def}\,\pi^{n/2}\frac{\Gamma(2-n/2)\Gamma^2(n/2-1)}
{\Gamma(n-2)}\,\langle |{\bf r}_\perp|^{n-4} \rangle\ .
\end{equation}
Using (\ref{EQN:PERPAVE}), with $\alpha = n-4$, gives (after some algebra)
\begin{equation}
C_{4\,sing}^N = -\xi^4\rho_{def}\,\pi^{n/2}\,\frac{\Gamma(n/2-1)\Gamma(1-n/2)
                \Gamma(d/2)}{\Gamma((d+n-4)/2)}\,r^{n-4}\ .
\label{EQN:CSING7}
\end{equation}

This result gives the {\em singular} part of $C_4^N$ for all $n>2$. The poles
in the numerator at even $n$ signal additional factors of $\ln r$ (actually
$\ln (r/L)$), as in the calculation of $C$. The amplitude of the logarithm
can be extracted by setting $n=2m+\epsilon$, with $m$ an integer, letting
$\epsilon$ go to zero, and picking up the term of order unity. This gives,
for even $n$ greater than 2,
\begin{equation}
C_{4\,sing}^N = -4\xi^4\rho_{def}\,(-\pi)^{n/2}\,\frac{\Gamma(d/2)}
                {(n-2)\Gamma((d+n-4)/2)}\,r^{n-4}\ln r\ .
\end{equation}

This singular short-distance behaviour implies, as usual, a power-law tail
in Fourier space, $S_4({\bf k},t) \sim k^{-(d+n-4)}$ for
$n>2$, in agreement with the prediction of the gaussian theories
(section \ref{SEC:C_4} and \cite{Bray93}). Again, there is a heuristic
argument for this \cite{Bray93}. The result
$\langle (1-\vec{\phi}^2)\rangle \sim 1/L^2$ for $n>2$ suggests the
scaling form $C_4^N = L^{-4}f(r/L)$, with Fourier transform
$L^{d-4}g(kL)$. Extracting the expected proportionality to $\rho_{def}
\sim L^{-n}$ for $kL \gg 1$ generates the required $k^{-(d+n-4)}$ tail.

The case $n=2$ is more complicated, due to an additional pole in
(\ref{EQN:CSING7}).  The same technique, however, can be used.
We put $n=2+\epsilon$ in (\ref{EQN:CSING7}),
let $\epsilon$ go to zero, and pick up the term of order unity. The result
is
\begin{equation}
C_{4\,sing}^N = \xi^4\rho_{def}\,\pi(d-2)\,\ln^2 r/r^2\ ,\ \ \ \ \ d>n=2\ .
\end{equation}
For $d=2$, we must set $d=2$ in (\ref{EQN:CSING7}) before taking
$\epsilon$ to zero. This gives
\begin{equation}
C_{4\,sing}^N = 2\pi \xi^4\rho_{def}\,\ln r/r^2\ ,\ \ \ \ \ \ d=n=2\ .
\end{equation}
The corresponding structure factor (the Fourier transform of
$C_{4\,sing}^N$) has a large-momentum tail of the form
$\xi^4\rho_{def}\,\ln^2(k\xi)/k^{d-2}$. These results are discussed in
more detail in reference \cite{BBunpub}. In particular, it is shown that
the result for $d=n=2$ is inconsistent with conventional scaling form.

\subsubsection{Defect-Defect Correlations}
The short-distance behaviour of the defect-defect correlation
functions, introduced in section \ref{SEC:DEFECT-DEFECT}, may
also be determined exactly, at least for extended defects.
The argument is a simple extension of that used to calculate
$C_4$ for scalar fields. From the first part of (\ref{EQN:rhorho}),
the correlator $\langle \rho(1)\rho(2)\rangle$ is just the joint
probability density $P(0,0)$ for points `1' and `2' both
lying on a defect. Clearly $P(0,0) =\rho_{def}P(2|1)$,
in the usual notation. But for $r \ll L$, $P(2|1)$ is dominated
by cases where `1' and `2' are in the {\em same} defect ({\em
provided} the defects are spatially extended, i.e. $n<d$). An
obvious generalisation of (\ref{EQN:C_4^N-4}) gives
\begin{eqnarray}
\langle \rho(1)\rho(2)\rangle & = & \rho_{def} \frac{S_{d-n}}{S_d r^n}
\nonumber \\
           & = & \frac{\rho_{def}}{\pi^{n/2}\,r^n}\,
                  \frac{\Gamma(d/2)}{\Gamma((d-n)/2)}\ .
\label{EQN:DEF-DEF2}
\end{eqnarray}
This result differs from (\ref{EQN:DEF-DEF1}), obtained by using
the gaussian approximation, but approaches it in the limit $d \to \infty$,
as we have by now come to expect.  In the exact result
(\ref{EQN:DEF-DEF2}), the amplitude of the $r^{-n}$ divergence vanishes
for point defects ($n=d$), an important physical feature that is missing
from the gaussian approximation (\ref{EQN:DEF-DEF1}).

\subsection{THE PROBABILITY DISTRIBUTION $P(m(1),m(2))$}
To conclude this section on short-distance behaviour, we compute the
exact form of $P(m(1),m(2))$ for scalar fields, valid for $|m(1)|$,
$|m(2)|$ and $r$ all much smaller than $L(t)$, and show explicitly
that it is not gaussian. Technically, the regime in which are working
corresponds to taking the limit $L(t) \to \infty$ with $|m(1)|$,
$|m(2)|$ and $r$ fixed but arbitrary. This situation is illustrated
in Figure 23, where the domain wall can be regarded as flat in the limit
of interest. The identity $P(m(1),m(2)) = P(m(1))\,P(m(2)|m(1))$,
where $P(x|y)$ indicates a conditional probability, gives
\begin{equation}
P(m(1),m(2)) = \rho\,\langle \delta(m(2)-m(1)-r\cos\theta)\rangle\ ,
\end{equation}
where the angled brackets indicate an isotropic average over $\theta$,
and we have used $P(m(1))=\rho$, the wall density, for $|m(1)| \ll L$.
Carrying out the angular average (with weight proportional to
$\sin^{d-2}\theta$) gives
\begin{equation}
P(m(1),m(2)) = \left[B\left(\frac{1}{2},\frac{d-1}{2}\right)\right]^{-1}\,
\frac{\rho}{r}\,\left(1 - \frac{(m(2)-m(1))^2}{r^2}\right)^{(d-3)/2}\ ,
\label{EQN:JOINTSHORT}
\end{equation}
for $|m(1)-m(2)|\le r$, and $P=0$ otherwise.
Clearly $P(m(1),m(2))$ is not gaussian for general $d$. However, it
approaches a gaussian for large $d$, as we now show.

It is clear from (\ref{EQN:JOINTSHORT}) that, in the limit $d \to \infty$,
$P(m(1),m(2))$ vanishes except when $|m(1)-m(2)|/r$ is of order $1/\surd{d}$.
Therefore we define
\begin{equation}
\Delta = \sqrt{d}\,(m(1) - m(2))/r\ .
\end{equation}
Now the large-$d$ limit can be taken at fixed $\Delta$. Taking $d$ large in
the beta function too gives
\begin{equation}
P(m(1),m(2)) \to \left(\frac{d}{2\pi}\right)^{1/2}\,
\frac{\rho}{r}\,\exp(-\Delta^2/2)\ ,\ \ \ \ \ d \to \infty\ .
\label{EQN:PSHORT1}
\end{equation}
Let's compare this result with that of the systematic approach, which we
argued is exact for large $d$, by  evaluating (\ref{EQN:JOINTPROB})
in the same limit. Equation (\ref{EQN:SYS7}) gives $S_0 = 4t/d$ and
$\gamma = \exp(-r^2/8t)$.
Inserting these in (\ref{EQN:JOINTPROB}), and taking the limit
$t \to \infty$ with $r$, $m(1)$ and $m(2)$ held fixed, gives
\begin{equation}
P(m(1),m(2)) = \frac{d}{4\pi r\sqrt{t}}\,\exp(-\Delta^2/2)\ .
\label{EQN:PSHORT2}
\end{equation}
Eqs.\ (\ref{EQN:PSHORT1}) and (\ref{EQN:PSHORT2}) agree if
$\rho = (d/8\pi t)^{1/2}$, which is just Eq.\ (\ref{EQN:GAUSS1}).
We conclude that exact result (\ref{EQN:JOINTSHORT}) is consistent
with the gaussian approximation in the limit $d \to \infty$, but not
for any finite $d$.

\section{GROWTH LAWS}
\label{SEC:GROWTH}

The exact short-distance singularities derived in the previous section,
together with the scaling hypothesis, provide a basis for deriving
exact growth laws for all phase-ordering systems with purely dissipative
dynamics.

Although the growth laws for both nonconserved and conserved
scalar systems, and conserved fields in general, have been derived by
a number of methods, there has up until now been no simple, general
technique for obtaining $L(t)$. In particular, the growth laws for
non-conserved vector fields have, until recently, been somewhat
problematical. Here we describe a very general approach, recently developed
by Bray and Rutenberg (BR) \cite{BR}, to obtain $L(t)$ consistently by
comparing the global rate of energy change to the energy dissipation from
the local evolution of the order parameter. This method allows the
explicit derivation of growth laws for $O(n)$ models, but the results can
be also be applied to other systems with similar defect structures.

The BR approach is based on the dissipation of energy that occurs as the
system relaxes towards its ground state. The energy dissipation
is evaluated by considering the motion of topological defects, when they
exist. The defect contribution either dominates the dissipation or gives
a contribution that scales with time in the same way as the total dissipation.
The global rate of energy change, computed from the time derivative of the
total energy, is equated to the energy dissipation from the local evolution
of the order parameter. For systems with a single characteristic scale
$L(t)$, this approach self-consistently determines the time-dependence
of $L(t)$.

\subsection{A USEFUL IDENTITY}
\label{SEC:IDENTITY}
We begin by writing down the equation of motion for the
Fourier components $\vec{\phi}_{\bf k}$:
\begin{equation}
\label{EQN:DYNAMICS}
\partial_t \vec{\phi}_{\bf k} =
-k^{\mu}\,(\partial F/\partial \vec{\phi}_{\bf -k}),
\end{equation}
The conventional non-conserved (model A) and conserved (model B)
cases are $\mu=0$  and $\mu=2$, respectively. (Elsewhere in this article,
the symbol $\mu$ has been used for the chemical potential: the meaning
should be clear from the context).

Integrating the rate of energy dissipation from each Fourier mode,
and then using the equation of motion (\ref{EQN:DYNAMICS}), we find
\begin{eqnarray}
\label{EQN:LHSRHS}
d\epsilon/dt &=& \int_{\bf k} \left< (\partial F/\partial\vec{\phi}_{\bf k})
	\cdot \partial_t \vec{\phi}_{\bf k} \right>  \nonumber \\
&=& - \int_{\bf k} k^{-\mu}\,
	\left< \partial_t \vec{\phi}_{\bf k} \cdot
	\partial_t \vec{\phi}_{\bf -k}
	\right>\ ,
\end{eqnarray}
where $\epsilon = \left< F \right>/V$ is the mean energy density, and
$\int_{\bf k}$ is the momentum integral
$\int d^d k /(2\pi)^d$. We will relate the scaling behaviour of both
sides of (\ref{EQN:LHSRHS}) to that of appropriate integrals over the
structure factor, $S({\bf k},t)$, and its two-time generalisation.
Either the integrals converge, and the dependence on the scale $L(t)$ can
be extracted using the scaling form (\ref{EQN:STRUCT}) (or its two-time
generalisation (\ref{EQN:TWOTIME})), or the integrals diverge in the
ultraviolet (UV) and have to be cut off at $k_{max} \sim 1/\xi$,
corresponding to a dominant contribution from the core scale.
It is just this small-scale structure that is
responsible for the generalised Porod law (\ref{EQN:GENPOROD})
for the structure factor, and the time-dependence of any integrals
controlled by the core scale can be extracted from a knowledge of the
defect structure.

\subsubsection{The Energy Integral}
\label{SEC:ENERGY}
To see how this works, we first calculate the scaling behaviour
of the energy density, $\epsilon$, which  is captured by
that of the gradient term in (\ref{EQN:HAMILTONIAN}):
\begin{eqnarray}
\label{EQN:GRADSQUARE}
\epsilon &\sim& \left< (\nabla \vec{\phi})^2 \right> \nonumber \\
         &=& \int_{\bf k} k^2\,L^d\,g(kL)\ ,
\end{eqnarray}
where we have used the scaling form (\ref{EQN:STRUCT})
for the structure factor. For $n >2$ the integral in UV convergent,
and a change of variables yields $\epsilon \sim L^{-2}$.
For $n \le 2$, when the integral in UV divergent, we use Porod's law
(\ref{EQN:GENPOROD}) and impose a UV cutoff at $k \sim 1/\xi$, to obtain
\cite{Toy92}
\begin{eqnarray}
\label{EQN:E}
\epsilon & \sim & L^{-n}\,\xi^{n-2}\ ,\ \ \ \ \ \ \ \ \ \ \ \ n<2\ ,\nonumber
\\
         & \sim & L^{-2}\,\ln(L/\xi)\ ,\  \ \ \ \ \ \ \ n=2\ ,\nonumber \\
         & \sim & L^{-2}\ ,\ \ \ \ \ \ \ \ \ \ \ \ \ \ \ \ \ \ n>2\ .
\end{eqnarray}
We see that the energy is dominated by the defect core density,
$\rho_{\rm def} \sim L^{-n}$, for $n<2$, by the defect field at all
length scales between $\xi$ and $L$ for $n=2$,
and by variations of the order parameter at scale $L(t)$ for $n>2$.

\subsubsection{The Dissipation Integral}
\label{SEC:DISSIPATION}
We now attempt to evaluate the right side of (\ref{EQN:LHSRHS}) in a
similar way. Using the scaling hypothesis for the two-time function,
\begin{equation}
\left< \vec{\phi}_{\bf k} (t) \cdot \vec{\phi}_{-{\bf k}} (t') \right>
= k^{-d} g(kL(t),kL(t')\,)\ ,
\end{equation}
which is the spatial Fourier transform of (\ref{EQN:TWOTIME}), we find
\begin{eqnarray}
\label{EQN:TWOTIMESS}
	\left< \partial_t \vec{\phi}_{\bf k} \cdot \partial_t
	\vec{\phi}_{-{\bf k}} \right> &=& \left. \frac{\partial^2}
	{\partial t \partial t'} \right|_{t=t'}
	\left< \vec{\phi}_{\bf k}(t) \cdot \vec{\phi}_{-{\bf k}}(t') \right>
	\nonumber \\
	&=& \dot{L}^2 L^{d-2} h(kL)\ ,
\end{eqnarray}
where $\dot{L} \equiv dL/dt$.

When the momentum integral on the right of (\ref{EQN:LHSRHS}) in UV
convergent we obtain, using (\ref{EQN:TWOTIMESS}),
$d\epsilon/dt \sim - \dot{L}^2\,L^{\mu-2}$.
If, however, the integral is UV divergent, it will be dominated by the
behaviour of the integrand near the upper limit, so we need to know the
form of the scaling function $h$ in (\ref{EQN:TWOTIMESS}) for $kL \gg 1$.
It turns out that, in general, the large-$kL$ form is quite complicated,
with many different cases to consider \cite{BRunpub}.
However, we only need the result for those cases where the dissipation
integral requires a UV cut-off, otherwise simple power counting
is sufficient. For those cases, one additional assumption, which can
be verified {\em a posteriori}, yields a simple and rather general
result (Eq.\ (\ref{EQN:RHSLARGEK}) below).

\subsection{EVALUATING THE DISSIPATION INTEGRAL}
\label{SEC:PROBLEMS}
\subsubsection{An Illustrative Example}
\label{SEC:EXAMPLE}
To see what difficulties arise, and how to circumvent them, it is
instructive to consider a scalar field. We want to calculate
$\left< \partial_t \phi_{\bf k} \, \partial_t \phi_{-{\bf k}} \right>$
in the limit $kL \gg 1$. It is clear that
$\partial_t\phi$ is appreciably different from zero only near interfaces.
In fact, since $d\phi/dt=0$ in a frame comoving with the interface, we
have, near an interface, $\partial_t\phi = -{\bf v}.\nabla\phi$,
where ${\bf v}$ is the interface velocity. In real space, therefore,
\begin{equation}
\left< \partial_t \phi(1) \, \partial_t \phi(2) \right>
= \left< {\bf v}(1)\cdot\nabla\phi(1)\,{\bf v}(2)\cdot\nabla\phi(2)\right>\ .
\end{equation}
The large $kL$ behaviour in Fourier space is obtained from the short-distance
($r \ll L$) behaviour in real space. For $r \ll L$, the points `1' and `2'
must be close to the {\em same} interface. For a typical interface, with radius
of curvature of order $1/L$, the speed $v$ is slowly varying along the
interface. Furthermore, the interface may be regarded as `flat' for the
calculation of the short-distance correlation, just as in the derivation of
Porod's law. It follows that the averages over the interface velocity
and position can be carried out independently, giving
\begin{equation}
\left< \partial_t \phi(1) \, \partial_t \phi(2) \right>
= (1/d)\,\langle v^2 \rangle\,\langle \nabla\phi(1)\cdot\nabla\phi(2)\rangle\ .
\end{equation}
Fourier transforming this result gives
\begin{eqnarray}
\left< \partial_t \phi_{\bf k} \, \partial_t \phi_{-{\bf k}} \right>
& = & (\langle v^2 \rangle/d)\,k^2\,S({\bf k},t)\ ,\ \ \ \ \ \ kL \gg 1
\nonumber \\
& \sim & \langle v^2 \rangle\ /Lk^{d-1}\ ,\ \ \ \ \ \ kL \gg 1\ ,
\label{EQN:NAIVE}
\end{eqnarray}
where the Porod result (\ref{EQN:POROD}) was used in the final line.
We will see that Eq.\ (\ref{EQN:NAIVE}) requires a careful interpretation.

The next step is to evaluate $\langle v^2 \rangle$. Since the
characteristic interface velocity is $\dot{L}$, we expect
$\langle v^2 \rangle \sim \dot{L}^2$. This assumes, however, that the
average is dominated by `typical' values. This, as we shall see, is the
key question. Consider a small spherical domain of radius $r$ in a
non-conserved system. The interface velocity (see Eq.\ (\ref{EQN:RADIUS}))
is $v \sim 1/r$. For a conserved system, Eq.\ (\ref{EQN:COPRADIUS})
gives $v \sim 1/r^2$. Thus the relation $v(r) \sim 1/r^{z-1}$
where $z=2$ and 3 for nonconserved and conserved systems respectively
(and the growth law is $L \sim t^{1/z}$) covers both cases. The fact
that $v$ diverges at small $r$ raises the possibility that
$\langle v^2 \rangle$ is dominated by small domains.
The domain-size distribution function $n(r)$ has the scaling form
$n(r) = L^{-(d+1)}f(r/L)$ (in order that $\int dr\,n(r)
\sim L^{-d}$, consistent with scaling). The  important small-$x$
behaviour of the function $f(x)$ can be determined as follows.
Consider a small time interval $\Delta t$. The domains that will have
disappeared  after this time interval are those with radius smaller than
$r_{\Delta t} \sim (\Delta t)^{1/z}$. The number of such domains is
of order $L^{-(d+1)}\int_0^{r_{\Delta t}} dr\,f(r/L)$. The requirement
that this be linear in $\Delta t$ forces $f(x) \sim x^{z-1}$ for $x \to 0$.
Using $v \sim r^{-(z-1)}$ we can estimate the contribution to
$\langle v^2 \rangle$ from short scales:
\begin{eqnarray}
\langle v^2 \rangle & \sim & \frac{\int_\xi^L dr\,r^{d-1}\,n(r)\,v^2(r)}
{\int_\xi^L dr\,r^{d-1}\,n(r)}\nonumber \\
& \sim & \int_\xi^L dr\,r^{d-z}/\int_\xi^L dr\,r^{d+z-2}\ .
\label{EQN:v^2}
\end{eqnarray}
The integral in the denominator converges at short scales, giving a
result of order $L^{d+z-1}$. For nonconserved fields ($z=2$), the numerator
converges for all $d>1$, giving $L^{d-1}$ for the numerator, and
$\langle v^2 \rangle \sim 1/L^2$. Since $\dot{L} \sim 1/L$ for this case,
we have $\langle v^2 \rangle \sim \dot{L}^2$ as expected. For conserved
fields ($z=3$), however, the numerator only converges at short scales
for $d>2$. For those cases, one again finds
$\langle v^2 \rangle \sim \dot{L}^2$. For $d=2$, though, the numerator
is of order $\ln (L/\xi)$. This gives
$\langle v^2 \rangle \sim L^{-4}\,\ln (L/\xi)$, i.e.\ there are
contributions from all scales, and $\langle v^2 \rangle \sim \dot{L}^2$
no longer holds. Putting this into (\ref{EQN:NAIVE}) gives
\begin{equation}
\left< \partial_t \phi_{\bf k} \, \partial_t \phi_{-{\bf k}} \right>
\stackrel{?}{\sim} \ln(L/\xi)/L^5k\ ,\ \ \ \ \ kL \gg 1\ ,
\ \ \ \ \ \ (d=2,\ {\rm conserved})\ ,
\label{EQN:v^2,d=2}
\end{equation}
We will now show that this result is wrong!

The factor $1/L$ in (\ref{EQN:NAIVE}) represents the total interfacial area
density: (\ref{EQN:NAIVE}) implies that interfaces contribute additively to
$\left< \partial_t \phi_{\bf k} \, \partial_t \phi_{-{\bf k}} \right>$.
In the derivation of Eq.\ (\ref{EQN:NAIVE}), however, we
explicitly assumed that only interfaces of typical curvature,
of order $1/L$,  contribute.  For a piece of interface of local
curvature $1/R$, the condition that the interface be regarded as
locally flat on the scale $1/k$ requires $kR \gg 1$, not simply $kL \gg 1$.
For fixed $k \gg 1/L$, sharply curved interfaces, with
$R \stackrel{<}{\sim} 1/k$, {\em do not contribute to the Porod tail}.
This means that, as far as the computation of the large $kL$ behaviour is
concerned, there is an {\em effective short-distance cut-off} at $1/k$:
only interfaces with radius of curvature $R \gg 1/k$ should be included.
For the calculation of the usual Porod tail in
$\langle \phi_{\bf k}\,\phi_{\bf -k}\rangle $ this makes no difference,
because interfaces with $R \stackrel{<}{\sim} 1/k$ make a negligible
contribution to the total interfacial area as $kL \to \infty$.
For the calculation of
$\left< \partial_t \phi_{\bf k} \, \partial_t \phi_{-{\bf k}} \right>$,
however, it can make a big difference, because of the extra
factor of $v^2$ inside the average. This means that,
in evaluating $\langle v^2 \rangle$, $1/k$ rather than $\xi$ is the
appropriate short-distance cut-off.  Applying this to conserved scalar
fields in $d=2$ gives $\langle v^2 \rangle \sim L^{-4}\,\ln (kL)$, and
\begin{equation}
\left< \partial_t \phi_{\bf k} \, \partial_t \phi_{-{\bf k}} \right>
\stackrel{?}{\sim} \ln(kL)/L^5k\ ,\ \ \ \ \ kL \gg 1,
\ \ \ \ \ \ (d=2,\ {\rm conserved})\ ,
\label{EQN:v^2,d=2,RIGHT}
\end{equation}
instead of (\ref{EQN:v^2,d=2}).

The final step is to insert Eq.\ (\ref{EQN:v^2,d=2,RIGHT}) into the
dissipation integral (\ref{EQN:LHSRHS}), with $d=2=\mu$.
One immediately sees that the integral is UV convergent:
the $L$-dependence can be extracted
trivially by a change of variable, $d\epsilon/dt \sim -1/L^4$.
So we did not actually need the asymptotic form of
$\left< \partial_t \phi_{\bf k} \, \partial_t \phi_{-{\bf k}} \right>$
after all (except to show that the dissipation integral is UV convergent)!
Note that the final result, $d\epsilon/dt \sim -1/L^4$, is consistent
with $\epsilon \sim 1/L$ (Eq.\ (\ref{EQN:E}) with $n=1$) and the result
$\dot{L} \sim 1/L^2$ for conserved (i.e.\ $\mu=2$) scalar fields.

There is, however, one last complication. Eq.\ (\ref{EQN:v^2,d=2,RIGHT})
is still not quite correct! This is because the expression $v(r) \sim 1/r^2$
for the velocity of a small drop (i.e.\ with $r \ll L$) breaks down for
$d=2$ due to the singular nature of the Greens function for the Laplacian.
For this case one finds instead \cite{BRunpub} $v(r) \sim 1/r^2\ln(L/r)$.
Using this gives finally $\left< \partial_t \phi_{\bf k}\,
\partial_t \phi_{-{\bf k}} \right> \sim \ln[\ln(kL)]/L^5k$ for $kL \gg 1$,
instead of (\ref{EQN:v^2,d=2,RIGHT}). This does not alter, of course, the
conclusion that the dissipation integral is UV convergent, and can
therefore be evaluated simply by a change of variable.

\subsubsection{The Way Forward}
\label{SEC:FORWARD}
We have gone through this one case in some detail, because we can extract
from it a general principle that avoids treating every case separately
(although this can be done \cite{BRunpub}). The central point, given
extra emphasis by the discussion above,  is that we only need to know
the asymptotics of  $\left< \partial_t \vec{\phi}_{\bf k} \cdot \partial_t
\vec{\phi}_{-{\bf k}} \right>$ in those cases where the dissipation
integral is UV divergent. The main result (with exceptions that can be
enumerated) is that in all such cases the `naive' estimates, obtained by
using $\langle v^2 \rangle \sim \dot{L}^2$, are correct.

To make further progress we introduce the additional assumption, which can
be checked {\em a posteriori}, that the dissipation is dominated by the
motion of defect structures of `characteristic scale' $L(t)$. By the
`characteristic scale' we mean the typical radius of curvature for extended
defects ($n<d$), or the typical defect separation for point defects ($n=d$).
That is, we are assuming that the dissipation is dominated by the motion
of typical defect structures, and not by the disappearance of small domain
bubbles, small vortex loops, or by the annihilation of defect-antidefect
pairs. If the latter were true, the dissipation would be dominated by
structure at the core scale, and the arguments given below would fail.
We recall that for the case $d=2=\mu$ discussed above, the final
${\bf k}$-integral for the dissipation was convergent, implying that
the dominant $k$-values are order $1/L$, and the worries about the possible
importance of small-scale structure were ultimately groundless.
If the final ${\bf k}$-integral {\em were} UV divergent, {\em and}
the large $kL$ limit of
$\left< \partial_t \vec{\phi}_{\bf k} \cdot \partial_t
\vec{\phi}_{-{\bf k}} \right>$ had important contributions from
short scales, then the dissipation {\em would} be dominated by
structure at the core scale, violating our assumption.
Therefore, when our assumption holds, {\em either}
$\left< \partial_t \vec{\phi}_{\bf k} \cdot \partial_t
\vec{\phi}_{-{\bf k}} \right>$ is dominated by defect structures of
scale $L$, {\em or} the final integral is UV convergent, or both.

For the required cases where the final integral is UV divergent,
the large-$kL(t)$ limit of (\ref{EQN:TWOTIMESS}) can be extracted from the
physical/geometrical arguments used to obtain the generalised Porod
law (\ref{EQN:GENPOROD}). According to our assumption, we can treat
the defects as locally flat (or well separated, for point defects)
for $kL \gg 1$. From (\ref{EQN:TWOTIMESS}), we are interested in the
behaviour of the two-time structure factor, $S({\bf k},t,t') \equiv
\left< \vec{\phi}_{\bf k}(t) \cdot \vec{\phi}_{-{\bf k}}(t') \right>$, in
the limit that the two times are close together. In the limit $kL \gg 1$,
this will be proportional to the total density $L^{-n}$ of defect core.
Introducing $L = (L(t)+L(t'))/2$ and $\Delta = (L(t)-L(t'))/2$, we obtain
\begin{equation}
\left< \vec{\phi}_{\bf k}(t) \cdot \vec{\phi}_{-{\bf k}}(t') \right>
\sim \frac{1}{L^n\,k^{d+n}}\,a(k\Delta)\ ,\ \ \ \ \ kL \gg 1\ ,
\end{equation}
where consistency with Porod's law for $t=t'$ requires $a(0) = {\rm const}$.
Using this in (\ref{EQN:TWOTIMESS}) gives
\begin{equation}
\left< \partial_t \vec{\phi}_{\bf k} \cdot \partial_t
\vec{\phi}_{-{\bf k}} \right>  \sim \frac{\dot{L}^2}{L^n\,k^{d+n-2}}\ ,
\ \ \ \ \ kL \gg 1\ .
\label{EQN:RHSLARGEK}
\end{equation}
This reduces to (\ref{EQN:NAIVE}) for $n=1$
(with $\langle v^2 \rangle \sim \dot{L}^2$).
It should be stressed that we are {\em not} claiming that
(\ref{EQN:RHSLARGEK}) is a general result, only that it is valid
when we need it, i.e.\ when the dissipation integral
(\ref{EQN:LHSRHS}) requires a UV cut-off. There are three possibilities:
(i) The integral is UV convergent,
its dependence on $L(t)$ can be extracted by a change of variable, and the
large-$kL$ behaviour of $\left< \partial_t \vec{\phi}_{\bf k} \cdot
\partial_t \vec{\phi}_{-{\bf k}} \right>$ is not required.
(ii) The integral is UV divergent, but the dissipation is still
dominated by structures of scale $L(t)$. Then we can use
(\ref{EQN:RHSLARGEK}). (iii) The dissipation has significant contributions
from structures with local curvature (or spacing) of order the core scale.
Then one cannot treat the contributions from different defect core elements
as independent, (\ref{EQN:RHSLARGEK}) no longer holds, and the present
approach is not useful. For the moment we will proceed on the assumption
that (i) or (ii) obtain. We will show that these possibilities cover
nearly all cases. Examples of when (iii) holds will also be given. These
include the physically interesting case $d=n=2$.

\subsection{RESULTS}
\label{SEC:RESULTS}
Putting (\ref{EQN:RHSLARGEK}) into the dissipation integral (\ref{EQN:LHSRHS})
shows that  the integral is UV convergent for $kL \gg 1$ when $n+\mu>2$.
Otherwise the integral is dominated by $k$ near the upper cut-off $1/\xi$.
This gives
\begin{eqnarray}
\label{EQN:RHS}
\int_{\bf k} k^{-\mu}\,
\left< \partial_t \vec{\phi}_{\bf k} \cdot \partial_t \vec{\phi}_{\bf -k}
\right> & \sim & \dot{L}^2\,L^{-n}\,\xi^{n+\mu-2}\ ,\ \ \ \ n+\mu<2\ ,
\nonumber \\
& \sim & \dot{L}^2\,L^{-n}\,\ln (L/\xi)\ ,\ \ \ n+\mu=2\ , \nonumber \\
& \sim & \dot{L}^2\,L^{\mu-2}\ ,\ \ \ \ \ \ \ \ \ \ \ n+\mu>2\ .
\end{eqnarray}

The final step is to equate the dissipation rate (\ref{EQN:RHS}) to the
time derivative of the energy density (\ref{EQN:E}), as required by
(\ref{EQN:LHSRHS}), and solve for $L(t)$.
The results are summarized in Figure 24, as a function of $n$ and $\mu$,
for systems with purely short-ranged interactions. The two straight lines
separating regimes of different behaviour are the lines $n=2$ and $n+\mu=2$
at which the energy and dissipation integrals change their form.
Note that conservation of the order parameter
(which applies in a global sense for any $\mu >0$) is irrelevant to the
growth law for $\mu < 2-n$, where $n$ is treated here as a
continuous variable. At the marginal values, logarithmic factors are
introduced. The growth laws obtained are independent of the spatial
dimension $d$ of the system.

For non-conserved fields ($\mu=0$), we find $L \sim t^{1/2}$ for all
systems (with $d>n$ or $n>2$). Leading corrections in the $n=2$
case are interesting: the $\ln L$ factors in (\ref{EQN:E}) and
(\ref{EQN:RHS}) will in general have different effective cutoffs,
of order the core size $\xi$.
This leads to a logarithmic correction to scaling,
$L \sim t^{1/2}(1 + O(1/\ln t))$, and
may account for the smaller exponent ($\sim 0.45$) seen in
simulations of $O(2)$ systems  \cite{MG92,Toy91a,BBunpub}.
Note that for nonconserved scalar fields, the energy (\ref{EQN:E})
and the dissipation (\ref{EQN:RHS}) have the {\em same} dependence
on the core size $\xi$ (i.e.\ both contain a factor $\xi^{-1}$), so
this dependence cancels from $L(t)$. The fact that the correct $t^{1/2}$
growth is obtained from naive power counting on the linear terms in the
equation of motion should therefore be regarded as fortuitous. For example,
with long-range interactions, this `cancellation of errors' no longer
occurs, and naive power counting gives an
incorrect result for nonconserved scalar fields \cite{ABLR,BR}.

For conserved fields ($\mu>0$) our results agree with
an earlier RG analysis \cite{Bray89,Bray90}, with additional logarithmic
factors for the marginal cases $n=2$ and $n+\mu=2$.
Note that the conservation law is only relevant for $n+\mu \ge 2$.
Therefore for vector fields
($n \ge 2$), any $\mu>0$ is sufficient to change the
growth law, while for scalar fields ($n=1$) the conservation law
is irrelevant for $\mu<1$, in agreement with the RG analysis
\cite{Bray90} and earlier work of Onuki \cite{OnukiLR}.

Siegert and Rao \cite{SR} have performed extensive simulations for
$n=2$, $d=3$ and $\mu=2$. In their original paper they fitted
$L(t)$ to a simple power, and obtained a growth exponent slightly larger
than $1/4$. Recently, however, Siegert has shown that a very
much better fit is obtained when the predicted logarithmic correction
is included \cite{Siegert}.

\subsubsection{Exceptional Cases: $n=d \le 2$}
\label{SEC:EXCEPTIONS}
In what cases is our key assumption, that dissipation is dominated
by the motion of defect structures of characteristic scale $L$,
correct? Certainly for any $n>2$, the energy density (\ref{EQN:E})
itself, and hence dissipation, is dominated by variations at scale $L(t)$.
Therefore, we limit the discussion to the case $n \le 2$.

For $n \leq 2$, the energy density is proportional to the defect
core volume (with an extra factor $\ln(L/\xi)$ for $n=2$, see (\ref{EQN:E})),
but we will show that, in general, dissipation is still dominated by defect
structures with length scales of order $L$. To see this, we investigate
the contribution to the energy dissipation from small-scale structures
(e.g.\ small domains, vortex loops, or defect-antidefect pairs):
\begin{eqnarray}
d\epsilon/dt & = & \partial_t \int_\xi^\infty dl\,n(l,t)\,\epsilon(l)
\nonumber \\
& = & -\int_\xi^\infty dl\,\partial_l j(l,t)\,\epsilon(l)\nonumber \\
& = & j(\xi)\epsilon(\xi) + \int_\xi^\infty dl\,j(l)\,
\partial\epsilon/\partial l \ ,
\label{EQN:SHORTE}
\end{eqnarray}
where $n(l,t)$ is the number density of defect features of scale $l$,
$\epsilon(l) \sim l^{d-n}$ is the energy of a defect feature (with an
extra $\ln (l/\xi)$ factor for $n=2$), and $j(l,t)$ is the number flux
of defect features. We have used the continuity equation,
$\partial_t n + \partial_l j=0$ to obtain the second line of
(\ref{EQN:SHORTE}), and the $t$-dependence has been suppressed in the
final line. The total number of defect features, $N$, scales as
$N \sim L^{-d}$, and so $N$ does not change significantly over times
smaller than $\dot{L}/L$. Since defects only vanish at the core scale, we have
$\dot{N} = j(\xi)$. It follows $j(l)$ has a finite, non-zero,
short-distance limit of order $\dot{N} \sim -\dot{L}/L^{d+1}$. We can use
this to examine the convergence (at short-distance) of the final integral
in (\ref{EQN:SHORTE}).

For $d>n$, the integral in (\ref{EQN:SHORTE}) is well-behaved at small $l$,
because $\epsilon(l) \sim l^{d-n}$ ($\times \ln (l/\xi)$ for $n=2$), and
the integral dominates the $j(\xi)\epsilon(\xi)$ term. The integral can
be estimated by setting $j(l)\simeq j(\xi)$ and introducing a
large-distance cut-off at $l \sim L$. This gives $d\epsilon/dt \sim
j(\xi)L^{d-n} \sim -\dot{L}/L^{n+1}$ ($\times \ln(L/\xi)$ for $n=2$).
This is just what one gets from differentiating (\ref{EQN:E}) (for the
cases $n\le 2$ considered here), verifying the consistency of the
calculation.

For $d=n$ and $n<2$, however, $\epsilon(l) \sim {\rm constant}$, since the
(point) defects only interact weakly through the tails of the defect profile.
(The one physical example is the $d=1$ scalar system).
The leading contribution to the energy of a defect pair is just the core
energy of the individual defects, and dissipation is dominated by the
$j(\xi)\epsilon(\xi)$ term in (\ref{EQN:SHORTE}), which describes defect
pairs annihilating. Since the dissipation occurs at separations
$l \sim \xi \ll L$, the derivation of (\ref{EQN:RHSLARGEK}) no longer
holds. In fact since, the energy of a defect pair depends only weakly on
the separation for $l \gg L$, the system will be disordered, with an
equilibrium density of defects at any non-zero temperature. At $T=0$,
we expect slow growth that depends on the details of the potential
$V(\phi)$ \cite{BR1d}. These cases, including the $d=1$ scalar system,
are at their lower critical dimension, and are beyond the scope of the
simplified approach presented here (see \cite{BR1d} for a fuller
discussion).

The 2d planar system ($n=d=2$) is a special case. The logarithm in the
energy of a vortex pair, $\epsilon \sim \ln (l/\xi)$, leads to a
logarithmically divergent integral in (\ref{EQN:SHORTE}), i.e.\
vortex pairs with separations between $\xi$ and $L$ contribute
significantly to the energy dissipation. In this case
Eq.\ (\ref{EQN:RHSLARGEK}), which depends on the $kL \gg 1$ limit being
a single defect property, is again questionable. As a result the present
method cannot address this case. Indeed, the contributions to the
dissipation from all length scales suggest a possible breakdown of
scaling.

%

\subsubsection{Systems Without Defects}
\label{SEC:NODEFECTS}
Since systems without topological defects ($n > d+1$) will have
convergent momentum integrals for $kL \gg 1$,
we obtain $L \sim t^{1/(2+\mu)}$ for these cases.
We can also apply this result to  systems
with topological textures ($n=d+1$), even though the appropriate Porod's
law is not known.  Since defects with $n>d$ must be spatially
extended and without a core, they will have a smaller large-$k$
tail to their structure factor $S({\bf k},t)$ than
any defects with cores.  So for $n > 2$, when the energy
dissipation clearly occurs at length scales of order $L(t)$
(see (\ref{EQN:E})) and the momentum integrals for defects with cores
converge, our results should apply (\cite{BRunpub} contains a fuller
discussion of this point). Consequently the results in Figure 24
will apply for {\em any} system, apart from those systems explicitly
excluded above.

Of course, all this is subject to the caveat that
the two-time structure factor exhibits the scaling form
(\ref{EQN:TWOTIMESS}), on which the whole of this section is built.
One explicit counterexample is the $d=1$, $n=2$ system discussed in
section \ref{SEC:XY}, for which (\ref{EQN:TWOTIMESS}) explicitly fails.
As a result, the growth law characterising equal-time correlations is
{\em not} $L \sim t^{1/2}$, as suggested by Figure 24, but
$L \sim t^{1/4}$. In fact one use the two-time result (\ref{EQN:1dXY})
to calculate $\left< \partial_t \vec{\phi}_{\bf k} \cdot
\partial_t \vec{\phi}_{\bf -k} \right>$ explicitly for this system
\cite{Kay}, and show that it is consistent with $L \sim t^{1/4}$ growth.

\subsubsection{Other Systems}
\label{SEC:OTHERSYSTEMS}
The strength of this approach is that it can be applied to systems
with more complicated order parameters than $n$-component vectors,
provided they have purely dissipative dynamics. Then an equation of the
form (\ref{EQN:LHSRHS}) can be written down.
The details of the energy functional (\ref{EQN:HAMILTONIAN})
are unimportant \cite{Note1}.
The important ingredients are the existence of an `elastic energy',
associated with spatial gradients of the order parameter, the conservation
law (if any), characterised by $\mu$, and the defect structure if any.
The derivation is independent of the initial conditions, and so,
e.g., applies equally to critical and off-critical
quenches {\em as long as the system scales at late times}.
We simply choose a Porod's law (\ref{EQN:POROD}) to represent the
dominant defect type, which is the one responsible for the asymptotic
tails of the structure factor scaling function, i.e.\ the
one with the smallest `$n$'. When the energy density is dominated by
defects, i.e.\ when the energy integral (\ref{EQN:GRADSQUARE}) is UV
divergent, the relation (\ref{EQN:GRADSQUARE}) between the energy density
and the structure factor, shows that the `dominant' defects will
also be the ones which dominate the energy density. As examples, we
consider nematic liquid crystals and Potts models.

In bulk  nematic liquid crystals, the `dominant' defects (in the above
sense) are strings, giving a Porod tail of (\ref{EQN:POROD}) with $n=2$,
which with no conservation law implies $L \sim t^{1/2}$, consistent with
recent experiments \cite{WWY,WWLY} and simulations \cite{BB92}.

The $q$-state Potts model has $q$ equivalent equilibrium phases, which
give rise to $q(q-1)/2$ different types of domain wall. These can indexed
$\alpha\beta$, where $\alpha$, $\beta = 1,\ldots q$, are the phases
separated by the wall. Three domain walls of type $\alpha\beta$,
$\alpha\gamma$ and $\beta\gamma$ can meet at a point ($d=2$) or line
($d=3$), which represents a new type of defect. It is clear, however,
that the Porod tail and energy density are dominated by the walls, so that
the Potts model behaves as an $n=1$ system. As a result,
$L(t) \sim t^{1/2}$ and $t^{1/3}$ for nonconserved and conserved order
parameter respectively. Recent numerical results \cite{PottsSimsNC,PottsSimsC}
support these predictions, after initial suggestions that the growth was
slower. The $t^{1/3}$ growth for conserved systems is also predicted by
the Renormalisation Group approach of section \ref{SEC:RG}.

It should be emphasized that the classification of nematic liquid crystals
and Potts models as `$n=2$-like' and `$n=1$-like' respectively, pertain
only to the Porod tails and the growth laws. As far as scaling functions
(e.g.\ for pair correlations) are concerned, these systems belong to their
own universality classes. Similarly, for off-critical quenches of
conserved systems, the growth law is independent of the volume fractions
of the phases, but the scaling functions are not.

\section{RENORMALIZATION GROUP RESULTS}
\label{SEC:RG}
As with any other scaling phenomenon, it is tempting to try to apply
Renormalisation Group (RG) concepts to the late stages of phase ordering.
The basic idea is to associate the scaling behaviour with a fixed point
of the equation of motion under a RG procedure consisting of a
coarse-graining step combined with a simultaneous rescaling of length
and time. Such a procedure, if successful, would indeed provide a
first principles derivation of the scaling behaviour itself, which has,
up to now, been lacking (except for specific soluble models discussed
in section \ref{SEC:SOLUBLE}).

Underlying such an idea is the schematic RG flow for the temperature,
depicted in Figure 4. The critical point $T_c$ corresponds to a
fixed point of the RG transformation. At temperatures above (below)
$T_c$, coarse graining the system leads to a system which is more
disordered (ordered), corresponding to a system at a higher (lower)
temperature. This schematic flow is indicated by the arrows in
Figure 4. It follows from this that a quench from any $T>T_c$ to
any $T < T_c$ should give the same asymptotic scaling behaviour.
Any short-range correlations present at the initial temperature will
become irrelevant when $L(t) \gg \xi_0$, where $\xi_0$ is the
correlation length for the initial condition. A different universality
class is obtained, however, when the initial condition contains
sufficiently {\em long-range} (power-law) spatial correlations, e.g.\
following a quench from $T_c$. Such cases will be discussed below.
It follows from the previous section that the initial conditions
{\em cannot} affect the growth law (provided that they still
yield scaling behaviour).

According to Figure 4, asymptotic scaling is controlled by the
zero-temperature (or {\em strong coupling}) fixed point, justifying
the neglect of thermal noise in the equations of motion. We will see
below how this works out in practice. A classification of systems
according to the role of thermal noise has been given by Lai
et al.\cite{LMV}. In some systems, such as the Cahn-Hilliard (CH)
systems considered in this section, thermal noise can simply be neglected.
For kinetic Ising models (with conserved dynamics), where freezing
occurs for $T$ strictly zero, the temperature modifies the bare
transport coefficient $\lambda$, but in a scale-independent way that
does not change the growth law. In systems with quenched disorder,
however, there is a {\em scale-dependent} renormalisation of the kinetic
coefficients that leads to logarithmic growth with $T$-dependent
amplitudes. This case will be discussed in detail in section
\ref{SEC:RGRANDOM}.

The idea of using RG methods in this context is not new
(see for example \cite{LMV,MVZ}), but in
practical  applications the RG framework has been exploited mostly
in the numerical context via, for example, the Monte Carlo RG
\cite{MonteCarloRG}.

The difficulty with applying the RG to phase ordering is that,
due to the absence of a convenient small parameter, analogous to
$\epsilon = 4-d$ for critical phenomena, one cannot obtain explicit
RG recursion relations. However, one can still make some progress.
For conserved fields, a very simple and general result taken over
from critical phenomena can be used to determine growth exponents
exactly \cite{Bray89,Bray90}, without the need to construct
explicit RG recursion relations for the entire set of parameters
specifying the equation of motion. Without such explicit
recursion relations, of course, the very existence of a fixed point has
to be taken on trust. This is tantamount to assuming the validity of the
scaling hypothesis {\em ab initio}, and inferring the existence of an
underlying RG fixed point. This is the approach we will adopt.
It will take us surprisingly far.

\subsection{THE RG PROCEDURE}
\label{SEC:PROCEDURE}
\subsubsection{Equation of Motion}
We start by recalling the Cahn-Hilliard equation (\ref{EQN:MODELB}) for
a conserved order parameter, generalised to vector fields. Introducing
the transport coefficient $\lambda$ explicitly on the right-hand side
(we have previously absorbed $\lambda$ into the timescale) gives
\begin{equation}
\partial \vec{\phi}/\partial t
= \lambda\,\nabla^2(\delta F/\delta \vec{\phi})\ .
\end{equation}
Next we Fourier transform, and divide through by $\lambda k^2$.
For generality, and anticipating future requirements, we will write the
equation in the form
\begin{equation}
\left(\frac{1}{\lambda k^\mu} + \frac{1}{\Gamma}\right)\,
\frac{\partial\vec{\phi}_{\bf k}}{\partial t}
= -\frac{\delta F}{\delta \vec{\phi}_{-{\bf k}}} + \vec{\xi}_{\bf k}(t)\ .
\label{EQN:LANGEVIN}
\end{equation}
Here we have replaced $k^2$ by the more general $k^\mu$, as in section
\ref{SEC:GROWTH}, and included on the left-hand side a term
$(1/\Gamma)\partial\vec{\phi}_{\bf k}/\partial t$ appropriate to non-conserved
dynamics, which is recovered in the limit $\lambda \to \infty$. For any
finite $\lambda$, however, the order parameter is conserved by the
dynamics (\ref{EQN:LANGEVIN}). We include the extra term because it
will in any case be generated (along with many other terms) after one
step of the RG procedure.

A gaussian white noise term $\vec{\xi}_{\bf k}(t)$, representing thermal
noise, has also be included in (\ref{EQN:LANGEVIN}). We require that the
canonical distribution be recovered in equilibrium,
i.e.\ $P[\vec{\phi}] \propto \exp(-F[\vec{\phi}]/T)$. The usual
fluctuation-dissipation relation fixes the noise correlator,
\begin{equation}
\langle \vec{\xi}^i_{\bf k}(t_1) \vec{\xi}^j_{-{\bf k}}(t_2) \rangle
= 2T\,\delta_{ij}\,\delta(t_1-t_2)\,
\left(\frac{1}{\lambda k^\mu} + \frac{1}{\Gamma}\right)\ ,
\label{EQN:NOISE}
\end{equation}
where $i,j = 1,\ldots,n$ indicate Cartesian components in the internal
space. We have argued previously that thermal noise is irrelevant. The
RG approach shows this explicitly.

\subsubsection{The Coarse-Graining Step}
\label{SEC:COARSE}
One RG step consists of the following four stages:
(i) The Fourier components $\vec{\phi}_{\bf k}(t)$ for the `hard' modes
with $\Lambda/b < k < \Lambda$ are eliminated by solving (\ref{EQN:LANGEVIN})
for the time evolution of these modes, and substituting the solution into
the equation of motion for the remaining `soft' modes with $k < \Lambda/b$.
Here $\Lambda \sim 1/\xi$ is a UV cut-off, and $b$ is the RG rescaling
factor.

(ii) A scale change is made, via the change of variable $k=k'/b$, in order
to reinstate the UV cut-off for the soft modes to its original value
$\Lambda$. Additionally, time is rescaled via $t=b^z t'$. The requirement,
imposed by the scaling hypothesis, that the domain morphology is invariant
under this procedure, fixes $z$ as the reciprocal of the growth exponent,
i.e.\ $L(t) \sim t^{1/z}$. Finally, the field $\vec{\phi}_{\bf k}(t)$
for $k<\Lambda/b$ is rewritten as
\begin{equation}
\vec{\phi}_{\bf k}(t) = \vec{\phi}_{{\bf k}'/b}(b^zt')
                       = b^\zeta\,\vec{\phi}'_{{\bf k}'}(t')\ .
\label{EQN:RESCALE}
\end{equation}
The scaling form (\ref{EQN:STRUCT}) for the
structure factor becomes $S({\bf k},t) = t^{d/z}g(kt^{1/z})$. From the
definition of $S$ and equation (\ref{EQN:RESCALE}),
\begin{equation}
S({\bf k},t) = b^{2\zeta} \langle \vec{\phi}'_{{\bf k}'}(t')\cdot
\vec{\phi}'_{-{\bf k}'}(t') \rangle = b^{2\zeta}\,t'^{d/z}g(k't'^{1/z})
= b^{2\zeta -d}\,t^{d/z}\,g(kt^{1/z})\ ,
\label{EQN:zeta}
\end{equation}
from which we identify $\zeta = d/2$.

(iii) The new equation of motion for the soft modes is interpreted in terms
of a rescaled transport coefficient $\lambda'$ and free energy $F'$. In
addition, terms not originally present in (\ref{EQN:LANGEVIN}) will be
generated, and must be included in subsequent RG steps. Similarly, one
must allow for a more general structure for the thermal noise than
(\ref{EQN:NOISE}). Finally, the distribution
$P_0(\{\vec{\phi_{\bf k}(0)}\}$ of initial conditions will also be
modified by the coarse graining.

(iv) Scaling behaviour is associated with a fixed point in which both the
equation of motion and $P_0$ are invariant under the RG procedure. In
particular the fixed-point free energy is the that appropriate to the
`strong-coupling' fixed point, which is attractive for systems below $T_C$.
Note also that the fixed distribution $P_0$ contains the scaling
morphology.

\subsubsection{RG Recursion Relations}
Unfortunately, the above procedure cannot be carried out explicitly, due
to the absence of a small parameter \cite{Cardy}, and remains largely
a `gedanken RG'. Nevertheless, on the assumption that a fixed point
exists (equivalent to assuming scaling), the recursion relations for
the transport coefficient $\lambda$  and the temperature $T$ can be
written down exactly. This is sufficient to determine $z$ and to
test the stability of the {\em nonconserved} fixed point against
the conservation constraint. These results agree with those of
section \ref{SEC:GROWTH}. In addition, however, we can also identify
universality classes, clarify the role of the initial conditions in
determining the large-scale structure, and make strong predictions
about the effects of quenched disorder. All of these are beyond the
scope of the methods of section \ref{SEC:GROWTH}. In this sense, the
two approaches are complementary.

The observation that enables further progress is that the $1/\lambda k^\mu$
term in the equation of motion (\ref{EQN:LANGEVIN}) is {\em singular} at
$k=0$. Since no new large-distance singularities can be introduced by
the elimination of small length-scale degrees of freedom, it follows
that the coarse-graining step (i) of the RG does not contribute to the
renormalisation of $1/\lambda$ in (\ref{EQN:LANGEVIN}). (By contrast it
can, and does, contribute to the renormalisation of $1/\Gamma$
\cite{Bray90}, which is a non-singular term in (\ref{EQN:LANGEVIN})).
As a result, $1/\lambda$ is changed only by the rescaling step (ii).
Exactly the same argument at the critical point \cite{HH,HHM}
leads to the identity $z=4-\eta$ between the dynamic critical
exponent $z$ and the static critical exponent $\eta$. It is important
to recognize that the latter result is {\em non-perturbative}, and is
not restricted to the conventional Wilson-Fisher fixed point.

Since the strong-coupling fixed point is {\em attractive} (see Figure 4),
the free-energy functional scales {\em up} at this fixed point,
$F[\{\vec{\phi}_{{\bf k'}/b}\}] = b^y\,F[\{\vec{\phi}'_{{\bf k}'}\}]$,
where the exponent $y$ can be determined by elementary arguments.
Using this and (\ref{EQN:RESCALE}) in (\ref{EQN:LANGEVIN}) gives the
coarse-grained equation of motion
\begin{equation}
\left(b^{\zeta+\mu-z}\frac{1}{\lambda k'^\mu}
+ b^{\zeta-z}\frac{1}{\Gamma}\left[1+\cdots\right] + \cdots\right)
\frac{\partial\vec{\phi}'_{{\bf k}'}}{\partial t'} +\cdots
= -b^{y-\zeta}\,\frac{\delta F}{\delta \vec{\phi}'_{-{\bf k}'}} +
\vec{\xi}_{{\bf k}'/b}(b^zt')\ .
\end{equation}
where $\ldots$ indicates additional terms generated by the coarse
graining step. Dividing through by $b^{y-\zeta}$, to restore the right-hand
side to its previous form, gives
\begin{equation}
\left(b^{2\zeta+\mu-y-z}\frac{1}{\lambda k'^\mu}
+ b^{2\zeta-y-z}\frac{1}{\Gamma}\left[1+\cdots\right] + \cdots\right)
\frac{\partial\vec{\phi}'_{{\bf k}'}}{\partial t'} +\cdots
= -\frac{\delta F}{\delta \vec{\phi}'_{-{\bf k}'}} +
\vec{\xi}'_{{\bf k}'}(t')\ ,
\label{EQN:LANGEVIN'}
\end{equation}
where the new noise term is
\begin{equation}
\vec{\xi}'_{{\bf k}'}(t') = b^{\zeta-y}\vec{\xi}_{{\bf k}'/b}(b^zt')\ ,
\end{equation}
with correlator
\begin{equation}
\langle \vec{\xi}'^i_{{\bf k}'}(t'_1) \vec{\xi}'^j_{-{\bf k}'}(t'_2) \rangle
= b^{2\zeta-2y-z}\,2T\delta_{ij}\,\delta(t'_1-t'_2)\,
\left(b^\mu \frac{1}{\lambda k'^\mu}
+ \frac{1}{\Gamma}\left[1+\cdots\right]\right)\ .
\end{equation}

The absence of contributions to the $1/\lambda k'^\mu$ terms, either in
the equation of motion or the noise correlator, from the coarse graining
step, means that the recursion relations for $1/\lambda$ and $T$ can be
written down exactly:
\begin{eqnarray}
\label{EQN:RRlambda}
\left(\frac{1}{\lambda}\right)' & = &
b^{2\zeta+\mu-y-z}\,\left(\frac{1}{\lambda}\right) , \\
  T' & = & b^{-y}\,T\ .
\label{EQN:RRT}
\end{eqnarray}
The $T$-equation is just what one would expect. Since the free energy
functional scales {\em up} as $b^y$ at the strong-coupling fixed point,
rewriting the equation of motion, as in (\ref{EQN:LANGEVIN'}), in a form
in which the free energy functional is {\em unchanged} is equivalent to
scaling temperature {\em down} by a factor $b^{-y}$. At the same time,
the transport coefficient $\lambda$ renormalises as in (\ref{EQN:RRlambda}).
This last equation determines the growth exponent for all cases in which
the conservation constraint is {\em relevant}, in a sense to be clarified
below.

\subsection{FIXED POINTS AND EXPONENTS}
In the strong-coupling phase (i.e.\ for $T<T_c$), $T$ flows to zero
under repeated iteration of the RG procedure, implying $y>0$ in
(\ref{EQN:RRT}). If the dynamical fixed point controlling the late-stage
scaling regime is described by a non-zero value of $1/\lambda$ (i.e.\
when the conservation law is {\em relevant}), then (\ref{EQN:RRlambda})
implies
\begin{equation}
z = 2\zeta + \mu - y = d+\mu-y\ ,
\label{EQN:z}
\end{equation}
where we inserted the value $\zeta = d/2$ at the last step.
Eq.\ (\ref{EQN:z}) is exact, given the scaling assumption underlying the
RG treatment.

It is interesting to consider the same argument at the critical fixed
point. Then $T'=T=T_c$ implies $y=0$ and $z=2\zeta+\mu$. The structure
factor scaling relation reads, for this case,
$S({\bf k},t) = L^{2-\eta}g(kL) = t^{(2-\eta)/z}g(kt^{1/z})$, so the
analogue of Eq.\ (\ref{EQN:zeta}) fixes $\zeta = (2-\eta)/2$, and
$z=2+\mu-\eta$. For $\mu=2$, this is the familiar result $z=4-\eta$
for model B \cite{HH,HHM}, which we stress is an exact, nonperturbative
result. Eq.\ (\ref{EQN:z}) is just the generalisation of this result to
the strong-coupling fixed point.

To determine $y$ for the strong-coupling fixed point, we coarse-grain
the system on the scale $L(t)$. At this scale the system looks completely
disordered, so the excess energy per degree of freedom (i.e.\ per volume
$L(t)^d$) is of the order of the local excess energy density at that
scale, i.e.\ of order $L(t)^y$. The excess energy density on the
original scale therefore decreases as $\epsilon \sim L(t)^{y-d}$.
Comparing this with the result (\ref{EQN:E}) for $\epsilon$ obtained
in section \ref{SEC:GROWTH} gives
\begin{eqnarray}
y & = & d-n\ ,\ \ \ \ \ \ \ n \le 2\ , \nonumber \\
  & = & d-2\ ,\ \ \ \ \ \ \ n \ge 2\ .
\label{EQN:y}
\end{eqnarray}
Note, however, the extra logarithm in (\ref{EQN:E}) for $n=2$.

For the usual scalar ($n=1$) and vector ($n \ge 2$) fields, (\ref{EQN:y})
gives the usual results $y=d-1$ and $y=d-2$ respectively, familiar from
statics: $b^{d-1}$ is just the energy cost of a domain wall of linear
dimension $b$, while $b^{d-2}$ is the energy cost of imposing a slow
twist of the vector field over a region of size $b^d$. The extra
logarithm in (\ref{EQN:E}) for $n=2$ is due to the vortices, which dominate
over slow `spin-wave' variations for this case. As an amusing aside we note
that (\ref{EQN:y}) gives the `lower critical dimension' $d_l$, below which
long-range order is not possible for $T>0$, for the continuation
of the theory to real $n$. Since the existence of an ordered phase
requires $y >0$, we have $d_l=n$ for $n \le 2$ and $d_l=2$ for $n \ge 2$.
The result for $n<2$ recovers the known results for $n=1$
(the scalar theory) and $n=0$ (the self-avoiding walk).

Inserting (\ref{EQN:y}) into (\ref{EQN:z}) gives the final result for $z$:
\begin{eqnarray}
z & = & n + \mu\ ,\ \ \ \ \ \ \ \ n \le 2\ , \nonumber \\
  & = & 2 + \mu\ ,\ \ \ \ \ \ \ \ n \ge 2\ .
\label{EQN:zO(n)}
\end{eqnarray}
These results agree with those derived in section \ref{SEC:GROWTH},
which are summarised in Figure 24. For scalar model B ($n=1$, $\mu=2$)
we recover the usual $t^{1/3}$ Lifshitz-Slyozov growth, while for
vector model B  with $n>2$ we obtain $t^{1/4}$ growth. At the crossover
value $n=2$, there is an extra logarithm that the RG method does not
see (since it determines only the growth {\em exponent}), but which
{\em is} captured by our previous approach (section \ref{SEC:GROWTH}).

Comparison of (\ref{EQN:z}) with Figure 24 shows that (\ref{EQN:z}) is
not valid below the line $n+\mu=2$. How do we see this within the RG
approach? Recall that to derive (\ref{EQN:z}) we have to assume
that $(1/\lambda)$ is {\em non-zero} at the fixed point, i.e.\ that the
conservation constraint is {\em relevant}, in the RG sense. Consider now
the fixed point of the {\em non-conserved} system, with $\lambda = \infty$
in the equation of motion (\ref{EQN:LANGEVIN}). Let the corresponding
value of $z$ be $z_{nc}$. Now introduce the conservation law through an
infinitesimal $1/\lambda$. The recursion relation (\ref{EQN:RRlambda}) then
gives
\begin{equation}
\left(\frac{1}{\lambda}\right)' =
b^{z_c - z_{nc}}\,\left(\frac{1}{\lambda}\right)\ ,
\label{EQN:RRlambda1}
\end{equation}
where $z_c = 2\zeta+\mu-y$ is the value of $z$ (Eq.\ (\ref{EQN:z})) at
the conserved fixed point. Eq.\ (\ref{EQN:RRlambda1}) shows immediately
that $(1/\lambda)$ iterates to zero for $z_c<z_{nc}$, i.e.\ {\em the
conservation law is irrelevant when} $z_c<z_{nc}$.
Since $L(t) \sim t^{1/z}$, this means that the conserved system cannot
exhibit {\em faster} growth than the nonconserved system. This is
intuitively reasonable: an additional constraint cannot speed up the
dynamics. There is an interchange of stability of RG fixed points when
$z_c=z_{nc}$, the fixed point with the larger $z$ being the stable one.
It follows that $z={\rm max}\,(z_c,z_{nc})$, with $z_c$ given by
(\ref{EQN:z}) in general and by (\ref{EQN:zO(n)}) for the $O(n)$ model.
Since $z=2$ for the non-conserved $O(n)$ model (see \ref{SEC:GROWTH}),
this interchange of stability accounts for the crossover line
$n+\mu=2$ in Figure 24.
(We note that the same reasoning implies a similar interchange of
stability, and the result $z = {\rm max}\,(z_c,z_{nc})$, at the
{\em critical} fixed point \cite{BrayRG92}).

The generality of Eq.\ (\ref{EQN:z}) deserves emphasis. For any system
with purely dissipative conserved dynamics, one only needs to insert
the value of $y$, which can be determined from the energetics as in
the derivation of (\ref{EQN:y}) for the $O(n)$ model. As an example,
consider again the $q$-state Potts model. The energy density is
dominated by domain walls, so the energy density scales as
$\epsilon \sim 1/L \sim L^{y-d}$, giving $y=d-1$ just as for the Ising
model. Therefore, the usual Lifshitz-Slyozov $t^{1/3}$ growth is obtained
for $\mu=2$. Of course, we already obtained this result in section
\ref{SEC:GROWTH}. Recent numerical studies \cite{PottsSimsC} confirm this
prediction.

As a second example we note that, in agreement with our
findings in section \ref{SEC:GROWTH}, the growth law in independent
of the nature of the initial conditions (which played no role in the
derivation), provided scaling is satisfied. A case of experimental
interest is a conserved scalar field -- the $t^{1/3}$ Lifshitz-Slyozov
growth is obtained for all volume fractions of the two phases.
By contrast, the scaling functions {\em can} depend on the form of
the initial conditions. This should not be too surprising, since the
fixed point distribution for the initial conditions contains the scaling
morphology. This will be discussed in detail in section \ref{SEC:IC}.

\subsection{UNIVERSALITY CLASSES}
The present RG approach cannot, unfortunately, determine $z_{nc}$, since
it rests on $(1/\lambda)$ being non-zero at the fixed point. Neither can
it explicitly pick up the logarithms on the boundary lines $n=2$ and
$n+\mu=2$ of Figure 24. So does it have any advantages over the seemingly
more powerful energy scaling approach of section \ref{SEC:GROWTH}? The
answer is an unequivocal yes. The reason is that the RG identifies
{\em universality classes} as well as exponents. As an example, consider
a scalar $n=1$ system with $\mu<1$. The energy scaling method tells us
that $L(t) \sim t^{1/2}$, as for nonconserved dynamics, but tells us nothing
about correlation functions. The RG, by contrast, tells us that when the
conservation is irrelevant not only the exponents but also all correlation
scaling functions are the same as those of the nonconserved system, i.e.\
for $\mu<1$ the scalar system is in the {\em nonconserved universality
class}.

At first this result seems paradoxical: in the scaling form
for the structure factor, $S({\bf k},t) = L^d g(kL)$, the scaling
function $g(x)$ has a non-zero value at $x=0$ for nonconserved dynamics,
whereas for conserved dynamics $g(x)$ must vanish at $x=0$.
So how can a system with conserved dynamics be in the
nonconserved universality class? To understand this
one needs to remember that the scaling limit is defined by $k \to 0$,
$L \to \infty$, with $kL$ fixed but arbitrary. Onuki has argued that,
for $\mu<1$, $S({\bf k},t) \sim k^{2\mu} L^{d+2}$ for $k \to 0$ \cite{OnukiLR}.
If we imagine plotting $S({\bf k},t)$ in scaling form, i.e.\
$g(x) = L^{-d}S({\bf k},t)$ against $x=kL$, then Onuki's small-$k$ form
gives a $g(x)$ of order unity when $x \sim L^{1-1/\mu}$, which is
vanishingly small as $L \to \infty$ for $\mu<1$. In other words,
on a scaling plot the region of $kL$ where the nonconserved scaling
function is inaccurate shrinks to zero as $L \to \infty$.

\subsubsection{The Role of the Initial Conditions}
\label{SEC:IC}
To what extent do the scaling functions depend on the probability
distribution for the initial conditions? The RG answers this question
\cite{IC}. New universality classes are obtained when sufficiently
long-ranged (power-law) spatial correlations are present immediately
after the quench. These could either arise `physically', as in a quench
from $T_c$, or be put in `by hand' as initial conditions on the $T=0$
dynamics.


Consider initial conditions with a gaussian probability distribution
of variance
\begin{equation}
\langle \phi^i_{\bf k}(0)\phi^j_{-{\bf k}}(0) \rangle =
\Delta(k)\,\delta_{ij}\ ,
\label{EQN:LRIC}
\end{equation}
We recall the definitions, introduced in section \ref{SEC:TWOTIMES},
of the {\em response to} and {\em correlation with} the initial condition:
\begin{eqnarray}
\label{EQN:RESPONSE1}
G({\bf k},t) & = &
\left< \frac{\partial\phi^i_{\bf k}(t)}{\partial\phi^i_{\bf k}(0)}
\right> \\
C({\bf k},t) & = & \langle \phi^i_{\bf k}(t)\,\phi^i_{-{\bf k}}(0) \rangle
\end{eqnarray}
respectively, where $C({\bf k},t)$ is a shorthand for the two-time
structure factor $S({\bf k},t,0)$. The gaussian property of
$\{\phi^i_{\bf k}(0)\}$ means that these two functions are related by
\begin{equation}
C({\bf k},t) = \Delta(k)\,G({\bf k},t)\ ,
\label{EQN:LRRELATION}
\end{equation}
a trivial generalisation of (\ref{EQN:TWOTIMERELATION}) that
can be proved easily using integration by parts.

The RG treatment proceeds as in section \ref{SEC:COARSE}. The only
additional feature is that the scaling form (\ref{EQN:GSCALING}) for
$G({\bf k},t)$ implies that the initial condition $\vec{\phi}_{\bf k}(0)$
acquires an anomalous scaling dimension related to the exponent $\lambda$.
(The exponent $\lambda$ should not be confused with the transport coefficient:
the meaning should be clear from the context).
Therefore we write, analogous to Eq.\ (\ref{EQN:RESCALE}) for the
rescaling of the field at late times,
\begin{equation}
\vec{\phi}_{\bf k}(0) = \vec{\phi}_{{\bf k}'/b}(0)
                       = b^\chi\,\vec{\phi}'_{{\bf k}'}(0)\
\label{EQN:ICRESCALE}
\end{equation}
for the rescaling of the initial condition. This gives, analogous to
(\ref{EQN:zeta}),
\begin{equation}
G({\bf k},t) = b^{\zeta-\chi}\,
\left< \frac{\partial\phi^i_{{\bf k}'}(t')}{\partial\phi^i_{{\bf k}'}(0)}
\right> = b^{\zeta-\chi}\,t'^{\lambda/z}\,g_R(k't'^{1/z})
= b^{\zeta -\chi-\lambda}\,t^{\lambda/z}\,g_R(kt^{1/z})\ ,
\label{EQN:chi}
\end{equation}
from which we identify $\chi=\zeta-\lambda$. The scaling of the equal time
structure factor gives, as before, $\zeta=d/2$, so $\chi=d/2-\lambda$.

Under the RG transformation, the correlator $\Delta(k)$ of the initial
condition becomes $b^{2\chi}\langle \phi'^i_{{\bf k}'}(0)
\phi'^i_{-{\bf k}'}(0)\rangle = \Delta(k'/b) + \cdots $, where the dots
indicate the contribution from the coarse-graining step of the RG. So
the new correlator, $\Delta'(k') = \langle \phi'^i_{{\bf k}'}(0)
\phi'^i_{-{\bf k}'}(0)\rangle$, is given by
\begin{equation}
\Delta'(k') = b^{2\lambda-d}\,[\Delta(k'/b) + \cdots]\ .
\label{EQN:RRDelta}
\end{equation}

Consider now the case where $\Delta(k)$ has a piece corresponding to
long-range (power-law) correlations:
\begin{equation}
\Delta(k) = \Delta_{SR} + \Delta_{LR}\,k^{-\sigma}\ ,
\end{equation}
with $0<\sigma<d$. Then the real-space correlations decay as
$r^{-(d-\sigma)}$. From (\ref{EQN:RRDelta}) we can deduce the recursion
relations for the short- and long-range parts of the correlator:
\begin{eqnarray}
\label{EQN:RRDeltaSR}
\Delta'_{SR} & = & b^{2\lambda-d}\,[\Delta_{SR} + \cdots]\ , \\
\Delta'_{LR} & = & b^{2\lambda-d+\sigma}\,\Delta_{LR}\ .
\label{EQN:RRDeltaLR}
\end{eqnarray}
Note that the long-range part $\Delta_{LR}$, being the coefficient of
a singular (as $k \to 0$) term, picks up no contributions from
coarse-graining -- Eq.\ (\ref{EQN:RRDeltaLR}) is exact.

At the fixed point, both the equation of motion and the initial condition
distribution must be invariant under the RG transformation. It follows
from (\ref{EQN:RRDeltaLR}) that the long-range correlations are
irrelevant at the `short-range fixed point' (i.e.\ $\Delta_{LR}$
iterates to zero) if $\sigma < \sigma_c$, where
\begin{equation}
\sigma_c = d-2\lambda_{SR}\ ,
\end{equation}
and $\lambda_{SR}$ is the value of $\lambda$ for purely short-range
correlations.
When $\sigma>\sigma_c$, the invariance of $\Delta_{LR}$ at the
`long-range fixed point' fixes $\lambda=\lambda_{LR} = (d-\sigma)/2$, an
exact result. Thus there is an exchange of stability of fixed points
when $\sigma = \sigma_c$. The determination of $\lambda_{SR}$
itself is non-trivial, since it requires explicit computation of the
terms represented by the dots in (\ref{EQN:RRDeltaSR}).

For $\sigma > \sigma_c$, the scaling behaviour belongs to a new
universality class, in which the growth exponent is unchanged but the
scaling functions, e.g.\ $g_R(x)$, depend explicitly on $\sigma$.
Note that the function $C({\bf k},t)$, the correlation with the
initial condition, depends on $\sigma$ for {\em any} $\sigma >0$
through the $k^{-\sigma}$ term in the prefactor $\Delta(k)$ in
(\ref{EQN:LRRELATION}). Thus, in the scaling region,
\begin{equation}
C({\bf k},t) = \Delta_{LR}k^{-\sigma}L^\lambda g_R(kL)\ .
\end{equation}
Summing this over $k$ gives the autocorrelation function:
\begin{equation}
A(t) \equiv \langle\vec{\phi}({\bf r},t)\cdot\vec{\phi}({\bf r},0)\rangle
\sim L^{-(d-\sigma-\lambda)}\ .
\end{equation}
In the `long-range' regime, where $\lambda = (d-\sigma)/2$, this gives
$A(t) \sim L^{-(d-\sigma)/2}$. Consider, as an example, the $2d$ Ising model
quenched from the equilibrium state at $T_c$. Then $\sigma = 2-\eta =7/4$.
Measurements of $\lambda$ for the same model quenched from $T=\infty$,
with nonconserved dynamics,
give $\lambda_{SR} \simeq 0.75$ \cite{FH88,BHlambda}, as do experiments
on twisted nematic liquid crystals, which are in the same universality
class \cite{Mason93}. Therefore $\sigma > d-2\lambda_{SR}$ and this
system is in the `long-range' universality class. It follows that
$A(t) \sim L^{-1/8} \sim t^{-1/16}$, which has been confirmed by
numerical simulations \cite{IC,BHLR}.

The scaling function $g(x)$ for the equal-time structure-factor also has
a different form in the long-range regime. For {\em any} $\sigma>0$,
$S({\bf k},t)$ has a `long-range' contribution $S_{LR}({\bf k},t)$ varying
as $k^{-\sigma}$ at small $k$. It is given by the diagram of Figure 25,
where the circle represents $\Delta(k)$ and the lines are exact response
functions. Thus
$S_{LR}({\bf k},t) = \Delta_{LR}k^{-\sigma}G({\bf k},t)G(-{\bf k},t)$.
Using the scaling form (\ref{EQN:GSCALING}) for $G$ gives
$S_{LR}({\bf k},t) = \Delta_{LR}k^{-\sigma}L^{2\lambda}[g_R(kL)]^2$.
Comparing this to the general scaling form $S({\bf k},t)=L^d g(kL)$,
we see that when $\sigma<\sigma_c$ (i.e.\ in the `short-range'
regime), $S_{LR}$ is negligible in the scaling limit ($k \to 0$,
$L \to \infty$, with $kL$ fixed), and so does not contribute to the
scaling function. In fact, since the long-range correlations are
irrelevant in this case, the scaling function is identical to that for
purely short-range correlations.

For $\sigma>\sigma_c$, the contribution $S_{LR}$ survives in the scaling
limit, and the full scaling function is long-ranged, $g(x) \sim x^{-\sigma}$
for $x \to 0$. In real space, this means that the equal-time correlation
function decays with the same power-law as the initial-condition correlator,
i.e.\ $C({\bf r},t) \sim (L/r)^{d-\sigma}$ for $r \gg L$. For the $2d$
Ising model quenched from $T_c$, the predicted $(L/r)^{1/4}$ decay has
been seen in simulations \cite{BHLR}.

\subsubsection{Systems with Quenched Disorder}
\label{SEC:RGRANDOM}
The influence of quenched disorder on the motion of interfaces and other
defects is of considerable current interest in a variety of contexts.
The new ingredient when quenched disorder is present is that the defects
can become  pinned in energetically favourable configurations. At $T=0$
this leads to a complete cessation of growth. For $T>0$, thermal fluctuations
can release the pins, but in general growth is much slower than in `pure'
systems, typically logarithmic in time.

To see how logarithms arise, consider a single domain wall in a system
with quenched random bonds. The typical transverse displacement of the
wall over a length $l$, due to disorder roughening, is of order
$l^\zeta$, while the typical fluctuation of the wall energy around its
mean value is of order $l^\chi$. These exponents are related by the
scaling law \cite{HuseHenley} $\chi=2\zeta + d-3$, which can be obtained
by estimating the elastic energy of the deformed wall as
$l^{d-1}(l^\zeta/l)^2$, and noting that the pinning and elastic energies
should be comparable. The barrier to domain motion can be estimated
by arguing \cite{HuseHenley,Villain,Bruinsma,Nattermann} that the walls move
in sections of length $l$, where $l$ is the length scale at which the
walls `notice' their curvature, i.e.\ the disorder roughening $l^\zeta$
should be comparable to the distortion, of order $l^2/L(t)$, due to the
curvature of walls with typical radius of curvature $L(t)$. This gives
$l \sim L^{1/(2-\zeta)}$ and an activation barrier of order
$l^{\chi} \sim L^{\chi/(2-\zeta)}$ (assuming that the energy {\em barriers}
scale in the same way as the energy {\em fluctuations} between local
equilibrium positions of the wall). Equating this barrier to $T$ gives
a growth law
\begin{equation}
L(t) \sim (T \ln t)^{(2-\zeta)/\chi}\ .
\label{EQN:RBGROWTH}
\end{equation}
For $d=2$, the exponents $\zeta$ and $\chi$ are exactly known \cite{HHF}:
$\zeta = 2/3$ and $\chi=1/3$, giving $L(t) \sim (T \ln t)^4$. A number
of attempts to measure $L(t)$ in computer simulations have been
made \cite{RBSims,BHRB,PuriRB}, but it is difficult to obtain a large
enough range of $(\ln t)^4$ for a convincing test of the theoretical
prediction. Recent experimental studies of the two-dimensional
random-exchange Ising ferromagnet $Rb_2Cu_{0.89}Co_{0.11}F_4$, however,
suggest $L(t) \sim (\ln t)^{1/\psi}$ with $\psi=0.20 \pm 0.05$
\cite{Schins}, consistent with the theoretical prediction $\psi=1/4$.

Perhaps of greater interest than the growth law itself is the
universality class for the {\em scaling functions}. It can be argued
\cite{FH88} that, since $L \gg l \sim L^{1/(2-\zeta)}$ for
$L \to \infty$ (note that $\zeta<1$ for a system above its lower
critical dimension, otherwise disorder-induced roughening would
destroy the long-range order), on length scales of order $L$ the
driving force for domain growth is still the interface curvature:
the pinning at smaller scales serves merely to provide the
(scale-dependent) renormalization of the kinetic coefficient responsible
for the logarithmic growth. This leads to the conclusion \cite{FH88}
that the scaling functions should be identical to those of the pure
system, a prediction that is supported by numerical studies
\cite{BHRB,PuriRB}. The same prediction can be made for systems with
random-field (i.e.\ local symmetry-breaking) disorder \cite{FH88}, and
is supported by recent simulations \cite{ChakrabartiRFIM}.

It is interesting that the argument leading to (\ref{EQN:RBGROWTH})
makes no reference to whether the order parameter is conserved or not.
The time taken to surmount the pinning barriers dominates all other
timescales in the problem. The argument outlined above suggests a
scale-dependent kinetic coefficient
$\Gamma(L) \sim \exp(-L^{\chi/(2-\eta)}/T)$. Putting this into
the usual nonconserved growth law $L \sim [\Gamma(L)t]^{1/2}$ gives
$L \sim [T \ln(t/L^2)]^{\chi/(2-\zeta)}$,  which reduces to
(\ref{EQN:RBGROWTH}) asymptotically, since $\ln L \ll \ln t$ for
$t \to \infty$. For conserved dynamics, the same argument just gives
$t/L^3$ instead of $t/L^2$ inside the logarithm, and (\ref{EQN:RBGROWTH})
is again recovered asymptotically.

While this physically based argument is certainly plausible, the RG
makes a more powerful prediction: not only are the growth laws the
same for conserved and nonconserved dynamics, but they belong to the
{\em same universality class}! This means, {\em inter alia}, that
they have the {\em same scaling functions}! To see this, we simply note
that since the {\em fluctuations} in the free energy,
$\delta F \sim L^\chi$, are asymptotically negligible compared to the
{\em mean}, $\langle F \rangle \sim L^{d-1}$ (provided the system
supports an ordered phase at infinitesimal $T$), the strong-coupling
exponent $y$ is given by the same expression, $y=d-1$, as in the
pure system. (Alternatively, and equivalently, the extra length of
domain wall due to disorder roughening of the interfaces in a volume
$L^d$ scales as $\sim L^{d-3+2\zeta} \ll L^{d-1}$). It follows from
(\ref{EQN:z}) that, {\em provided the conservation law is relevant}, the
growth law is $L(t) \sim t^{1/3}$ (for $\mu=2$) as in the pure system.
Since, however, (\ref{EQN:RBGROWTH}) shows that $L(t)$ grows more
slowly than $t^{1/3}$ for the nonconserved system, our previous
arguments show that the conservation law is {\em irrelevant} for
systems with quenched disorder. Therefore conserved and nonconserved
systems are in the same universality class.

Numerical simulations \cite{PuriRBC,HayaRBC} allow us in principal to test
this prediction. They certainly show logarithmic growth, but with an
insufficient range of $L$ for a definitive test of (\ref{EQN:RBGROWTH}).
The most striking conclusion of the RG is that the scaling
functions are those of the nonconserved system. For example, a scaling plot
for the structure factor, i.e.\ a plot of $L^dS(k,t)$ against $kL$ should
give a non-zero intercept at $kL = 0$. For any fixed $L$, of course,
the conservation law requires that $S$ vanish at $k=0$, but in the scaling
limit ($k\to 0$, $L \to \infty$, with $kL$ fixed) the region of small $k$
where the conservation law is effective should shrink to zero faster than
$1/L$ as $L \to \infty$. There are indeed indications of this in the
small-$k$ data of Iwai and Hayakawa \cite{HayaRBC}, but the range of $L$
explored in not large enough to reach the true scaling limit. Indeed,
this will always be difficult with growth as slow as (\ref{EQN:RBGROWTH}).

\subsection{THE RG FOR BINARY LIQUIDS}
\label{SEC:RGBINLIQS}
As a final application of the RG approach, we return to phase separation
in binary liquids. The new element here is that the temperature $T$,
though formally irrelevant, can enter scaling functions in cases
where the minority phase consists of disconnected droplets, when
the nominally dominant linear growth, due to hydrodynamic flow,
is absent.

The analysis parallels that of model B, but including the extra
hydrodynamic term of Eq.\ (\ref{EQN:BINLIQS}). In order to discuss the
role of temperature, we have to include the thermal noise explicitly.
The presence of the hydrodynamic term in (\ref{EQN:BINLIQS}) implies,
via the fluctuation-dissipation theorem, that the noise correlator takes
the form \cite{KO}
\begin{eqnarray}
\langle \xi({\bf r},t) \xi({\bf r}',t') \rangle
& = & - 2\lambda T\,\nabla^2 \delta ({\bf r}-{\bf r}')\,\delta (t-t')
\nonumber \\
&   & + 2T\,\nabla\phi({\bf r})\cdot T({\bf r}-{\bf r}')\cdot
        \nabla'\phi({\bf r}')\,\delta (t-t')\ .
\label{EQN:BLNOISE}
\end{eqnarray}
Carrying out the RG step as before, the recurrence relations for $\lambda$
and the viscosity $\eta$ (implicit in the Oseen tensor (\ref{EQN:OSEEN}))
become
\begin{eqnarray}
\label{EQN:BLlambda}
\left(\frac{1}{\lambda}\right)' & =  &
b^{3-z}\,\left(\frac{1}{\lambda}\right) \\
\left(\frac{1}{\eta}\right)' & = &
b^{z-1}\,\left(\frac{1}{\eta}\right)\ .
\label{EQN:BLeta}
\end{eqnarray}
Eq.\ (\ref{EQN:BLlambda}) is the same as (\ref{EQN:RRlambda}) with
$\zeta=d/2$ and $y=d-1$ inserted explicitly. The renormalisation
of the noise again gives (\ref{EQN:RRT}), which we display again
for convenience (with $y=d-1$):
\begin{equation}
T' = b^{1-d}\,T\ .
\label{EQN:BLT}
\end{equation}

It is clear that the conventional conserved fixed point,
$\lambda = \lambda^*$, which has $z=3$, is unstable against the
introduction of hydrodynamics, since $1/\eta$ (which measures the strength
of the hydrodynamic interaction) scales up as $b^2$ at this fixed point.
Rather, the `hydrodynamic fixed point', with $\eta=\eta^*$, must have
$z=1$, recovering the dimension-analytic result obtained in section
\ref{SEC:BINLIQS}. At this fixed point $\lambda$ scales to zero, i.e.\
bulk diffusion of the order parameter is irrelevant at the largest scales.
Temperature is also irrelevant, as expected.

The physical arguments of section \ref{SEC:BINLIQS} \cite{Siggia,KO},
however, show that the linear growth (i.e.\ $z=1$), which is a
consequence of hydrodynamic flow along interfaces, is possible only
when the minority phase is continuous. What happens if the minority
phase consists of isolated droplets?  Then $z>1$, and the relevant
fixed point must be $\eta^* = 0$. Let's reconsider the usual conserved
(`model B') fixed point in this light. This fixed point, with
$\lambda = \lambda^*$ non-zero and finite, has $z=3$, i.e.\ $t^{1/3}$
growth. At this fixed point the recursion relations for $\eta$ and $T$
are $\eta' = b^{-2} \eta$ and $T'=b^{1-d}T$. Therefore, $\eta$ and $T$
both flow to zero, but their {\em ratio} remains fixed (for $d=3$).
Note that the ratio $T/\eta$ is exactly what appears in the hydrodynamic
part of the noise correlator (\ref{EQN:BLNOISE}). This means that, while
the temperature is technically irrelevant, the hydrodynamic part of the
noise cannot be discarded -- in fact it is just this part which drives
the brownian motion of the droplets that is responsible for coarsening
by droplet coalescence. The ratio $T/\eta$ is a {\em marginal} variable,
so in principle we expect scaling functions to depend on it, reflecting
the relative importance of evaporation-condensation and droplet coalescence
to the coarsening (see, however, the discussion below).

To be more precise, we can use dimensional arguments to construct the
important variables. The effect of thermal fluctuations on scales smaller
than the correlation length $\xi$ can be incorporated through the surface
tension $\sigma$ (which scales $F$) and the equilibrium order parameter
$M$ (which scales $\phi$). The length-scale associated with the
Lifshitz-Slyozov mechanism is then \cite{Huse86,Bray90}
$L(t) = (\lambda\sigma t/M^2)^{1/3}$, while the dimensionless marginal
variable is $k_BTM^2/\sigma\lambda\eta$. The general form for $L(t)$ is
therefore $L(t) = (\lambda\sigma t/M^2)^{1/3}f(k_BTM^2/\sigma\lambda\eta)$,
where $f(x)$ is a crossover function with $f(0)={\rm constant}$. For
large $x$, one must have $f(x) \sim x^{1/3}$ so that $L(t)$ is
independent of $\lambda$, giving $L(t) \sim (k_BTt/\eta)^{1/3}$ in this
regime, in agreement with the brownian motion argument of section
\ref{SEC:BINLIQS}. Note that the function $f$ also depends implicitly on
the volume fraction $v$ of the minority phase.

This marginal behaviour is specific to $d=3$. For general $d$,
(\ref{EQN:BLeta}) and (\ref{EQN:BLT}) give $(T/\eta)' = b^{z-d}\,(T/\eta)$.
So for $d>3$ the ratio $T/\eta$ is irrelevant at the Lifshitz-Slyozov (LS)
fixed point, and the evaporation-condensation mechanism dominates
asymptotically for all $T < T_c$, giving unique scaling functions for a
given volume fraction. For $d <3$, on the other hand, $T/\eta$ is
{\em relevant} at the LS fixed point. The dynamics is therefore controlled
by a `coalescence fixed point', with $T/\eta$ fixed, implying $z=d$,
and $\lambda$ is an irrelevant variable. This agrees with the $d=2$
result of San Miguel et al.\ \cite{SanMiguel}. To summarise, for $d>3$ the
LS mechanism dominates and $L \sim t^{1/3}$, for $d<3$ coalescence dominates
and $z=d$, i.e.\  $L \sim t^{1/d}$, while for the physically relevant case
$d=3$ both mechanisms operate and marginal behaviour is expected such
that, even for a given volume fraction, scaling functions will depend
continuously on the dimensionless ratio $k_BTM^2/\sigma\lambda\eta$.

It is important to note, however, that in the above discussion the
transport coefficient $\lambda$ in the equation of motion (\ref{EQN:BINLIQS}),
and the viscosity $\eta$ appearing in the Oseen tensor (\ref{EQN:OSEEN}),
have been treated as {\em independent variables}. While they can
certainly be treated as independent in numerical simulations, in real
binary liquids they are related \cite{Siggia}. Linearising (\ref{EQN:BINLIQS})
around one of the bulk phases gives equation (\ref{EQN:BULKLINEAR})
as in model B dynamics (the hydrodynamic term in (\ref{EQN:BINLIQS})
drops out at linear order). Inserting the transport coefficient on the
right-hand side (it was absorbed into the timescale in (\ref{EQN:BULKLINEAR}))
gives a bare diffusion constant $D_0 = \lambda V''(1) \simeq \lambda/\xi^2$,
where $\xi$ is the interface thickness. The diffusion constant for a drop
of size $L$ is $D(L) \sim D_0\xi/L \sim \mu k_BT \sim k_BT/\eta L$,
using the Einstein relation, and the usual relation $\mu(L) \sim 1/\eta L$
for the mobility. This gives $\lambda\eta \sim k_BT\xi$.  Using also
$\sigma \sim M^2/\xi$, which follows from (\ref{EQN:SIGMA}), for the
surface tension gives the crossover variable $x$ as
$k_BTM^2/\sigma\lambda\eta \simeq 1$. A more careful calculation
\cite{Siggia} shows that the evaporation-condensation and droplet
coalescence  mechanisms both lead to a mean droplet size $R(t)$ given
by $R^3 = k(k_BT/5\pi\eta)t$, with $k=0.053$ for the LS mechanism
and $k=12v$ for the droplet coalescence mechanism.

Effectively two-dimensional binary liquid systems can be achieved using
the Hele-Shaw geometry, where the fluid is confined between parallel
plates. The no-slip boundary conditions mean that the Navier-Stokes equation
(\ref{EQN:NS}) simplifies, and the results are different from those obtained
by simply putting $d=2$ in the previous paragraph, which correspond to
using `free' boundaries. In the Hele-Shaw geometry, $\nabla^2 {\bf v}$
is dominated by the term $\partial^2 {\bf v}/dz^2$, due to the rapid
variation of ${\bf v}$ perpendicular to the plates (the $z$-direction).
This leads to the `Darcy's Law' form of the Navier-Stokes equation
(the inertial terms can be neglected due to the frictional effect of
the boundaries):
\begin{equation}
{\bf v} = (d^2/12\eta)\,(-\nabla p - \phi\nabla\mu)\ ,
\end{equation}
where $d$ is the plate spacing and ${\bf v}(x,y)$ is now the velocity
averaged over the $z$-direction. Using the incompressibility condition
$\nabla\cdot{\bf v}=0$ to eliminate $p$ yields an equation of the form
(\ref{EQN:OSEEN}), but with the Oseen tensor replaced by its Hele-Shaw
equivalent
\begin{equation}
T^{HS}_{\alpha\beta}({\bf k}) = \frac{d^2}{12\eta}\,
\left(\delta_{\alpha\beta} - \frac{k_\alpha k_\beta}{k^2}\right)\ .
\end{equation}
Our starting point for the RG analysis is therefore equations (\ref{EQN:NS})
and (\ref{EQN:BLNOISE}), with the Oseen tensor $T$ replaced by $T^{HS}$.

Coarse-graining in the $xy$-plane, holding the plate spacing $d$ fixed,
gives the recurrence relations
\begin{eqnarray}
\lambda' & = & b^{z-3}\,\lambda \nonumber \\
\eta' & = & b^{3-z}\,\eta \nonumber \\
T' & = & b^{-1}\,T \ ,
\label{EQN:RGHS}
\end{eqnarray}
which have the same form as (\ref{EQN:BLlambda}) -- (\ref{EQN:BLT}), with
the extra factor $b^2$ in the $\eta$ equation corresponding to the extra
factor of $k^2$ in $T^{HS}$. Eqs.\ (\ref{EQN:RGHS}) still have the
LS fixed point $\lambda=\lambda^*$, with $z=3$ and $T$ irrelevant, but now
$\eta$ (or the product $\lambda\eta$) is marginal, suggesting again a
continuous family of universality classes reflecting the relative
importance of bulk diffusion and hydrodynamic flow.
In their numerical studies of critical quenches in the Hele-Shaw geometry,
Shinozaki and Oono \cite{SOHS} verify the $t^{1/3}$ growth and find that
the scaling functions do indeed depend systematically on the value of
$\lambda\eta$ (and propose essentially the same explanation).

For any off-critical quench, fluid flow along the interfaces terminates
when the droplets of minority phase become circular. Eventually, the
LS mechanism dominates the coarsening, with $L(t) \sim t^{1/3}$ still,
but unique scaling functions for a given volume fraction. In this geometry
the ratio $T/\eta$, representing the hydrodynamic noise, flows to zero at
the LS fixed point, $(T/\eta)' = b^{z-4}(T/\eta)$, so droplet
coalescence is subdominant asymptotically in time. If the transport
coefficient $\lambda$ is small enough, however, the coalescence fixed point,
with $z=4$ so that $T/\eta$ is fixed, will dominate the coarsening for a range
of times, giving $L(t) \sim (Tt/\eta)^{1/4}$.

This result may be derived heuristically by extending to the Hele-Shaw
geometry the argument given in the final paragraph of section
\ref{SEC:BINLIQS}. For droplets of size $R$ with areal number density
$n \sim v/R^2$, where $v$ is the volume fraction, the `coalescence time'
is given by the same expression, $t_c \sim R^2/vD$, as in the bulk case.
In the Hele-Shaw geometry, however, the mobility of a droplet of size
$R$ is $\mu \sim d/\eta R^2$, so the Einstein relation
$D=k_BT\mu$ for the diffusion constant gives $t_c \sim \eta R^4/vdk_BT$.
This implies a time-dependence $R \sim (vdk_BTt/\eta)^{1/4}$ for the
typical radius of a drop. Comparing this with the LS growth,
$R \sim (\lambda\sigma t/M^2)^{1/3}$, shows that the crossover from
coalescence dominated to LS dominated regimes occurs when
$R \sim vdk_BTM^2/\eta\lambda\sigma$. This length is
just $vd$ times the dimensionless crossover variable $x$ we identified in
the discussion of bulk binary liquids. For real binary liquids (as opposed
to computer simulations) we have seen that $x$ is of order unity, so the
crossover length is set by the product $vd$ of the volume fraction and
the plate spacing. Since this product is obviously less than $d$, it follows
that a $t^{1/4}$ coalescence regime (which requires $R \gg d$) should be
unobservable in real binary liquids.

\section{SUMMARY}
\label{SEC:SUMMARY}

In this article I have reviewed our current understanding of
the dynamics of phase ordering, and discussed some recent developments.
The concept of topological defects provides a unifying framework for
discussing the growth laws for the characteristic scale, and
motivates approximate treatments of the pair correlation function.

The most important consequence of the presence of topological defects
in the system is the `generalized Porod law', equation
(\ref{EQN:GENPOROD}), for the large $kL$ tail of the structure
factor. This power-law tail, whose existence has long been known
for scalar systems, has recently been observed in computer
simulations of various vector systems \cite{BBunpub,Toy91a,Toy91b}. It
should be stressed that the form of the tail depends only on the nature of
the dominant topological defects. In nematic liquid crystals, for example,
the presence of disclinations \cite{Kleman}, or `1/2-strings', implies a
structure factor tail described by (\ref{EQN:GENPOROD}) with $n=2$
\cite{BPBS}, i.e.\ a $k^{-5}$ tail for bulk systems.
This tail has been seen in simulations \cite{BB92,BPBS}, and is not
inconsistent \cite{BPBS} with experimental results \cite{WWY,WWLY}.

The Porod law (\ref{EQN:GENPOROD}), together with the scaling hypothesis,
leads to a powerful and general technique for deriving growth laws \cite{BR}.
The results are summarized in Figure 24. Again, the technique is more
general than the simple $O(n)$ models to which it has been applied here.
Nematic liquid crystals, for example, are described by the nonconserved
dynamics of a traceless, symmetric, tensor field. However, the presence
of dominant string defects implies the same growth law as for the $O(2)$
model, namely $L(t) \sim t^{1/2}$, consistent with the simulations
\cite{BB92} (allowing for the predicted logarithmic corrections to scaling)
and experiment \cite{WWY,WWLY}.

The dominant role of topological defects also motivates approximate
treatments of the pair correlation scaling function $f(x)$
\cite{BP,Toy92,LM,BH,OJK}, and the systematic treatment \cite{BHsys}
discussed in section \ref{SEC:APPROXSF}. All of these
theories lead to the same scaling function (\ref{EQN:BPT}), with the
OJK scaling function (\ref{EQN:ARCSIN}) corresponding to the special
case $n=1$. The form (\ref{EQN:BPT}) is a direct consequence of the
non-linear mapping $\vec{\phi}(\vec{m})$, with $\vec{\phi} \to \hat{m}$
for $|\vec{m}| \to \infty$, and the gaussian distribution assumed for
the field $\vec{m}$. The `OJK-type'  theories \cite{BP,Toy92,OJK,BHsys}
and the `Mazenko-type' theories \cite{LM,BH,Maz90,Maz91} differ only
in the equation for $\gamma$, the normalized pair correlation function
for $\vec{m}$.

These approximate scaling functions all give good fits to experiment and
simulation data (see, e.g., Figures 14 and 15). However, there is one
important caveat. When fitting data to theoretical scaling functions,
it is conventional to adjust the scale length $L(t)$ for the best fit.
An {\em absolute} test can, however, be obtained by calculating two
{\em different} scaling functions and plotting one against the other
\cite{BSB}. For example, the normalized correlator (\ref{EQN:C_4})
of the square of the field can also be calculated within `gaussian'
theories of the OJK or Mazenko type \cite{Bray93}. The result depends
only on $\gamma$, the normalised correlator of the gaussian auxiliary
field. Eliminating $\gamma$ between $C(12)$ and $C_4(12)$ gives an absolute
prediction for the function $C_4(C)$. When this prediction
is compared to simulation results, however, the agreement is found to
be rather poor (Figure 17): $C$ and $C_4$ can be fitted
separately, as functions of $r/L(t)$, by choosing the scale length $L(t)$
independently for each fit, but not simultaneously. However, the agreement
improves with increasing $d$, in agreement with the idea that these theories
based on a gaussian auxiliary field become exact at large $d$ \cite{BHsys}.
Including the $1/N$ correction in the systematic approach of section
\ref{SEC:SYSTEMATIC} will presumably improve the fit at fixed $d$.
Mazenko has recently introduced an alternative way of including
non-gaussian corrections \cite{Mazenko94}, and finds improved agreement
with the simulation results.

The calculation of scaling functions for {\em conserved fields} is a
significantly greater challenge, especially for scalar fields, where
even obtaining the correct $t^{1/3}$ growth law, within an approximate
theory for the pair correlation function, is not straightforward.
Mazenko has extended his approximate theory to conserved scalar fields
\cite{Maz91}, but the agreement with high quality simulation data is
not as good as for nonconserved fields \cite{Shin}. There is an additional
complication that a naive application of Mazenko's method gives $t^{1/4}$
growth, which Mazenko argues corresponds to surface diffusion only.
In order to recover the $t^{1/3}$ growth, he has to add an additional term
to incorporate the effect of bulk diffusion. For conserved vector fields,
the naive Mazenko approach gives the expected $t^{1/4}$ growth (see Figure
24), but without the logarithmic correction expected for $n=2$. The
approximate analytic treatment \cite{BH92,Rojas} (presented in section
\ref{SEC:BH92}) of the equation for $C(12)$, valid for $n \gg 1$, gives
good agreement with scaling functions extracted from simulations \cite{SR}.
A systematic approach for conserved fields, generalizing the treatment of
section \ref{SEC:SYSTEMATIC}, would be very welcome, although it is far
from straightforward. An even greater challenge is to develop good
approximate scaling functions for binary liquids.

To summarize, we have focussed on the role of topological defects as a
general way of deriving, through the Porod law (\ref{EQN:GENPOROD}) and
the scaling hypothesis (represented by equations (\ref{EQN:STRUCT}) and
(\ref{EQN:TWOTIME})) the forms of the growth laws for phase ordering in
various systems. The study of such defects also motivates, through the mapping
to an auxiliary field that varies smoothly through the defect, approximate
theories of scaling functions. For nonconserved fields, such methods are,
in principle, systematically improvable (section \ref{SEC:APPROXSF}).
One of the challenges for the future is to try to develop comparable
methods for conserved fields.

{}From a wider perspective, phase ordering dynamics is, perhaps, the simplest
example of a scaling phenomenon controlled by a `strong-coupling' RG fixed
point (Figure 4). It may not be too much to hope that techniques developed
here will find useful applications in other branches of physics.

\bigskip

\noindent{\bf Acknowledgements} \\
It is a pleasure to thank Rob Blundell, John Cardy, Jo\~{a}o Filipe,
Nigel Goldenfeld, Khurram Humayun, David Jasnow, David Huse, J\"{u}rgen
Kissner, Satya Majumdar, Gene Mazenko, Alan McKane, Mike Moore, Tim Newman,
Yoshi Oono, Sanjay Puri, Andrew Rutenberg, Andres Somoza, Neil Turok and
Martin Zapotocky for discussions. I also gratefully acknowledge the hospitality
of the Isaac Newton Institute for Mathematical Sciences, where this work
was completed.

\newpage

\begin{large}
\noindent{\bf Figure Captions} \\
\end{large}
\noindent\underline{Figure 1} Magnetization of the Ising model in zero
applied field as a function of temperature (schematic), showing spontaneous
symmetry-breaking at $T_C$. The arrow indicates a temperature quench, at
time $t=0$, from $T_I$ to $T_F$.

\medskip

\noindent\underline{Figure 2} Monte Carlo simulation of domain growth in
the $d=2$ Ising model at $T=0$ (taken from J. G. Kissner, Ph.D. thesis,
University of Manchester, 1992). The system size is 256 $\times$ 256, and
the snapshots correspond to 5, 15, 60 and 200 Monte Carlo steps per spin
after a quench from $T=\infty$.

\medskip

\noindent\underline{Figure 3} Typical form of the symmetric double-well
potential $V(\phi)$ in equation (\ref{EQN:MODELA}). The detailed
functional form of $V(\phi)$ is not important.

\medskip

\noindent\underline{Figure 4} Schematic Renormalization Group flow diagram,
with fixed points at $T=0$, $T_C$ and $\infty$. All $T>T_C$ are equivalent
to $T=\infty$ and all $T<T_C$ to $T=0$, as far as large length-scale
properties are concerned.

\medskip

\noindent\underline{Figure 5} Scaling function $f(x)$ for the pair correlation
function of the $d=2$ Ising model with nonconserved order parameter (from
reference \cite{BHlambda}). The time $t$ is the number of Monte Carlo steps
per spin.

\medskip

\noindent\underline{Figure 6} Domain-wall profile function $\phi(g)$
(schematic).

\medskip

\noindent\underline{Figure 7} Asymmetric potential $V(\phi)$ for a conserved
order parameter, showing the common-tangent construction that determines
the compositions of the separated phases.

\medskip

\noindent\underline{Figure 8} Sketch of the function $g(x)$, given by
Eq.\ (\ref{EQN:xflow}), for different $\gamma$, where $\gamma_0=4/27$.

\medskip

\noindent\underline{Figure 9} The `Mexican hat' potential $V(\vec{\phi})$
for the $O(n)$ model with $n=2$.

\medskip

\noindent\underline{Figure 10} Types of topological defect in the $O(n)$
model: (a) domain wall ($n=1$) (b) vortex ($n=2=d$) (c) string ($n=2$,
$d=3$) (d) monopole, or `hedgehog', ($n=3=d$) (e) antivortex.

\medskip

\noindent\underline{Figure 11} Cross section of $\pm 1/2$-string
configurations for a nematic liquid crystal.

\medskip

\noindent\underline{Figure 12} Scaling function $f(x)$ for the nonconserved
$d=2$ Ising model, showing Monte Carlo data (MC) from Figure 5, and the
approximations of OJK \cite{OJK} and Mazenko \cite{Maz90}. The scaling
lengths $L(t)$ for the theoretical curves were chosen to give the same
slope as the data in the linear `Porod' regime at small $x$ (from reference
\cite{BHLR}).

\medskip

\noindent\underline{Figure 13} Spatial variation (schematic) of the
order parameter $\phi$ and the auxiliary field $m$, defined by
Eq.\ (\ref{EQN:MAZENKO1}).

\medskip

\noindent\underline{Figure 14} Scaling plots for the pair correlation
function of nonconserved systems with an $O(n)$-symmetric vector
order parameter, plotted against $r\rho^{1/n}$ where $\rho$ is the
defect density (proportional to $\langle 1-\phi^2 \rangle$ for $n=1$).
The data are taken from reference \cite{BBunpub}. In (d), the length
scale $L(t)$ was chosen independently at each time to give the best
collapse. The continuous curves are `best fits by eye' of the BPT
prediction \cite{BP,Toy92}.

\medskip

\noindent\underline{Figure 15} Log-log scaling plots of the structure
factor for nonconserved systems with an $O(n)$-symmetric vector
order parameter. The data are taken from reference \cite{BBunpub}.
The continuous curves are the Fourier transforms of the corresponding
curves in Figure 14. The data exhibit the expected $k^{-(d+n)}$
tails for large $kL(t)$.

\medskip

\noindent\underline{Figure 16} Schematic forms of (a) the potential
$V_1(\phi)$and (b) the total potential $V(\phi)$ used to incorporate
external fields into the systematic approach (section
\ref{SEC:EXTERNAL-FIELDS}). The dashed lines indicate parts of the
potential that are irrelevant to the dynamics described by
Eq.\ (\ref{EQN:SYSH}).

\medskip

\noindent\underline{Figure 17} An `absolute test' for theories of
nonconserved dynamics based on an assumed gaussian auxiliary field.
Here $C$ and $C_4$ are the pair correlation functions for the
order parameter and it square, the latter normalised by its
large-distance limit (Eq.\ (\ref{EQN:C_4})). The data are for a scalar
order parameter in dimension (a) $d=2$, and (b) $d=3$. The continuous
curve (independent of $d$) is the prediction of gaussian theories
based on the OJK or Mazenko approaches. The broken lines give the
predicted short-distance behaviour (see \cite{BBunpub} and section
\ref{SEC:SHORT}).

\medskip

\noindent\underline{Figure 18} Same as Figure 17, but for vector
fields: (a) $d=2=n$, (b) $d=2$, $n=2$, and (c) $d=3=n$. The continuous
curves are the predictions of the gaussian theories.

\medskip

\noindent\underline{Figure 19(a)} Real-space pair correlation function
for a nematic liquid crystal within the `equal-constant approximation',
calculated using the `KYG' approach as described in the text. The scaling
variable $x$ is $r/\sqrt{8t}$. The data are the Monte Carlo simulations
from reference \cite{BB92}, with $L(t)$ fixed from the best fit to the
theory. Inset: short-distance behaviour of the theory, showing a leading
$x^2 \ln x$ singularity.

\medskip

\noindent\underline{Figure 19(b)} Log-log plot for the scaled structure
factor of a nematic liquid crystal. Continuous curve: the $O(2)$ theory;
data points: simulation data from reference \cite{BB92}, with $L(t)$
chosen as in Figure 19(a). Experimental data from reference \cite{WWLY}
are shown on the left, arbitrarily positioned: they can be moved left-right
and up-down. The straight line is a guide to the eye, with slope -5.

\medskip

\noindent\underline{Figure 20} Coordinate system employed for the
calculation of the amplitude of the Porod tail for $d=3$, $n=2$.

\medskip

\noindent\underline{Figure 21} Simulation data from Figure 15 replotted
to reveal the amplitude $A(n,d)$ of the Porod tail, defined by
$S({\bf k},t) \to A(n,d)\rho_{def}/k^{d+n}$ for $kL(t) \gg 1$. The
horizontal dashed lines are the prediction of Eq.\ (\ref{EQN:EXACTTAIL})
for the asymptotic limit.

\medskip

\noindent\underline{Figure 22} Relative positions of points `1' and `2'
for the small-$r$ limit of $C_4(12)$.

\medskip

\noindent\underline{Figure 23} Calculation of the short-distance limit
of $P\left(m(1),m(2)\right)$.

\medskip

\noindent\underline{Figure 24} Time-dependence of the characteristic
scale $L(t)$ for systems with purely dissipative dynamics. Exceptional
cases are discussed in the text.

\medskip

\noindent\underline{Figure 25} `Long-range' contribution to the structure
factor, for long-range correlations in the initial conditions. The circle
represents the initial condition correlator, Eq.\ (\ref{EQN:LRIC}),
while the external legs represent the response function
(\ref{EQN:RESPONSE1}).


\newpage

\begin{thebibliography}{99}
\bibitem{Lifshitz} I. M. Lifshitz, Zh.\ Eksp.\ Teor.\ Fiz.\ {\bf 42},
1354 (1962) [Sov.\ Fiz.\ JETP {\bf 15}, 939 (1962)].
\bibitem{LS} I. M. Lifshitz and V. V. Slyozov, J. Phys.\ Chem.\ Solids
{\bf 19}, 35 (1961).
\bibitem{W} C. Wagner, Z. Elektrochem.\ {\bf 65}, 581 (1961).
\bibitem{Gunton} J. D. Gunton, M. San Miguel and P. S. Sahni,
in {\em Phase Transitions and Critical Phenomena}, Vol.\ 8, eds.\ C. Domb
and J. L. Lebowitz (Academic, New York, 1983) p.267.
\bibitem{Binder} K. Binder, Rep.\ Prog.\ Phys.\ {\bf 50}, 783 (1987).
\bibitem{Furukawa} H. Furukawa, Adv.\ Phys.\ {\bf 34}, 703 (1985).
\bibitem{Langer} J. S. Langer, in {\em Solids Far From
Equilibrium}, ed. C. Godr\`{e}che (Cambridge, Cambridge, 1992).
\bibitem{HH} P. C. Hohenberg and B. I. Halperin, Rev. Mod. Phys.
{\bf 49}, 435 (1977).
\bibitem{Bray89} A. J. Bray, Phys.\ Rev.\ Lett.\ {\bf 62}, 2841 (1989).
\bibitem{Bray90} A. J. Bray, Phys.\ Rev.\ B {\bf 41}, 6724 (1990).
\bibitem{BS} K. Binder and D. Stauffer, Phys.\ Rev.\ Lett.\ {\bf 33}, 1006
(1974).
\bibitem{Marro} J. Marro, J. L. Lebowitz and M. H. Kalos, Phys.\ Rev.\ Lett.\
{\bf 43}, 282 (1979).
\bibitem{Furu78} H. Furukawa, Prog.\ Theor.\ Phys.\ {\bf 59}, 1072 (1978).
\bibitem{Furu79} H. Furukawa, Phys.\ Rev.\ Lett.\ {\bf 43}, 136 (1979).
\bibitem{Glauber} A. J. Bray, J. Phys.\ A {\bf 22}, L67 (1990);
J. G. Amar and F. Family, Phys.\ Rev.\ A {\bf 41}, 3258 (1990). See also
B. Derrida, C. Godr\`{e}che and I. Yekutieli, Phys.\ Rev.\ A {\bf 44}, 6241
(1991).
\bibitem{CZ} A. Coniglio and M. Zannetti, Europhys. Lett. {\bf 10}, 575
(1989).
\bibitem{Furukawa89} H. Furukawa, J. Phys.\ Soc.\ Jpn.\ {\bf 58}, 216
(1989).
\bibitem{Furukawa89a} H. Furukawa, Phys.\ Rev.\ B {\bf 40}, 2341 (1989).
\bibitem{FH88} D. S. Fisher and D. A. Huse, Phys.\ Rev.\ B {\bf 38}, 373
(1988). Note that our exponent $\bar{\lambda}$ is called $\lambda$ in this
paper.
\bibitem{NB90} T. J. Newman and A. J. Bray, J. Phys.\ A {\bf 23}, 4491 (1990).
\bibitem{Mason93} N. Mason, A. N. Pargellis, and B. Yurke, Phys.\ Rev.\ Lett.\
{\bf 70}, 190 (1993); for earlier work on twisted nematics see H. Orihara
and Y. Ishibashi, J. Phys.\ Soc.\ Jpn.\ {\bf 55}, 2151 (1986); T. Nagaya,
H. Orihara and Y. Ishibashi, {\em ibid.} {\bf 56}, 1898 (1987);
{\bf 56}, 3086 (1987); {\bf 59}, 377 (1990).
\bibitem{Porod51} G. Porod, Kolloid Z.\ {bf 124}, 83 (1951); {\bf 125},
51 (1952).
\bibitem{Debye} P. Debye, H. R. Anderson and H. Brumberger, J. Appl.\ Phys.\
{\bf 28}, 679 (1957); G. Porod, in {\em Small-Angle X-Ray Scattering},
edited by O. Glatter and O. Kratky (Academic, New York, 1982).
\bibitem{AC} S. M. Allen and J. W. Cahn, Acta.\ Metall.\ {\bf 27}, 1085 (1979).
\bibitem{Huse86} D. A. Huse, Phys.\ Rev.\ B {\bf 34}, 7845 (1986).
\bibitem{Sims} J. Amar, F. Sullivan and R. Mountain, Phys.\ Rev.\ B {\bf 37},
196 (1988); T. M. Rogers, K. R. Elder and R. C. Desai, Phys.\ Rev.\ B {\bf 37},
9638 (1988); R. Toral, A. Chakrabarti and J. D. Gunton, Phys.\ Rev.\ B
{\bf 39}, 4386 (1989);  C. Roland and M. Grant, Phys.\ Rev.\ B {\bf 39},
11971 (1989).
\bibitem{BRunpub} A. D. Rutenberg and A. J. Bray, submitted to Phys.\ Rev.\ E.
\bibitem{Yao} J. H. Yao, K. R. Elder, H. Guo and M. Grant, Phys.\ Rev. B
{\bf 47}, 14110 (1993).
\bibitem{RogersDesai} T. M. Rogers and R. C. Desai, Phys.\ Rev.\ B {\bf 39},
11956 (1989).
\bibitem{Marqusee} J. A. Marqusee and J. Ross, J. Chem.\ Phys.\ {\bf 80},
536 (1984).
\bibitem{Tokuyama} M. Tokuyama and K. Kawasaki, Physica {\bf 123A}, 386
(1984);  M. Tokuyama, K. Kawasaki and Y. Enomoto, {\em ibid.}
{\bf 134A}, 323 (1986); K. Kawasaki, Y. Enomoto and M. Tokuyama, {\em ibid.}
{\bf 135A}, 426 (1986); Y. Enomoto, M. Tokuyama and K. Kawasaki,
Acta Metall.\ {\bf 34}, 2119 (1986); M. Tokuyama and Y. Enomoto,
Phys.\ Rev.\ E {\bf 47}, 1156 (1993).
\bibitem{Voorhees} P. W. Voorhees and M. E. Glicksman, Acta Metall.\
{\bf 32}, 2001 (1984); P. W. Voorhees, J. Stat.\ Phys.\ {\bf 38},
231 (1985).
\bibitem{Beenakker} C. W. J. Beenakker, Phys.\ Rev.\ A {\bf 33}, 4482 (1986).
\bibitem{Ardell} A. J. Ardell, Phys.\ Rev.\ B {\bf 41}, 2554 (1990);
Acta Metall.\ {\bf 20}, 61 (1972).
\bibitem{Tsumuraya} K. Tsumuraya and Y. Miyata, Acta Metall.\ {\bf 31},
437 (1983).
\bibitem{Brailsford} A. D. Brailsford and P. Wynblatt, Acta Metall.\
{\bf 27}, 489 (1979).
\bibitem{Marder} M. Marder, Phys.\ Rev.\ Lett.\ {\bf 55}, 2953 (1985);
Phys.\ Rev.\ A {\bf 36}, 858 (1987).
\bibitem{KO} K. Kawasaki and T. Ohta, Physica {\bf 118A}, 175 (1983).
\bibitem{Ohta} T. Ohta, Ann.\ Phys.\ {\bf 158}, 31 (1984).
\bibitem{Siggia} E. D. Siggia, Phys.\ Rev.\ A {\bf 20}, 595 (1979).
\bibitem{BINLIQEXPTS} N. C. Wong and C. M. Knobler, J. Chem.\ Phys.\
{\bf 69}, 725 (1978); Phys.\ Rev.\ A {\bf 24}, 3205 (1981);
Y. C. Chou and W. I. Goldburg, Phys.\ Rev.\ A {\bf 20}, 2105 (1979);
{\bf 23}, 858 (1981).
\bibitem{BINLIQSIMS} S. Puri and B. Dunweg, Phys.\ Rev.\ A {\bf 45}, 6977
(1992); O. T. Valls and J. E. Farrell, Phys.\ Rev.\ E {\bf 47}, 36 (1993);
T. Koga and K. Kawasaki, Physica {\bf 196A}, 389 (1993).
\bibitem{ShinOono} A. Shinozaki and Y. Oono, Phys.\ Rev.\ E {\bf 48}, 2622
(1993).
\bibitem{Alexander} F. J. Alexander, S. Chen and D. W. Grunau,
Phys.\ Rev.\ B {\bf 48}, 634 (1993).
\bibitem{FuruInertial} H. Furukawa, Phys.\ Rev.\ A {\bf 31}, 1103 (1985).
\bibitem{2DBINLIQS} J. E. Farrell and O. T. Valls, Phys.\ Rev.\ B {\bf 40},
7027 (1989); {\bf 42}, 2353(1990); {\bf 43}, 630 (1991).
\bibitem{AndrewNote} I thank Andrew Rutenberg for a useful discussion of
this point.
\bibitem{Kleman} For a general discussion of topological defects, see e.g.\
M. Kl\'{e}man, {\em Points, Lines and Walls, in Liquid Crystals, Magnetic
Systems, and Various Ordered Media} (Wiley, New York, 1983).
\bibitem{Ostlund} See, for example, S. Ostlund, Phys.\ Rev.\ B {\bf 24},
485 (1981).
\bibitem{BLOGGS} A. N. Pargellis, P. Finn, J. W. Goodby, P. Pannizza,
B. Yurke and P. E. Cladis, Phys.\ Rev.\ A {\bf 46}, 7765 (1992).
\bibitem{Pargellis} B. Yurke, A. N. Pargellis, T. Kovacs and D. A. Huse,
Phys.\ Rev.\ E {\bf 47}, 1525 (1993).
\bibitem{Turok} I am grateful to N. Turok for a useful discussion of this
approach.
\bibitem{Musny} C. D. Musny and N. A. Clark, Phys.\ Rev.\ Lett.\ {\bf 68},
804 (1992); H. Pleiner, Phys.\ Rev.\ A {\bf 37}, 3986 (1988); P. E. Cladis,
W. van Sarloos, P. L. Finn and A. R. Kortan, Phys.\ Rev.\ Lett.\ {\bf 58},
222 (1987); G. Ryskin and M. Kremenetsky, Phys.\ Rev.\ Lett.\ {\bf 67},
1574 (1991); A. Pargellis, N. Turok and B. Yurke, Phys.\ Rev.\ Lett.\
{\bf 67}, 1570 (1991).
\bibitem{Pismen} L. M. Pismen and B. Y. Rubinstein, Phys.\ Rev.\ Lett.
{\bf 69}, 96 (1992).
\bibitem{Bray93} A. J. Bray, Phys.\ Rev.\ E {\bf 47}, 228 (1993).
\bibitem{BH93} A. J. Bray and K. Humayun, Phys.\ Rev.\ E {\bf 47}, R9 (1993).
\bibitem{BP} A. J. Bray and S. Puri, Phys.\ Rev.\ Lett.\ {\bf 67}, 2670 (1991).
\bibitem{Toy92} H. Toyoki, Phys.\ Rev.\ B {\bf 45}, 1965 (1992).
\bibitem{LM} Fong Liu and G. F. Mazenko, Phys.\ Rev.\ B {\bf 45}, 6989 (1992).
\bibitem{BH} A. J. Bray and K. Humayun, J. Phys. A {\bf 25}, 2191 (1992).
\bibitem{YurkeStrings} I. Chuang, R. Durrer, N. Turok and B. Yurke,
Science {\bf 251}, 1336 (1991); I. Chuang, N. Turok and B. Yurke, Phys.\
Rev.\ Lett.\ {\bf 66}, 2472 (1991); B. Yurke, A. N. Pargellis, I. Chuang
and N. Turok, Physica {\bf 178B}, 56 (1992).
\bibitem{WWY} A. P. Y. Wong, P. Wiltzius and B. Yurke, Phys.\ Rev.\ Lett.\
{\bf 68}, 3583 (1992).
\bibitem{WWLY} A. P. Y. Wong, P. Wiltzius, R. G. Larson and
B. Yurke, Phys.\ Rev.\ E {\bf 47}, 2683 (1993).
\bibitem{DeGennes} See, e.g.\ , P. G. de Gennes, {\em The Physics of
Liquid Crystals} (Clarendon, Oxford, 1974).
\bibitem{YurkePreprint} A. N. Pargellis, S. Green and B. Yurke, Phys.\ Rev.\
E {\bf 49}, 4250 (1994).
\bibitem{BPBS} A. J. Bray, S. Puri, R. E. Blundell and A. M. Somoza,
Phys.\ Rev.\ E {\bf 47}, 2261 (1993).
\bibitem{BB92} R. E. Blundell and A. J. Bray, Phys.\ Rev.\ A {\bf 46}, R6154
(1992).
\bibitem{Large-n} G. F. Mazenko and M. Zannetti, Phys.\ Rev.\ Lett.\
{\bf 53}, 2106 (1984); Phys.\ Rev.\ B {\bf 32}, 4565 (1985); M. Zannetti
and G. F. Mazenko, Phys.\ Rev.\ B {\bf 35}, 5043 (1987); F. de Pasquale
in {\em Nonequilibrium Cooperative Phenomena in Physics and Related
Topics}, p. 529, edited by M. G. Velarde (New York, Plenum, 1984);
F. de Pasquale, D. Feinberg and P. Tartaglia, Phys.\ Rev.\ B {\bf 36},
2220 (1987).
\bibitem{CRZ} A. Coniglio, P. Ruggiero and M. Zannetti, Phys.\ Rev.\ E
{\bf 50}, 1046 (1994). This paper contains a rather complete discussion
of growth kinetics in the large-$n$ limit, for both nonconserved and
conserved dynamics.
\bibitem{NBM} T. J. Newman, A. J. Bray, and M. A. Moore, Phys.\ Rev.\ B
{\bf 42} 4514, (1990).
\bibitem{Kiss93} J. G. Kissner and A. J. Bray, J. Phys.\ A {\bf 26}, 1571
(1993). Note that this paper corrects an error in reference \cite{NB90}.
\bibitem{Janssen} A similar exponent has been introduced in the context
of nonequilibrium {\em critical} dynamics by H. K. Janssen, B. Schaub
and B. Schmittman, Z. Phys.\ {\bf 73}, 539 (1989).
\bibitem{Note3} Note that the exponent $\lambda$ defined here (and in
earlier papers by the author) differs from that defined in reference
\cite{FH88} and in the papers of Mazenko: the exponent $\lambda$
defined in these papers is our $\bar{\lambda}$.
\bibitem{CZOS} A. Coniglio, Y. Oono, A. Shinozaki and M. Zannetti,
Europhys.\ Lett.\ {\bf 18}, 59 (1992).
\bibitem{MG} M. Mondello and N. Goldenfeld, Phys.\ Rev.\ E {\bf 47},
2384 (1993).
\bibitem{SR} M. Siegert and M. Rao, Phys.\ Rev.\ Lett.\ {\bf 70}, 1956 (1993).
\bibitem{CR} M. Rao and A. Chakrabarti, Phys.\ Rev.\ E {\bf 49}, 3727 (1994).
\bibitem{Kay} M. Kay, A. D. Rutenberg and A. J. Bray, unpublished.
\bibitem{BH92} A. J. Bray and K. Humayun, Phys.\ Rev.\ Lett.\ {\bf 68}, 1559
(1992).
\bibitem{Nagai} T. Nagai and K. Kawasaki, Physica {\bf 134A}, 483 (1986);
A. D. Rutenberg and A. J. Bray, Phys.\ Rev.\ E {\bf 50}, 1900 (1994).
\bibitem{BDG} A. J. Bray, B. Derrida and C. Godr\`{e}che, Europhys.\ Lett.\
{\bf 27}, 175 (1994).
\bibitem{DBG} B. Derrida, A. J. Bray and C. Godr\`{e}che, J. Phys.\ A
{\bf 27}, L357 (1994).
\bibitem{BR} A. J. Bray and A. D. Rutenberg, Phys.\ Rev.\ E {\bf 49}, R27
(1994).
\bibitem{IC} A. J. Bray, K. Humayun and T. J. Newman, Phys.\ Rev.\ B
{\bf 43}, 3699 (1991).
\bibitem{OJK} T. Ohta, D. Jasnow and K. Kawasaki, Phys.\ Rev.\ Lett.\
{\bf 49}, 1223 (1982).
\bibitem{KYG} K. Kawasaki, M. C. Yalabik and J. D. Gunton, Phys.\ Rev.\
A {\bf 17}, 455 (1978).
\bibitem{Maz89} G. F. Mazenko, Phys.\ Rev.\ Lett.\ {\bf 63}, 1605 (1989).
\bibitem{Maz90} G. F. Mazenko, Phys.\ Rev.\ B {\bf 42}, 4487 (1990).
\bibitem{Maz91} G. F. Mazenko, Phys.\ Rev.\ B {\bf 43}, 5747 (1991).
\bibitem{BHsys} A. J. Bray and K. Humayun, Phys.\ Rev.\ E {\bf 48}, 1609
(1993).
\bibitem{YeungJasnow} C. Yeung and D. Jasnow, Phys.\ Rev.\ B {\bf 42},
10523 (1990).
\bibitem{Oono88} Y. Oono and S. Puri, Mod.\ Phys.\ Lett.\ B {\bf 2}, 861
(1988).
\bibitem{Suzuki} M. Suzuki, Prog.\ Theor.\ Phys.\ {\bf 56}, 77 (1976);
{\bf 56}, 477 (1976).
\bibitem{Puri90} S. Puri and C. Roland, Phys.\ Lett.\ A {\bf 151}, 500 (1990).
\bibitem{BHLR} K. Humayun and A. J. Bray, Phys.\ Rev.\ B {\bf 46},
10594 (1992).
\bibitem{LM91} F. Liu and G. F. Mazenko, Phys.\ Rev.\ B {\bf 44}, 9185
(1991).
\bibitem{BHlambda} K. Humayun and A. J. Bray, J. Phys.\ A {\bf 24},
1915 (1991).
\bibitem{BHvector} A. J. Bray and K. Humayun, J. Phys.\ A {\bf 23},
5897 (1990).
\bibitem{LMHighLow} F. Liu and G. F. Mazenko, Phys.\ Rev. B {\bf 45},
4656 (1992).
\bibitem{Tomita} H. Tomita, Prog.\ Theor.\ Phys.\ {\bf 72}, 656 (1984);
{\bf 75}, 482 (1986).
\bibitem{YSO} C. Yeung, Y. Oono and A. Shinozaki, Phys.\ Rev.\ E {\bf 49},
2693 (1994).
\bibitem{Mazenko94} G. F. Mazenko, Phys.\ Rev.\ E {\bf 49}, 3717 (1994).
\bibitem{Note2} Eq.\ (\ref{EQN:SYSPOT}) only fixes $V(\phi)$ for
$\phi^2 \leq 1$. Note that, for $T=0$, $\phi^2({\bf x},0) \leq 1$ everywhere
implies $\phi^2({\bf x},t) \leq 1$ everywhere, so $\phi({\bf x},t)$ does not
depend on the form of $V(\phi)$ for $\phi^2 > 1$. Of course, for stability
against thermal fluctuations the points $\phi=\pm 1$ must be global minima of
$V(\phi)$.
\bibitem{Abramowitz} M. Abramowitz and I. Stegun, {\em Handbook of
Mathematical Functions} (Dover, New York, 1968).
\bibitem{BBunpub} R. E. Blundell and A. J. Bray, Phys.\ Rev.\ E {\bf 49},
4925 (1994).
\bibitem{Kissner92} A. J. Bray and J. G. Kissner, J. Phys.\ A {\bf 25},
31 (1992).
\bibitem{BSB} R. E. Blundell, A. J. Bray and S. Sattler, Phys.\ Rev.\ E
{\bf 48}, 2476 (1993).
\bibitem{CDS} Y. Oono and S. Puri, Phys.\ Rev.\ Lett.\ {\bf 58}, 836 (1987);
Phys.\ Rev.\ A {\bf 38}, 434 (1988).
\bibitem{Defect-Defect} F. Liu and G. F. Mazenko, Phys.\ Rev.\ B
{\bf 46}, 5963 (1992).
\bibitem{Nozaki} T. Ohta and H. Nozaki, in {\em Space-Time Organization in
Macromolecular Fluids}, ed.\ F. Tanaka et al.\ (Springer-Verlag, 1989).
\bibitem{Tomita93} Prog.\ Theor.\ Phys.\ {\bf 90}, 521 (1993).
\bibitem{Kramer} E. Kramer and G. F. Mazenko, unpublished.
\bibitem{Mazenko94a} G. F. Mazenko, preprint.
\bibitem{Shin} A. Shinozaki and Y. Oono, Phys.\ Rev.\ Lett.\ {\bf 66}, 173
(1991).
\bibitem{Rojas} F. Rojas Iniguez and A. J. Bray, Phys.\ Rev.\ E, in press.
\bibitem{Yeung88} C. Yeung, Phys.\ Rev.\ Lett.\ {\bf 61}, 1135 (1988).
\bibitem{Furukawa89b} H. Furukawa, Prog.\ Theor.\ Phys.\ Suppl.\
{\bf 99}, 358 (1989).
\bibitem{Tomita91} H. Tomita, Prog.\ Theor.\ Phys.\ {\bf 85}, 47 (1991).
\bibitem{Fratzl91} P. Fratzl, J. L. Lebowitz, O. Penrose and J. Amar,
Phys.\ Rev.\ B {\bf 44}, 4794 (1991).
\bibitem{Fratzl89} P. Fratzl and J. L. Lebowitz, Acta Metall.\ {\bf 37},
3245 (1989).
\bibitem{LMOffCritical} G. F. Mazenko, unpublished.
\bibitem{MZ} I am grateful to M. Zapotocky for a useful discussion of
this point.
\bibitem{Textures} A. J. Bray and S. Puri, unpublished; A. D. Rutenberg
and A. J. Bray, unpublished.
\bibitem{HumayunUnpub} K. Humayun and A. J. Bray, unpublished.
\bibitem{Toyoki94} H. Toyoki, Mod.\ Phys.\ Lett.\ B {\bf 7}, 397 (1993);
and in {\em Formation, Dynamics and Statistics of Patterns} vol.\ 2,
edited by K. Kawasaki, M. Suzuki and A. Onuki (World Scientific,
Singapore 1994).
\bibitem{Onuki} A. Onuki, Phys.\ Rev. A, {\bf 45}, 3384 (1992).
\bibitem{MG92} M. Mondello and N. Goldenfeld, Phys.\ Rev.\ A {\bf 45},
657 (1992).
\bibitem{Toy91a} H. Toyoki, J.\ Phys.\ Soc.\ Jpn.\ {\bf 60}, 1433 (1991).
\bibitem{ABLR} A. J. Bray, Phys.\ Rev.\ E {\bf 47}, 3191 (1993).
\bibitem{OnukiLR} A. Onuki, Prog.\ Theor.\ Phys.\ {\bf 74}, 1155 (1985).
\bibitem{Siegert} M. Siegert, private communication.
\bibitem{BR1d} A. D. Rutenberg and A. J. Bray, Phys.\ Rev.\ E, in press.
\bibitem{Note1} Of course, this means that the present approach will not
address systems with a potential-dependent growth law, e.g. $d=n$
for $n<2$. We also do not address quenches in which thermal noise
is essential, such as systems with static disorder (see, however,
section \ref{SEC:RGRANDOM}) or quenches to a $T>0$ critical point.
\bibitem{PottsSimsNC} C. Roland and M. Grant, Phys.\ Rev.\ B {\bf 41},
4663 (1990).
\bibitem{PottsSimsC} C. Jeppesen and O. G. Mouritsen, Phys.\ Rev.\ B
{\bf 47}, 14724 (1993).
\bibitem{LMV} Z. W. Lai, G. F. Mazenko and O. T. Valls, Phys.\ Rev.\ B
{\bf 37}, 9481 (1988).
\bibitem{MVZ} G. F. Mazenko, O. T. Valls and F. C. Zhang, Phys.\ Rev.\ B
{\bf 32}, 5807 (1985).
\bibitem{MonteCarloRG} J. Vin\~{a}ls, M. Grant, M. San Miguel, J. D. Gunton
and E. T. Gawlinski, Phys.\ Rev.\ Lett.\ {\bf 54}, 1264 (1985);
S. Kumar, J. Vin\~{a}ls and J. D. Gunton, Phys.\ Rev.\ B {\bf 34}, 1908 (1986);
C. Roland and M. Grant, Phys.\ Rev.\ Lett.\ {\bf 60}, 2657 (1988).
\bibitem{Cardy} J. L. Cardy, J. Phys.\ A {\bf 25}, 2765 (1992), has
however developed a perturbative RG treatment for a system described by
the TDGL equation with potential $V(\phi)=g\phi^4$.
\bibitem{HHM} B. I. Halperin, P. C. Hohenberg and S.-K. Ma, Phys.\ Rev.\ B,
{\bf 13}, 4119 (1976).
\bibitem{BrayRG92} A. J. Bray, Phys.\ Rev.\ Lett.\ {\bf 66}, 2048 (1991).
\bibitem{HuseHenley} D. A. Huse and C. L. Henley, Phys.\ Rev.\ Lett.\
{\bf 54}, 2708 (1985).
\bibitem{Villain} J. Villain, Phys.\ Rev.\ Lett.\ {\bf 52}, 1543 (1984).
\bibitem{Bruinsma} R. Bruinsma and G. Aeppli, Phys.\ Rev.\ Lett.\
{\bf 52}, 1543 (1984).
\bibitem{Nattermann} T. Nattermann, Phys.\ Status Solidi b {\bf 132},
125 (1985).
\bibitem{HHF} D. A. Huse, C. L. Henley and D. S. Fisher, Phys.\ Rev.\
Lett.\ {\bf 5}, 2924 (1985).
\bibitem{RBSims} G. S. Grest and D. J. Srolovitz, Phys.\ Rev.\ B {\bf 32},
3014 (1985); D. Chowdhury, M. Grant and J. D. Gunton, Phys.\ Rev.\ B
{\bf 35}, 6792 (1985); D. Chowdhury and S. Kumar, J. Stat.\ Phys.\ {\bf 49},
855 (1987); J. H. Oh and D. I. Choi, Phys.\ Rev.\ B {\bf 33}, 3448 (1986);
D. Chowdhury, J. Physique {\bf 51}, 2681 (1990).
\bibitem{BHRB} A. J. Bray and K. Humayun, J. Phys.\ A {\bf 24}, L1185 (1991).
\bibitem{PuriRB} S. Puri, D. Chowdhury and N. Parekh, J. Phys.\ A {\bf 24},
L1087 (1991).
\bibitem{Schins} A. G. Schins, A. F. M. Arts and H. W. de Wijn,
Phys.\ Rev.\ Lett.\ {\bf 70}, 2340 (1993).
\bibitem{ChakrabartiRFIM} M. Rao and A. Chakrabarti, Phys.\ Rev.\ Lett.\
{\bf 71}, 3501 (1993).
\bibitem{PuriRBC} S. Puri and N. Parekh, J. Phys.\ A {\bf 15}, 4127 (1992).
\bibitem{HayaRBC} T. Iwai and H. Hayakawa, J. Phys.\ Soc.\ Jpn.\ {\bf 62},
1583 (1993); H. Hayakawa and T. Iwai, in {\em Pattern Formation in Complex
Dissipative Systems}, edited by S. Kai (World Scientific, 1992).
\bibitem{SanMiguel} M. San Miguel, M. Grant and J. D. Gunton, Phys.\
Rev.\ A {\bf 31}, 1001 (1985).
\bibitem{SOHS} A. Shinozaki and Y. Oono, Phys.\ Rev.\ A {\bf 45}, R2161
(1992).
\bibitem{Toy91b} H. Toyoki, J.\ Phys.\ Soc.\ Jpn.\ {\bf 60}, 1153 (1991).
\end{thebibliography}
\end{document}